\documentclass[aps,prl,twocolumn,nofootinbib,notitlepage]{revtex4-2}

\usepackage{graphicx,color,tcolorbox}
\usepackage{amsmath,amsthm,amssymb,amsfonts,mathtools,latexsym,cancel,braket}
\usepackage[none]{hyphenat}
\def\red#1{{\color{red}#1}}
\def\blue#1{{\color{blue}#1}}
\usepackage[pdftex,colorlinks=true,linkcolor=Cerulean,citecolor=Cerulean,urlcolor=Cerulean]{hyperref}
\hypersetup{linkcolor=blue,citecolor=blue,urlcolor=blue}

\usepackage{array,booktabs,longtable,tabularx}
\usepackage{enumitem}
\usepackage{pifont}
\usepackage{changepage}
\usepackage{stackengine}
\stackMath

\newcommand{\norm}[1]{\left\lVert#1\right\rVert}
\DeclareFontFamily{OMS}{oasy}{\skewchar\font48}
\DeclareFontShape{OMS}{oasy}{m}{n}{
	<-5.5>		oasy5     <5.5-6.5>	oasy6
	<6.5-7.5>	oasy7     <7.5-8.5>	oasy8
	<8.5-9.5>	oasy9     <9.5->	oasy10
}{}
\DeclareFontShape{OMS}{oasy}{b}{n}{
	<-6>	oabsy5
	<6-8>	oabsy7
	<8->	oabsy10
}{}
\DeclareSymbolFont{oasy}{OMS}{oasy}{m}{n}
\SetSymbolFont{oasy}{bold}{OMS}{oasy}{b}{n}
\DeclareMathSymbol{\smallleftrightarrow}{\mathrel}{oasy}{"24}

\hypersetup
{
	pdfauthor={
		J. Enrique V\'azquez-Lozano (enrique.vazquez@unavarra.es) \&
		I\~nigo Liberal (inigo.liberal@unavarra.es)},
	pdfsubject={
		Department of Electrical, Electronic and Communications Engineering, Institute of Smart Cities (ISC), Universidad P\'ublica de Navarra (UPNA)},
	pdftitle={
		Incandescent temporal metamaterials},
	pdfkeywords={thermal radiation, time-varying media, fluctuation-dissipation theorem, macroscopic quantum electrodynamics, black-body spectrum, metamaterials}
}

\begin{document}


\title{Incandescent temporal metamaterials}
\author{J. Enrique V\'azquez-Lozano}
\email{enrique.vazquez@unavarra.es}
\affiliation{Department of Electrical, Electronic and Communications Engineering, Institute of Smart Cities (ISC), Universidad P\'ublica de Navarra (UPNA), 31006 Pamplona, Spain}
\author{I\~nigo Liberal}
\email{inigo.liberal@unavarra.es}
\affiliation{Department of Electrical, Electronic and Communications Engineering, Institute of Smart Cities (ISC), Universidad P\'ublica de Navarra (UPNA), 31006 Pamplona, Spain}

\date{\today}

\begin{abstract}
Regarded as a promising alternative to spatially shaping matter, time-varying media can be seized to control and manipulate wave phenomena, including thermal radiation. Here, based upon the framework of macroscopic quantum electrodynamics, we elaborate a comprehensive quantum theoretical formulation that lies the basis for investigating thermal emission effects in time-modulated media. Our theory unveils new physics brought about by time-varying media: nontrivial correlations between thermal fluctuating currents at different frequencies and positions, thermal radiation overcoming the black-body spectrum, and quantum vacuum amplification effects at finite temperature. We illustrate how these features lead to striking phenomena and novel thermal emitters, specifically, showing that the time-modulation releases strong field fluctuations confined within epsilon-near-zero (ENZ) bodies, and that, in turn, it enables a narrowband (partially~coherent) emission spanning the whole range of wavevectors, from near to far-field regimes.
\end{abstract}

\maketitle
\sloppy

\label{Sect.I}

\begin{figure*}[t!]
	\centering
	\includegraphics[width=0.9125\linewidth]{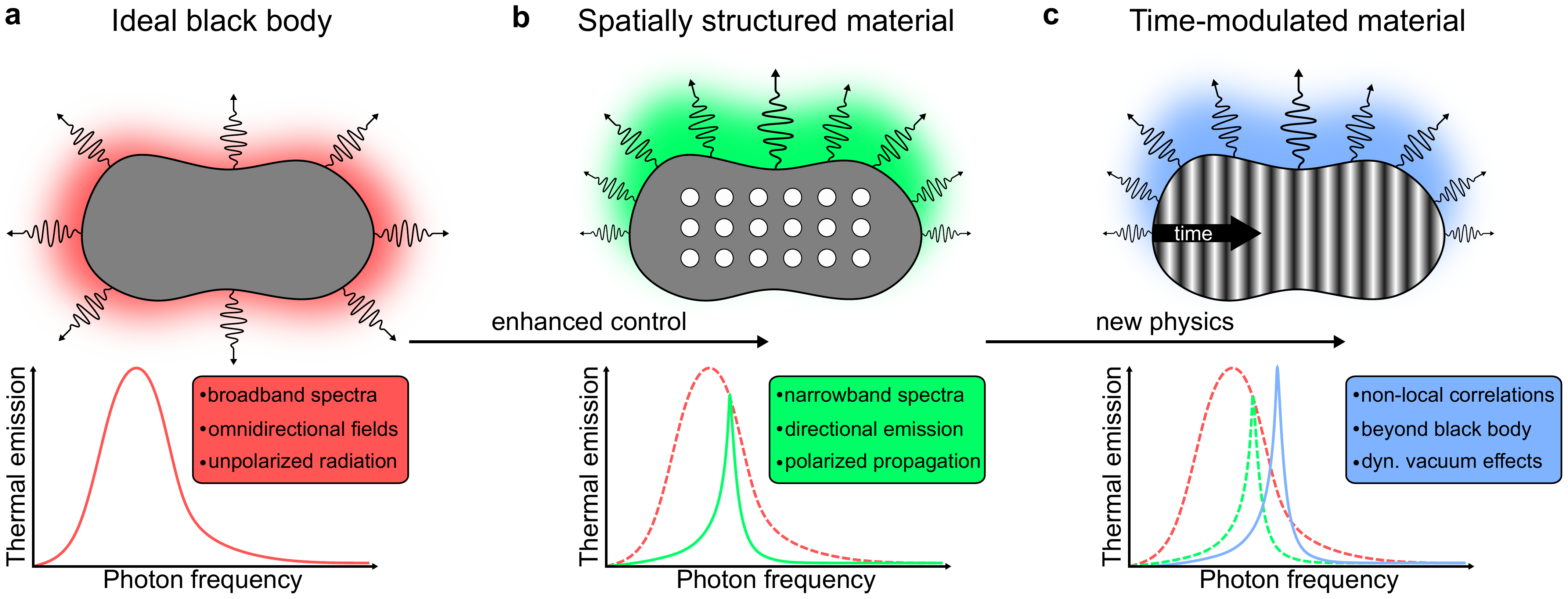}
	\caption{\textbf{Breakthroughs in thermal emission.} \textbf{a},~According to Planck's law, the broadband emission spectrum (tied to isotropic and unpolarized thermal radiation) of a black body in thermal equilibrium only depends on the temperature. \textbf{b},~Thermal emission can be controlled by structuring the matter, enabling narrowband, directive, and polarized radiation. \textbf{c},~Modulating temporally the optical properties of a medium enables more sophisticated ways to obtain similar effects, yielding the emergence of new physics, such as non-local correlations, overcoming the black-body spectrum, or dynamical vacuum effects.}
	\label{Fig.01}
\end{figure*}

On the basis of the latest scientific and technological breakthroughs in nanophotonics and material science, the development of metamaterials have brought forth an ideal playground for engineering innovative forms of light-matter interactions~\cite{Zheludev2015}. In the quest for reaching an increasing control over wave phenomena, a new burgeoning approach consists in harnessing {\em time} as an additional degree of freedom to be exploited~\cite{Engheta2021}. This revival of {\em time-varying~media}~\cite{Galiffi2022}, has in turn boosted the discovery of new physics associated with time-dependent optical phenomena, ultimately giving rise to the emerging field of {\em temporal~metamaterials}~\cite{Yin2022,Yuan2022}.

Another research area where structuring matter and shaping their optical properties is attracting a great deal of attention is the engineering of {\em thermal~emission}~\cite{Li2018,Baranov2019,Li2021B}. As a basic mechanism of heat transfer~\cite{Modest}, whereby an incandescent object at finite temperature emits (thermal) light~\cite{Boriskina2017}, thermal radiation is of fundamental interest. Likewise, it is also the basis of multiple technological applications including heat and energy management~\cite{Raman2014,Bierman2016}, light sources~\cite{Ilic2016}, sensing~\cite{Lochbaum2017}, and communications~\cite{Li2012}. In sharp contrast to the behavior of (non-thermal) light emanating from coherent sources, thermal light is characterized for displaying a broadband spectrum and isotropic field distribution, as well as for being unpolarized and temporally incoherent. Due to these properties, the control and manipulation of thermal fields has long been (and continues to be) a challenging issue. Significant efforts have been made to explore and stretch out the physical limits of thermal radiation (imposed by {\em Planck's}~\cite{Biehs2016,Hurtado2018} and {\em Kirchhoff's} radiation laws~\cite{Hadad2016,Greffet2018}) by looking into the distinctive features occurring at the nanoscale~\cite{Boriskina2016,Inoue2015,Shen2014,Ikeda2008}. Practical implementations have mainly been~based~on metamaterials~\cite{Liu2011}, metasurfaces~\cite{Liu2015}, photonic crystals~\cite{Luo2004}, or subwavelength structures such as gratings~\cite{Greffet2002}, to name a few. In this sense, besides affording a better far-field thermal emission performance (e.g., via the so-called {\em thermal~extraction~schemes}~\cite{Yu2013}), nanophotonic engineering has stimulated the unveiling of a plethora of novel {\em near-field~thermal~effects}~\cite{Cuevas2018}. Akin to customary coherent optical sources, the near-field radiation of thermal emitters greatly differs from that in the far-field regime~\cite{Carminati1999,Shchegrov2000,Joulain2005}. This is essentially due to the existence of evanescent modes, which are dominant in the near-field and negligible in the far-field~\cite{Novotny}. Apart from modifying the spectral distribution, the evanescent contribution gives access to additional channels over the frequency-wavevector~($\omega$-$k$) space, thus strengthening thermal fields by several orders of magnitude~\cite{Greffet2002,Carminati1999,Shchegrov2000,Joulain2005}.

Just like in the field of nanophotonics, it seems natural to think that passing from spatially structured to time-modulated materials could revolutionize the field of thermal engineering [see~\hyperref[Fig.01]{Fig.~1}]. A clear example comes through the grating structures~\cite{Greffet2002}, whose temporal analog could similarly open new opportunities~\cite{Galiffi2020}. Moreover, the temporal dimension owns in itself some fundamental attributes tied to the principle of causality~\cite{Solis2021}. Nonetheless, the topic of thermal emission in time-varying media is at a very incipient stage~\cite{Buddhiraju2020,Alcazar2021,Coppens2017} and the underlying physics is not fully understood as yet.

Upon this ground, here we put forward a quantum formalism to address near- and far-field thermal emission in time-varying media. Noteworthily, this theoretical formulation would also be extensible to purely quantum phenomena (e.g., Casimir forces~\cite{Gong2021} and vacuum amplification effects~\cite{Nation2012,Dodonov2020,Sloan2021}) at finite temperature. Our formalism allows us to unveil new thermal physics associated with time-modulated materials, including fluctuating currents with non-local (space and frequency) correlations, and far-field thermal emission beyond the black-body spectrum. In turn, these properties lead to novel thermal phenomena empowered by the time-modulation, such as the releasing of fluctuations trapped within a material body, or the narrowband near-to-far field thermal linking.

\section{A semiclassical approach to thermal radiation from fluctuating currents}
\label{Sect.II}

\begin{figure}[b!]
	\centering
	\includegraphics[width=0.85\linewidth]{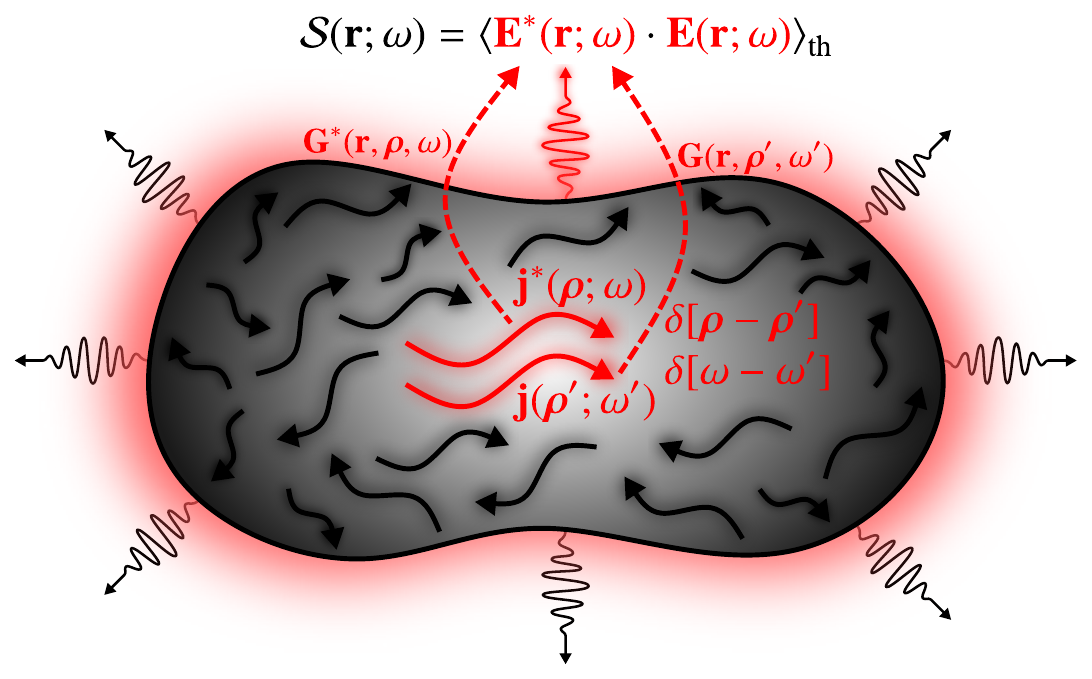}
	\caption{\textbf{Thermal emission from EM fluctuations.} Schematic representation of the fluctuating density currents moving inside a hot body. The solid red arrows represent two particular electric currents which are locally correlated, releasing so, via the corresponding Green's function (dashed red arrows), the emission of thermal fields out the material.}
	\label{Fig.02}
\end{figure}

Theoretical modeling of thermal emission is typically carried out within the framework of {\em fluctuational~electrodynamics}~\cite{Rytov}. According to this semiclassical approach, the emergence of thermal radiation emanating from a body at temperature $T$ can be understood as a result of the radiation emitted by the fluctuating electromagnetic (EM) currents (mathematically characterized by means of the current density correlations)~[see~\hyperref[Fig.02]{Fig.~2}]. The~theoretical cornerstone of this formalism is the {\em fluctuation-dissipation~theorem}~(FDT)~\cite{Nyquist1928,Callen1951,Kubo1966}, which provides with a relationship between the correlations of fluctuating systems and the linear response of the system associated to its dissipative features:
\begin{align}
\nonumber \braket{\bf{j}^*(\boldsymbol{\rho};\omega)\cdot\bf{j}(\boldsymbol{\rho}';\omega')}_{\rm th}=&
4\pi\varepsilon_0\varepsilon''(\boldsymbol{\rho},\omega)\hbar\omega^2\\
&\quad\Theta(\omega,T)
\delta{[\omega\!-\!\omega']}\delta{[\boldsymbol{\rho}\!-\!\boldsymbol{\rho}']},
\label{Eq.01}
\end{align}
where the brackets $\braket{\cdots}_{\rm th}$ denote a thermal ensemble average, $\Theta(\omega,T)=[e^{\hbar \omega/(k_BT)}-1]^{-1}$, and $\varepsilon({\bf r},\omega)=\varepsilon'({\bf r},\omega)+i\varepsilon''({\bf r},\omega)$ is the lossy and dispersive permittivity of the body. A very important feature of Eq.~\eqref{Eq.01} is that thermally fluctuating currents at different frequencies and positions are uncorrelated, describing the stochastic nature of thermal fields.

Once the fluctuating currents are known, the spectral energy density, $\mathcal{S}({\bf r};\omega)=\braket{\bf{E}^*({\bf r};\omega)\cdot\bf{E}({\bf r};\omega)}_{\rm th}$, at a given position, ${\bf r}$, and frequency, $\omega$, can be directly found from the connection between fields and currents ${\bf E}({\bf r};\omega)=i\omega\mu_0\int{d^3\boldsymbol{\rho}{\bf G}({\bf r},\boldsymbol{\rho},\omega)\cdot{\bf j}(\boldsymbol{\rho};\omega)}$ via the dyadic Green's function of the body, ${\bf G}({\bf r},\boldsymbol{\rho},\omega)$ [see~\hyperref[Fig.02]{Fig.~2}].

Fluctuational electrodynamics is a very successful theory that has made possible breakthrough advances in engineering thermal fields with photonic nanostructures. At the same time, it is a semiclassical theory that does not allow for the simultaneous modeling of quantum vacuum and thermal fluctuations, particularly in their interaction with dynamical systems. In addition, its extension and applicability to time-varying media is not rigorously justified \cite{Buddhiraju2020,Alcazar2021}. In the following, from a first-principles approach, we introduce a full-quantum formalism to thermal emission in time-varying media, which enables the calculation of quantum and thermally fluctuating current correlations, and thermal fields, without the need of any additional assumptions beyond those implicitly set in the Hamiltonian of the~system.

\section{A quantum approach to thermal emission in time-varying media}
\label{Sect.III}

To establish a general theoretical framework for extending the study of thermal emission and the behavior of fluctuating thermal currents and fields in time-varying media, we make use of {\em macroscopic~quantum~electrodynamics}~\cite{Vogel,Scheel2008}. Within this framework, instead of bare photons, one actually deals with elementary excitations, namely, EM field-matter coupled states modeled by a continuum of harmonic oscillators~\cite{Rivera2020}. These modes are described by polaritonic operators, $\hat{\bf f}({\bf r},\omega_f;t)$, which, in the Heisenberg picture, obey the equal-time commutation relations: $[\hat{\bf f}({\bf r},\omega_f;t),\hat{\bf f}({\bf r}',\omega_f';t)]=[\hat{\bf f}^\dagger({\bf r},\omega_f;t),\hat{\bf f}^\dagger({\bf r}',\omega_f';t)]=0$, and $[\hat{\bf f}({\bf r},\omega_f;t),\hat{\bf f}^\dagger({\bf r}',\omega_f';t)]=\hat{\mathbb{I}}\delta{[{\bf r}-{\bf r}']}\delta{[\omega_f-\omega_f']}$, where $\hat{\mathbb{I}}$ is the identity operator.

To describe the dynamical behavior of the time-varying quantum system we assume a perturbative approach with a Hamiltonian given by $\hat{\mathcal{H}}=\hat{\mathcal{H}}_{0}+\hat{\mathcal{H}}_{\rm T}$ \cite{Sloan2021,Vogel,Scheel2008,Rivera2020,SupplementaryInformation}, where $\hat{\mathcal{H}}_{0}$ represents the macroscopic body without time-modulation,
\begin{equation}
\hat{\mathcal{H}}_{0}=\int{d^3{\bf r}\int_{0}^{+\infty}{d\omega_f \hbar\omega_f \hat{\bf f}^\dagger({\bf r},\omega_f;t)\cdot\hat{\bf f}({\bf r},\omega_f;t) } },
\label{Eq.02}
\end{equation}
while $\hat{\mathcal{H}}_{\rm T}$ accounts for the perturbation describing the changes induced by the time-modulation,
\begin{equation}
\hat{\mathcal{H}}_{\rm T}=-\int{d^3{\bf r}\hat{\boldsymbol{\mathcal{P}}}({\bf r};t)\cdot\hat{\boldsymbol{\mathcal{E}}}({\bf r};t) }.
\label{Eq.03}
\end{equation}
Here, the polarization field operator, given by $\hat{\boldsymbol{\mathcal{P}}}({\bf r};t)\equiv\int_{0}^{t}{d\tau\Delta\chi({\bf r},t,\tau)\hat{\boldsymbol{\mathcal{E}}}({\bf r};\tau)}$, is tied to the time-varying susceptibility of the medium $\Delta\chi({\bf r},t,\tau)$~\cite{Sloan2021}, and the electric field operator $\hat{\boldsymbol{\mathcal{E}}}({\bf r};t)=\hat{\boldsymbol{\mathcal{E}}}^{(+)}({\bf r};t)+\hat{\boldsymbol{\mathcal{E}}}^{(-)}({\bf r};t)$, whose positive-frequency component reads as,
\begin{equation}
\!\!\hat{\boldsymbol{\mathcal{E}}}^{(+)}({\bf r};t)=\int_{0}^{\infty}{d\omega_{f}\int{d^{3}\boldsymbol{\rho}\,{\bf G}_{\rm E}({\bf r},\boldsymbol{\rho},\omega_f)\cdot \hat{\bf f}(\boldsymbol{\rho},\omega_f;t) } },\!\!
\label{Eq.04}
\end{equation}
with ${\bf G}_{\rm E}({\bf r},\boldsymbol{\rho},\omega_f)\!=\!i\sqrt{\hbar\varepsilon''(\boldsymbol{\rho},\omega_f)/\pi\varepsilon_0}(\omega_f/c)^2{\bf G}({\bf r},\boldsymbol{\rho},\omega_f)$ being the response function characterizing the background medium, ${\bf G}({\bf r},\boldsymbol{\rho},\omega_f)$ the dyadic Green's function for the unmodulated system, and noticing that $\hat{\boldsymbol{\mathcal{E}}}^{(-)}({\bf r};t)=[\hat{\boldsymbol{\mathcal{E}}}^{(+)}({\bf r};t)]^\dagger$. Similar perturbation Hamiltonians are adopted for modeling other nonlinear quantum processes~\cite{Loudon,Boyd}. Moreover, it is implicitly assumed that the susceptibility function, $\Delta\chi$, is small enough so that it could be regarded as a perturbation to the background structure, and does not significantly affect to the quantization procedure~\cite{Loudon}.

The emission spectrum, both in the far- and near-fields, is given by~\cite{Delga2014,Liberal2019,Mandel,Glauber1963,Mandel1965}
\begin{equation}
\mathcal{S}({\bf r};\omega)=\braket{\hat{\bf E}^{\rm (+)}({\bf r};\omega)^\dagger\cdot\hat{\bf E}^{\rm (+)}({\bf r};\omega)}_{\rm th},
\label{Eq.05}
\end{equation}
where $\hat{\bf E}^{\rm (+)}({\bf r};\omega)=\mathcal{L}_{\omega}[\hat{\boldsymbol{\mathcal{E}}}({\bf r},t)]$ is the Laplace's transform of the electric field operator. By solving the {\em Heisenberg equations of motion} for the polaritonic operators, $i\hbar\partial_t\hat{\mathcal{O}}=[\hat{\mathcal{O}},\hat{\mathcal{H}}]$, performing an integral in the complex frequency plane, and rearranging the terms~\cite{SupplementaryInformation}, $\hat{\bf E}^{\rm (+)}({\bf r};\omega)$ can be compactly written as follows:
\begin{equation}
\hat{\bf E}^{\rm (+)}({\bf r};\omega)=i\omega\mu_0 \int{d^{3}\boldsymbol{\rho}\,{\bf G}({\bf{r}},\boldsymbol{\rho},\omega)\cdot\hat{\bf j}(\boldsymbol{\rho};\omega)},
\label{Eq.06}
\end{equation}
with the current density operator,
\begin{equation}
\!\!\!\!\hat{\bf j}(\boldsymbol{\rho};\omega)\!=\!2\omega\!\left[\sqrt{\pi\hbar\varepsilon_0\varepsilon''(\boldsymbol{\rho},\omega)}\,\hat{\bf f}_0\!-\!i\mathcal{L}_{\omega}{[\Delta\tilde{\chi}(\boldsymbol{\rho},t)\hat{\boldsymbol{\mathcal{E}}}(\boldsymbol{\rho};t)]}\right]\!\!,\!\!\!\!
\label{Eq.07}
\end{equation}
where $\hat{\bf f}_0\equiv\hat{\bf f}({\bf r},\omega;t=0)$. The first term in Eq.~\eqref{Eq.07} corresponds to the currents associated with a system without time-modulation, while the second term represents the currents excited due to the time-modulation of the permittivity. In order to obtain Eq.~\eqref{Eq.07}, we have conducted a sharp but routine assumption, whereby the time-varying susceptibility exhibits a modulation which actually is local in time, i.e., $\Delta\chi({\bf r},t,\tau)=\Delta\tilde{\chi}({\bf r},t)\delta{[t-\tau]}$~\cite{Galiffi2022,Yin2022,Sloan2021,Pendry2022,Caloz2022}. This simplifies the mathematical treatment, and allows us to use the aforementioned equal-time commutation relationships. Despite this particularization, it should be noted that the formalism is completely general, and a time-modulation with an arbitrary form would be feasible with the proper adoption of time-dependent commutation relations~\cite{Vogel}.

Equations~\eqref{Eq.05}--\eqref{Eq.07} provide a quantum framework for the computation of thermal emission spectra that is conceptually similar to the semiclassical treatment sketched above: fluctuating EM currents give rise to fluctuating thermal fields by means of the corresponding propagator (the dyadic Green's function) through the entire medium. Notwithstanding, the use of a quantum formulation generalizes the semiclassical treatment, enabling the evaluation of thermal currents and fields in time-varying media, the calculation of purely quantum phenomena such as vacuum amplification effects, and, ultimately, a sound, self-consistent, and systematic formulation that directly arises from the assumptions on the Hamiltonian of the system, without the need of any semiclassical additions to the theory.

\section{Fluctuating currents in time-varying macroscopic bodies}
\label{Sect.IV}

\begin{table*}[t!]
\renewcommand*{\arraystretch}{1.5}
\caption{Fluctuation-dissipation theorem for time-varying systems: First and second-order current density correlations.}
\label{Tab.01}
\begin{tabularx}{\linewidth}{l}
\toprule
\toprule
$\braket{\hat{\bf j}_0^\dagger(\boldsymbol{\rho};\omega)\cdot\hat{\bf j}_{\vphantom{^\dagger}1}(\boldsymbol{\rho}';\omega')}_{\rm th}=\displaystyle\omega'\left[\frac{\mu_0}{\pi}\right]\int_{\mathcal{V}}{d^3\boldsymbol{\rho}''\int{d\omega''\omega''\Delta\tilde{\chi}(\boldsymbol{\rho}',\omega'-\omega''){\bf G}(\boldsymbol{\rho}',\boldsymbol{\rho}'',\omega'')\braket{\hat{\bf j}_0^\dagger(\boldsymbol{\rho};\omega)\cdot\hat{\bf j}_{\vphantom{^\dagger}0}(\boldsymbol{\rho}'';\omega'')}_{\rm th}}}$;\\[10pt]
$\braket{\hat{\bf j}_0^\dagger(\boldsymbol{\rho};\omega)\cdot\hat{\bf j}_{\vphantom{^\dagger}2}(\boldsymbol{\rho}';\omega') }_{\rm th}=\displaystyle\omega'\left[\frac{\mu_0}{\pi}\right]\int_{\mathcal{V}}{d^3\boldsymbol{\rho}''\int{d\omega''\omega''\Delta\tilde{\chi}(\boldsymbol{\rho}',\omega'-\omega''){\bf G}(\boldsymbol{\rho}',\boldsymbol{\rho}'',\omega'')\braket{\hat{\bf j}_0^\dagger(\boldsymbol{\rho};\omega)\cdot\hat{\bf j}_{\vphantom{^\dagger}1}(\boldsymbol{\rho}'';\omega'')}_{\rm th} }}$\\[10pt]
$\braket{\hat{\bf j}_1^\dagger(\boldsymbol{\rho};\omega)\cdot\hat{\bf j}_{\vphantom{^\dagger}1}(\boldsymbol{\rho}';\omega') }_{\rm th}=\displaystyle\omega\omega'\left[\frac{\mu_0}{\pi}\right]^2\iint_{\mathcal{V}}{d^3\tilde{\!\boldsymbol{\rho}}d^3\tilde{\!\boldsymbol{\rho}}'\iint{d\tilde{\omega}d\tilde{\omega}'\tilde{\omega}\tilde{\omega}'\Delta\tilde{\chi}^*(\boldsymbol{\rho},\omega-\tilde{\omega})\Delta\tilde{\chi}(\boldsymbol{\rho}',\omega'-\tilde{\omega}')}}$\\[10pt]
$\quad\quad\quad\quad\quad\quad\qquad\qquad\,\,\,\,\left[{\bf G}^*(\boldsymbol{\rho},\tilde{\!\boldsymbol{\rho}},\tilde{\omega}){\bf G}(\boldsymbol{\rho}',\tilde{\!\boldsymbol{\rho}}',\tilde{\omega}')\braket{\hat{\bf j}_0^\dagger(\tilde{\!\boldsymbol{\rho}};\tilde{\omega})\cdot\hat{\bf j}_{\vphantom{^\dagger}0}(\tilde{\!\boldsymbol{\rho}}';\tilde{\omega}')}_{\rm th}+{\bf G}(\boldsymbol{\rho},\tilde{\!\boldsymbol{\rho}},\tilde{\omega}){\bf G}^*(\boldsymbol{\rho}',\tilde{\!\boldsymbol{\rho}}',\tilde{\omega}')\braket{\hat{\bf j}_{\vphantom{^\dagger}0}(\tilde{\!\boldsymbol{\rho}};\tilde{\omega})\cdot\hat{\bf j}_0^\dagger(\tilde{\!\boldsymbol{\rho}}';\tilde{\omega}')}_{\rm th}\right]$\\
\bottomrule
\bottomrule
\end{tabularx}
\end{table*}

Next, we use this formalism to obtain a general form of the fluctuating currents excited in a macroscopic body whose permittivity is modulated in time, providing an extension to usual forms justified through the fluctuation-dissipation theorem. To this end, we first note that Eq.~\eqref{Eq.07} is an implicit equation, where the current density operator is defined as a function of the electric field operator, which is itself generated by the current density operator. This fact makes a clear signature of the sharply intertwined dynamic of the system. At any rate, such an equation may be solved iteratively leading to a solution in the form of a series expansion of current density operators at different orders~\cite{SupplementaryInformation}:
\begin{equation}
\hat{\bf j}({\bf r};\omega)=\sum_{n=0}^{\infty}{\,\hat{\bf j}_n({\bf r};\omega)}.
\label{Eq.08}
\end{equation}

Accordingly, the first three elements of the series can be explicitly written as~\cite{SupplementaryInformation}:
\begin{subequations}
\begin{align}
&\!\!\hat{\bf j}_{0}=\omega\sqrt{4\pi\hbar\varepsilon_0\varepsilon''({\bf r},\omega)}\,\hat{\bf f}({\bf r},\omega;t=0);\!\!
\label{Eq.09a}\\
\nonumber&\!\!\hat{\bf j}_{1}\propto\!\!\int{\!d^3\boldsymbol{\rho}'}\!\!\int{\!d\omega'} \omega'\Delta\tilde{\chi}({\bf r},\omega\!-\!\omega'){\bf G}({\bf r},\boldsymbol{\rho}',\omega')\hat{\bf j}_{0}(\boldsymbol{\rho}';\omega')\!\!\\
&\!\!+\!\!\int{\!d^3\boldsymbol{\rho}'}\!\!\int{\!d\omega'}\omega'\Delta\tilde{\chi}({\bf r},\omega\!-\!\omega'){\bf G}^*({\bf r},\boldsymbol{\rho}',\omega')\hat{\bf j}^\dagger_{0}(\boldsymbol{\rho}';\omega');\!\!
\label{Eq.09b}\\
&\!\!\hat{\bf j}_{2}\propto\!\!\int{\!d^3\boldsymbol{\rho}'}\!\!\int{\!d\omega'}\omega'\Delta\tilde{\chi}({\bf r},\omega\!-\!\omega'){\bf G}({\bf r},\boldsymbol{\rho}',\omega')\hat{\bf j}_{1}(\boldsymbol{\rho}';\omega').\!\!
\label{Eq.09c}
\end{align}
\end{subequations}
These expressions for the current density operators have a clear physical meaning: the current density operators of successive orders result from the fields generated by the preceding ones. Specifically, a current density operator of order $n$ at position~$\boldsymbol{\rho}'$ and frequency~$\omega'$, i.e., $\hat{\bf{j}}_n(\boldsymbol{\rho}';\omega')$, generates a field at position $\bf{r}$ via the propagation of the Green's function $\bf{G}({\bf r},\boldsymbol{\rho}',\omega')$. Roughly speaking, at such a position, the action of the field over the time-varying susceptibility, $\Delta\tilde{\chi}({\bf r},\omega\!-\!\omega')$, generates a higher-order current at frequency $\omega$, i.e., $\hat{\bf{j}}_{n+1}(\bf{r},\omega)$. As we will show, the interplay between sources of different order result in nontrivial correlations between fluctuating currents at different frequencies and points of space. Moreover, a crucial aspect of the current density operators is that they mix creation,~$\hat{\bf j}_{n}^{\dagger}$, and annihilation, $\hat{\bf j}_{n}$, polaritonic operators, akin to {\em Bogoliubov}~(or squeezing) {\em transformations}~\cite{Hopfield1958}. The difference between creation and annihilation operators, not present in semiclassical treatments, allows for the prediction of {\em dynamical vacuum effects}.

Subsequently, the correlation between fluctuating currents at different frequencies and points of space can also be written in a series form:
\begin{equation}
\braket{\hat{\bf j}^\dagger({\bf r};\omega)\cdot\hat{\bf j}({\bf r}';\omega')}_{\rm th}=\sum_{l,m}{\braket{\hat{\bf j}^\dagger_l({\bf r};\omega)\cdot\hat{\bf j}_m({\bf r}';\omega')}_{\rm th}}.
\label{Eq.10}
\end{equation}
With this in mind, the corresponding $n$th-order correlations can be obtained from a direct evaluation of $\braket{\hat{\bf j}^\dagger_l({\bf r};\omega)\cdot\hat{\bf j}_m({\bf r}';\omega')}_{\rm th}$, where $l+m=n$, and $\braket{\cdots}_{\rm th}\equiv\text{Tr}{[\cdots\hat{\rho}_{\rm th}]}$, with $\hat{\rho}_{\rm th}$ being the {\em thermal~density~operator} that yields the thermal fields at a given temperature~$T$~\cite{Vogel,Scheel2008,Loudon}.

As expected, the zeroth-order correlation of the current density, $\braket{\hat{\bf j}^\dagger_0\cdot\hat{\bf j}_{\vphantom{^\dagger}0}}_{\rm th}$, coincides with the original version of the FDT given in Eq.~\eqref{Eq.01}. In other words, our quantum formalism correctly recovers the semiclassical case for a steady (non-time-modulated) system. At the same time, it generalizes this result via the higher-order correlations. Indeed, proceeding iteratively, one can find closed form expressions for such higher-order contributions to the current density correlations~\cite{SupplementaryInformation}. The corresponding results for the first and second-order contributions are presented in the Table~\ref{Tab.01}.

Comparing this result with the original form of the FDT for stationary systems, one may realize that, even truncating the expansion at the second order, time-varying media bring new physics to fluctuational electrodynamics. First feature concerns to the breakdown of locality, since higher-order fluctuating currents are correlated at different frequencies and position of space. For conventional (non-time-modulated) thermal emitters, the correlations are local both in position and frequency. This becomes evident at a glance from the involvement of the Dirac delta functions [see Eq.~\eqref{Eq.01}], and is physically understood as a consequence of the random nature of the thermal fields. However, higher-order terms include integrals over frequencies and positions. In this manner, the time-modulation enables the possibility of correlating (or connecting) different frequencies appearing at different locations of the material system, thus underscoring its non-local character and, consequently, the potential to enhance the coherence of thermal fields.

There is an additional feature that affect to the photon distribution. Indeed, in the $\braket{\hat{\bf j}^\dagger_1\cdot\hat{\bf j}_{\vphantom{^\dagger}1}}_{\rm th}$ term, the black-body spectrum, $\Theta(\omega,T)$, appears inside a frequency integral. For conventional thermal emitters, the spectrum of thermal radiation is given by $I_{\rm real}(\omega,T)=\alpha(\omega)I_{\rm BB}(\omega,T)$, where $\alpha(\omega)=\epsilon(\omega)\leq1$ is the spectral absorptivity (related to~the~emissivity by the Kirchhoff's radiation law), and $I_{\rm BB}\propto\Theta(\omega,T)$, refers to the black-body emission spectrum. Herein, $\Theta(\omega,T)$ acts as a fixed frequency window that ultimately sets an upper limit for the radiative heat transfer. Thus, customary spatial-like nanophotonic engineering of emission spectra have so far been limited to the control of the optical absorptivity (or the emissivity) of materials~\cite{Jacob2010,Mason2011}, within the limits imposed by the black-body~spectrum~[see~\hyperref[Fig.01]{Fig.~1(b)}]. By~contrast, our analysis reveals that time-modulated thermal emitters can affect to the black-body's photon distribution, thus making tunable the accessible window of frequencies~[see~\hyperref[Fig.01]{Fig.~1(c)}]. This suggests an additional and unprecedented way to engineer the density of states.

Another crucial aspect of the $\braket{\hat{\bf j}^\dagger_1\cdot\hat{\bf j}_{\vphantom{^\dagger}1}}_{\rm th}$ term, is that it contains anti-normally ordered correlations $\braket{\hat{\bf j}_{\vphantom{^\dagger}0}\cdot\hat{\bf j}_0^\dagger}_{\rm th}$~\cite{Agarwal1975,Mandel1966}, indicating so the eventual occurrence of vacuum~amplifications effects~\cite{Nation2012}. In fact, this term is the dominant contribution in the zero-temperature limit~($T\to0$). In this sense, going beyond a semiclassical approach, our quantum formalism would allow for unifying quantum photon production effects (such as the dynamical Casimir effect \cite{Dodonov2020,Sloan2021}, parametric~amplification, ...) and thermal emission processes. In turn, it paves the way to study vacuum amplification effects at a finite temperature, which might be required for the analysis of realistic experimental configurations. Finally, our quantum formalism for fluctuating currents sets the basis for calculating quantum vacuum forces in time-varying media, such as Casimir forces~\cite{Gong2021} and quantum friction~\cite{Pendry1997,Volokitin2011,Intravaia2014,Manjavacas2010A,Manjavacas2010B,Zhao2012}.

\section{New thermal emission effects in time-varying media}
\label{Sect.V}

\begin{figure}[t!]
	\centering
	\includegraphics[width=0.85\linewidth]{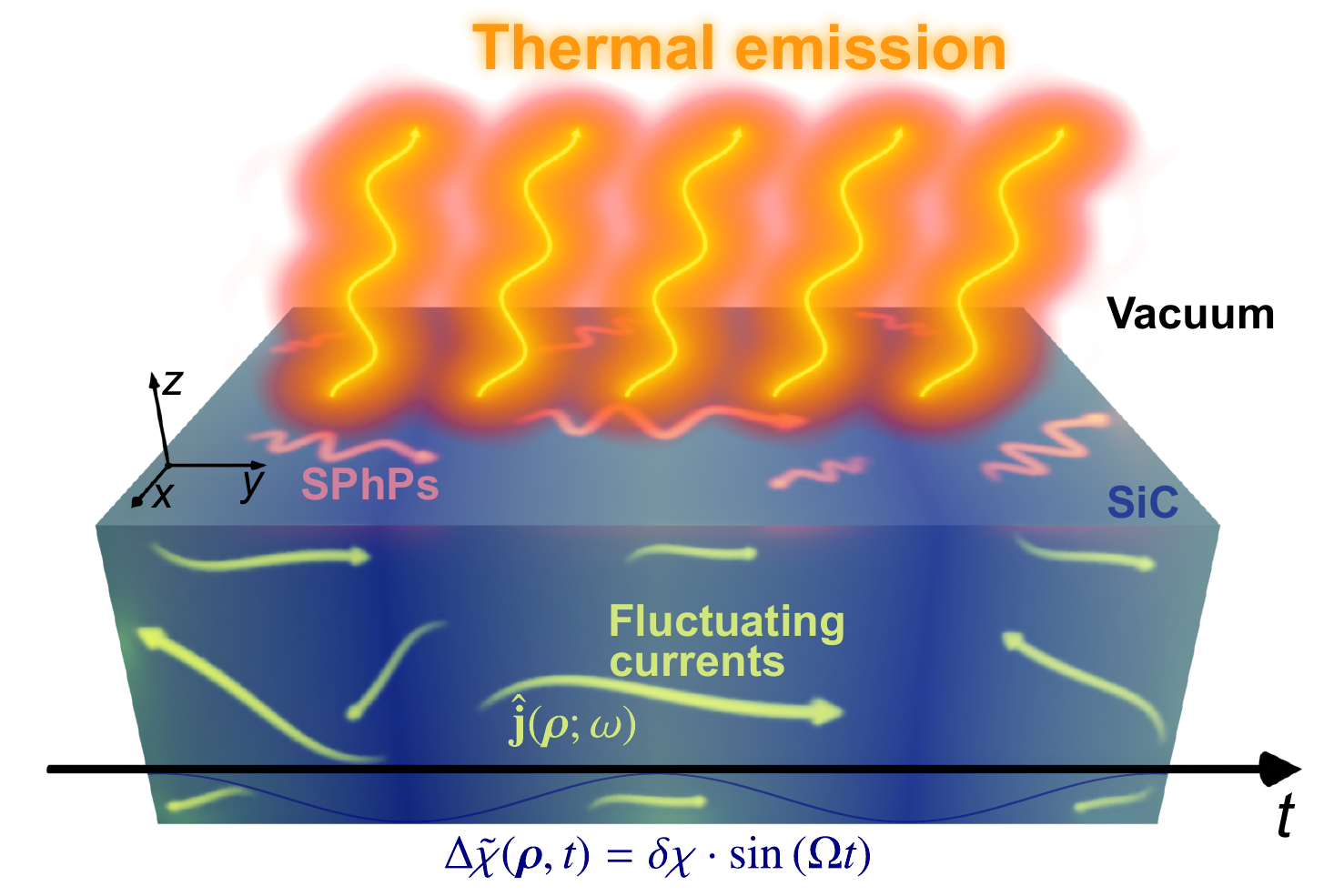}
	\caption{\textbf{Thermal emission from a semi-infinite planar slab of SiC subjected to a harmonic time-modulation.} Schematic depiction of a semi-infinite slab of SiC at temperature $T$ with a random distribution of fluctuating currents moving inside. The horizontal axis represents the time, so that each section displays a different~susceptibility.}
	\label{Fig.03}
\end{figure}

Thus far we have sketched out the theoretical model of thermally fluctuating currents and their correlations in time-varying media. Next, we illustrate some of the main consequences of such a time-modulation in connection with the emergence of new thermal emission effects. To this end, we revisit a historical example that helped initiating the field of nanophotonic engineering of thermal emission~\cite{Carminati1999,Shchegrov2000,Joulain2005}. It consists of a silicon carbide~(SiC) substrate~($z<0$), in contact with vacuum~($z>0$)~[see~\hyperref[Fig.03]{Fig.~3}]. The frequency-dependent permittivity of SiC is described by a Drude-Lorentz model, so that $\varepsilon(\omega)=\varepsilon_\infty(\omega_L^2-\omega^2-i\gamma\omega)/(\omega_T^2-\omega^2-i\gamma\omega)$, where $\varepsilon_\infty=6.7$, $\omega_L=29.1$~THz, $\omega_T=23.8$~THz, and $\gamma=0.14$~THz~\cite{Shchegrov2000}, stand, respectively, for the high-frequency-limit permittivity, the longitudinal and transverse optical phonon frequencies, and the damping factor (or characteristic collision frequency). Because of such frequency dispersion, a SiC substrate supports nontrivial far and near-field thermal fluctuations, including the thermal excitation of surface phonon polaritons. Here, instead on introducing a grating to outcouple near-field thermal waves \cite{Greffet2002}, we consider a time-harmonic modulation of the susceptibility: $\Delta\tilde{\chi}({\bf r},t)=\varepsilon_0\delta\chi\sin{\Omega t}$. As we will show, it leads to new thermal wave phenomena, associated with the release of strong epsilon-near-zero (ENZ) thermal fluctuations trapped within the material body.

\subsection{Evaluating the dyadic Green's functions}
\label{Sect.V.A}

One of the major technical difficulties in evaluating thermal emission from a time-modulated system lies in computing the product of multiple dyadic Green's functions, which represent the interaction between fluctuating currents and the radiated fields. Due to the translational symmetry of the proposed system, one can take advantage of an {\em angular~spectrum representation}~\cite{Novotny,Mandel}, whereby the dyadic Green's function is expressed~as~a superposition of plane waves: ${\bf G}({\bf r},\boldsymbol{\rho},\omega)=\frac{k_0^2}{4\pi^2}\iint_{\mathcal{V}}{d^2\boldsymbol{\kappa}_{\parallel} \hat{\bf G}(\kappa_x,\kappa_y;\omega|z,\rho_z)e^{ik_0\boldsymbol{\kappa}_{\parallel} \boldsymbol{\rho}_{\parallel}}}$, where $k_0=\omega/c$, $\boldsymbol{\kappa}_{\parallel}=(\kappa_x,\kappa_y,0)$, and $\boldsymbol{\rho}_{\parallel}=(\rho_x,\rho_y,0)$. This formalism provides with valuable physical intuition by separating the modes into propagating and evanescent, from the character of their wavevector. Indeed, for each half-space, ${\bf k}_i=k_0(\kappa_x,\kappa_y,\sqrt{\varepsilon_i(\omega)}\kappa_{z,i})$, so $k_{z,i}=k_0\tilde{k}_{z,i}=k_0\sqrt{\varepsilon_i(\omega)}\kappa_{z,i}$, where $\kappa_{z,i}=\sqrt{1-\kappa_R^2/\varepsilon_i(\omega)}$, and $\kappa_R^2=\kappa_x^2+\kappa_y^2$~\cite{Joulain2005}. Thus, for lossless media (i.e., those where $\varepsilon_i(\omega)$ is real), it is possible to set the usual correspondence of $\kappa_R^2\leq \varepsilon_i$ and $\kappa_R^2>\varepsilon_i$, with propagating and evanescent modes.

Next, note that the currents and field, linked by the Green's functions, are in general at different locations, which may be placed either in different~half-spaces, or both in the same medium (cf.~Refs.~\cite{Joulain2005,Novotny,Mandel,Sipe1987}). In the former case, the specific form of the Green's tensor relating currents in the lower half-space (made of a dispersive material, labeled as medium $2$, with permittivity $\varepsilon_2=\varepsilon(\omega)$ and coordinates $\rho_z\leq0$) to fields in the upper half-space (being the vacuum, labeled as medium $1$, with $\varepsilon_1=1$ and coordinates $z>0$), reads as
\begin{equation}
\!\!\!\hat{\bf G}_{1\leftarrow2}({\bf k};\omega|z,\rho_z)\!=\!\frac{i}{2k_{z,2}}[t^{\rm (s)}_{1\leftarrow2}\hat{\bf t}^{\rm (s)}_{1\leftarrow2}+t^{\rm (p)}_{1\leftarrow2}\hat{\bf t}^{\rm (p)}_{1\leftarrow2}]\Gamma_{1\leftarrow2},\!\!\!
\label{Eq.11}
\end{equation}
where $\hat{\bf t}^{\rm (s)}_{1\leftarrow2}=\hat{\bf s}\otimes\hat{\bf s}$ and $\hat{\bf t}^{\rm (p)}_{1\leftarrow2}=\hat{\bf p}_1^{+}\otimes\hat{\bf p}_2^{+}$ stand for the dyadic product of the normalized polarization-vector basis, being $\hat{\bf s}\equiv(\sin{\kappa_{\varphi}},-\cos{\kappa_{\varphi}},0)$ and $\hat{\bf p}_i^{\pm}\equiv(-\kappa_{z,i}\cos{\kappa_{\varphi}},-\kappa_{z,i}\sin{\kappa_{\varphi}},\pm\kappa_R/\sqrt{\varepsilon_i})$, $t^{\rm (s/p)}_{1\leftarrow2}$ are the corresponding Fresnel transmission coefficients~associated to the $s$ and $p$ polarizations~\cite{Novotny}, and $\Gamma_{1\leftarrow2}=e^{i(k_{z,1}z-k_{z,2}\rho_z)}$ is the field propagator for this particular case~\cite{SupplementaryInformation}. On the other hand, in the case of two points lying in the lower medium $2$, it follows that
\begin{equation}
\hat{\bf G}_{2\leftarrow2'}({\bf k};\omega|\rho_z,\rho_z')=\hat{\bf R}_{2\leftarrow2'}+\hat{\bf T}_{2\leftarrow2'}+\hat{\bf Z}_{2\leftarrow2'},
\label{Eq.12}
\end{equation}
with,
\begin{subequations}
	\begin{align}
	&\!\!\!\!\hat{\bf R}_{2\leftarrow2'}\!=\!\frac{i}{2k_{z,2}}[r^{\rm (s)}_{2\leftarrow1}\hat{\bf r}^{\rm (s)}_{2\leftarrow2'}\!+\!r^{\rm (p)}_{2\leftarrow1}\hat{\bf r}^{\rm (p)}_{2\leftarrow2'}]e^{-ik_{z,2}(\rho_z+\rho_z')},\!\!\!\!
	\label{Eq.13a}\\
	&\!\!\!\!\hat{\bf T}_{2\leftarrow2'}\!=\!\frac{i}{2k_{z,2}}[\hat{\bf t}^{\rm (s)}_{2\leftarrow2'}\!+\!\hat{\bf t}^{\rm (p)}_{2\leftarrow2'}]e^{-ik_{z,2}|\rho_z-\rho_z'|},\!\!\!\!
	\label{Eq.13b}\\
	&\!\!\!\!\hat{\bf Z}_{2\leftarrow2'}\!=\!-\frac{1}{k_0^2\varepsilon_2}[\hat{\bf z}\otimes\hat{\bf z}]\delta{[\rho_z-\rho_z']},\!\!\!\!
	\label{Eq.13c}
	\end{align}
\end{subequations}
where $\hat{\bf r}^{\rm (s)}_{2\leftarrow2'}=\hat{\bf t}^{\rm (s)}_{2\leftarrow2'}=\hat{\bf s}\otimes\hat{\bf s}$, $\hat{\bf r}^{\rm (p)}_{2\leftarrow2'}=\hat{\bf p}^{-}_2\otimes\hat{\bf p}^{+}_2$, $\hat{\bf t}^{\rm (p)}_{2\leftarrow2'}=\hat{\bf p}^{\mp}_2\otimes\hat{\bf p}^{\pm}_2$, with the signs $+$ and $-$ being properly arranged according to the terms appearing in the absolute value of the exponential characterizing the field propagation~\cite{SupplementaryInformation}, and $r^{\rm (s/p)}_{2\leftarrow1}$ are the Fresnel reflection coefficients~\cite{Novotny}.

These are all the ingredients needed to calculate the spectra of thermal emission of a time-modulated semi-infinite planar slab. Despite the geometrical simplicity of the system, complexity arises as a consequence of the need to consider the interactions inside the medium, which shall lead to the occurrence of nontrivial correlations between fluctuating currents.

\subsection{Zeroth-order correlations: Ground contribution}
\label{Sect.V.B}

\begin{figure}[t!]
	\centering
	\includegraphics[width=0.85\linewidth]{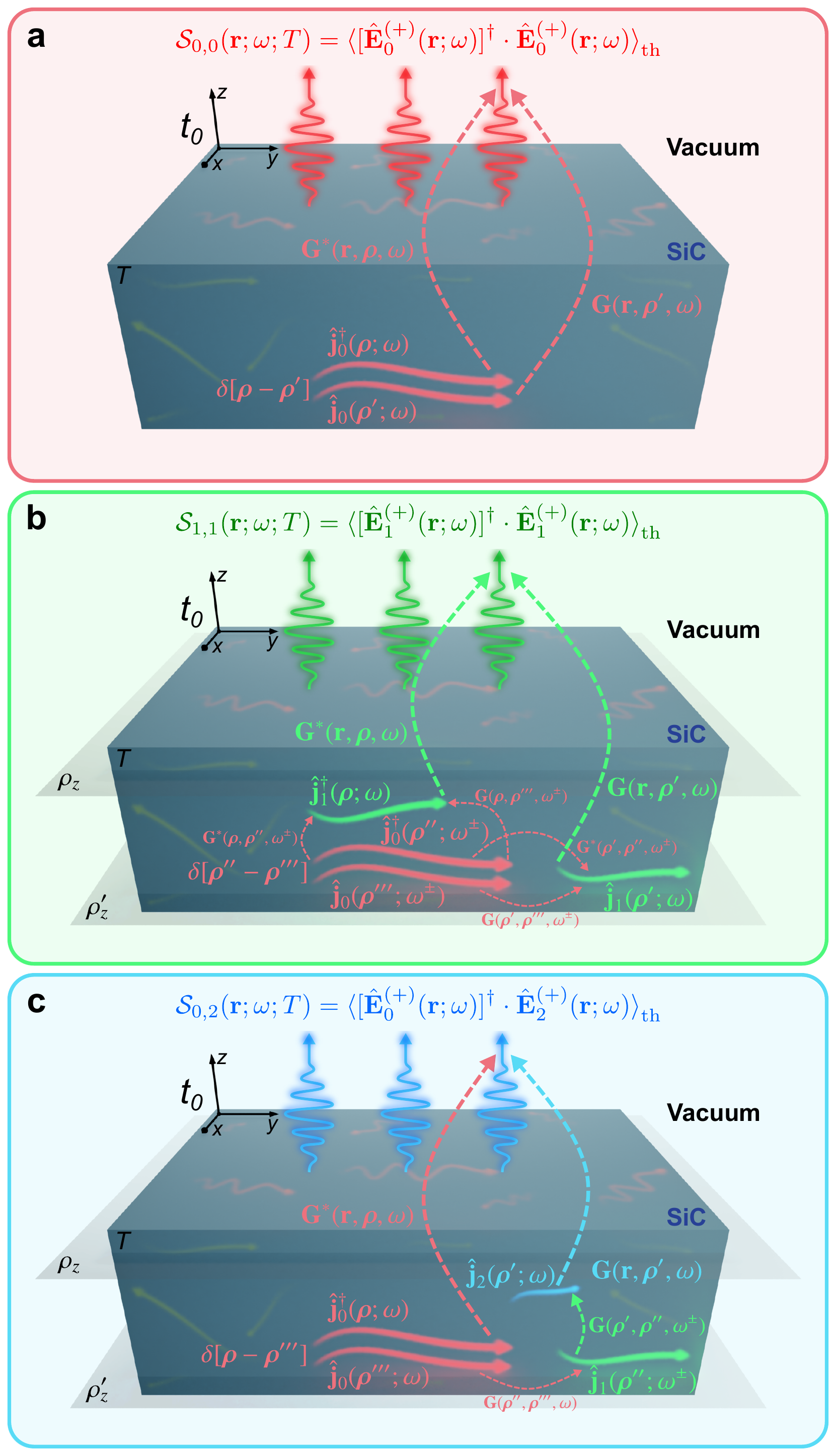}
	\caption{\textbf{Physical picture of the field correlations.} \textbf{a},~Conceptual representation of the local processes contributing to the zeroth-order of correlations ($\mathcal{S}_{0,0}$). \textbf{b},~Depiction of the non-local processes giving rise to the $\mathcal{S}_{1,1}$ contribution to the second-order of correlations. \textbf{c},~Contribution $\mathcal{S}_{0,2}$ to the second-order of correlations. The~two planes at $\rho_z$ and $\rho_z'$ separate different regions to distinguish among the possible cases that may occur. Red,~green, and blue colors represent zeroth, first, and second order features (fields, currents, or propagators), respectively.}
	\label{Fig.04}
\end{figure}

Using this mathematical formalism, the zeroth-order contribution to the thermal emission spectrum can be obtained from the corresponding electric field correlation:
\begin{equation}
\mathcal{S}_{0}({\bf r};\omega;T)=\frac{4\pi}{\varepsilon_0}\frac{\omega^3}{c^4}\varepsilon''(\omega)\hbar\omega\Theta(\omega,T)\mathcal{G}_{0,0}({\bf r},\omega),
\label{Eq.14}
\end{equation}
with
\begin{equation}
\!\!\!\!\mathcal{G}_{0,0}\!=\!\frac{k_0^2}{2\pi}\int_{-\infty}^{0}{\!\!d\rho_z\int_{0}^{+\infty}{\!\!d\kappa_R \kappa_R\norm{\hat{\bf G}_{1\leftarrow2}({\bf k};\omega|z,\rho_z)}^2_{\mathcal{F}}}},\!\!\!\!
\label{Eq.15}
\end{equation}
where the subscript $\mathcal{F}$ stands for the {\em Frobenius~norm}. As anticipated, this zeroth-order term recovers the spectrum of a steady (non-time-modulated) material~\cite{Shchegrov2000,Joulain2005}. 

\hyperref[Fig.04]{Figure~4(a)} schematically depicts a physical picture of this zeroth-order correlation-emission process. Since zeroth-order fluctuating currents are uncorrelated in space and frequency (meaning that the correlations are local, accounted by $\delta{[\boldsymbol{\rho}-\boldsymbol{\rho}']}$ and $\delta{[\omega - \omega']}$), the spectrum of thermal radiation can be reconstructed by adding the individual contributions of the fluctuating currents $\hat{\bf j}_{0}(\boldsymbol{\rho};\omega)$  at each point of space $\boldsymbol{\rho}$, for a fixed frequency~$\omega$~[see \hyperref[Fig.04]{Fig.~4(a)}]. Such a procedure is characterized by means of the (norm of the) dyadic Green's function, integrated all along the lower half-space $\rho_z\leq0$, as indicated in Eq.~\eqref{Eq.15}.

\subsection{Second-order correlations: Time-modulated term}
\label{Sect.V.C}

Taking into account that the first-order term does not contribute to the thermal emission spectrum~\cite{SupplementaryInformation}, the second-order correlations are the firsts contributions to the thermal emission spectrum yielding an explicit dependence of the time-modulation:
\begin{equation}
\mathcal{S}_2({\bf r};\omega;T)=\mathcal{S}_{1,1}({\bf r};\omega;T)+2\text{Re}{[\mathcal{S}_{0,2}({\bf r};\omega;T)]},
\label{Eq.16}
\end{equation}
noticing that $\mathcal{S}_{2,0}({\bf r};\omega;T)=[\mathcal{S}_{0,2}({\bf r};\omega;T)]^*$.

The first term in Eq.~\eqref{Eq.16} is associated with the $\braket{\hat{\bf j}^\dagger_1\cdot\hat{\bf j}_{\vphantom{^\dagger}1}}_{\rm th}$ correlation function, and is given by~\cite{SupplementaryInformation}:
\begin{equation}
\mathcal{S}_{1,1}({\bf r};\omega;T)=\mathcal{S}_{1,1}^{+}+\mathcal{S}_{1,1}^{-},
\label{Eq.17}
\end{equation}
with
\begin{equation}
\mathcal{S}_{1,1}^{\pm}=\frac{4\pi\delta\chi^2}{\varepsilon_0}\frac{\omega^5}{c^8}\varepsilon''(\omega^\pm)\hbar\omega^\pm\Theta(\omega^\pm,T)(\omega^\pm)^2\mathcal{G}_{1,1}^{(\pm)},
\label{Eq.18}
\end{equation}
and
\begin{equation}
\mathcal{G}_{1,1}^{(\pm)}\equiv(1+\tilde{\Omega})[\bar{\mathcal{G}}_{1,1}^{(\pm)}+\tilde{\mathcal{G}}_{1,1}^{(\pm)}+[\Theta(\omega^\pm,T)]^{-1}\tilde{\mathcal{G}}_{1,1}^{(\pm)}],
\label{Eq.19}
\end{equation}
Here, we have defined the shifted frequency $\omega^\pm\equiv\omega\pm\Omega$, and $\bar{\mathcal{G}}_{1,1}^{(\pm)}\!\equiv\!\iiint_{\mathcal{V}}\!d^3\boldsymbol{\rho}d^3\boldsymbol{\rho}'d^3\boldsymbol{\rho}''\text{Tr}{[{\bf G}^*({\bf r},\boldsymbol{\rho},\omega){\bf G}^*(\boldsymbol{\rho},\boldsymbol{\rho}'',\omega^\pm)}$ ${\bf G}({\bf r},\boldsymbol{\rho}',\omega){\bf G}(\boldsymbol{\rho}',\boldsymbol{\rho}'',\omega^\pm)]$, $\tilde{\mathcal{G}}_{1,1}^{(\pm)}\equiv\iiint_{\mathcal{V}}d^3\boldsymbol{\rho}d^3\boldsymbol{\rho}'d^3\boldsymbol{\rho}''$ $\text{Tr}\;{[\,{\bf G}^*({\bf r},\boldsymbol{\rho},\omega)\;{\bf G}(\boldsymbol{\rho},\boldsymbol{\rho}'',\omega^\pm)}\;{\bf G}({\bf r},\boldsymbol{\rho}',\omega)\;{\bf G}^*(\boldsymbol{\rho}',\boldsymbol{\rho}'',\omega^\pm)\,]$, and $\tilde{\Omega}\equiv\Omega/\omega$.

Similarly, the second term in Eq.\,(\ref{Eq.16}) is associated with the $\braket{\hat{\bf j}^\dagger_0\cdot\hat{\bf j}_{\vphantom{^\dagger}2}}_{\rm th}$ correlation function, and it is given by~\cite{SupplementaryInformation}:
\begin{equation}
\mathcal{S}_{0,2}({\bf r};\omega;T)=\mathcal{S}_{0,2}^{+}+\mathcal{S}_{0,2}^{-},
\label{Eq.20}
\end{equation}
where
\begin{equation}
\mathcal{S}_{0,2}^\pm=\frac{4\pi\delta\chi^2}{\varepsilon_0}\frac{\omega^5}{c^8}\varepsilon''(\omega)\hbar\omega\Theta(\omega,T)(\omega^\pm)^2\mathcal{G}_{0,2}^{(\pm)},
\label{Eq.21}
\end{equation}
with $\mathcal{G}_{0,2}^{(\pm)}\equiv\iiint_{\mathcal{V}}d^3\boldsymbol{\rho}d^3\boldsymbol{\rho}'d^3\boldsymbol{\rho}''\text{Tr}{[{\bf G}^*({\bf r},\boldsymbol{\rho},\omega){\bf G}({\bf r},\boldsymbol{\rho}',\omega)}$ ${\bf G}(\boldsymbol{\rho}',\boldsymbol{\rho}'',\omega^\pm){\bf G}(\boldsymbol{\rho}'',\boldsymbol{\rho},\omega)]$.

While the expressions for the second-order correlations are mathematically involved, they have a clear physical meaning. Indeed, such a correlation-emission process is schematically illustrated in \hyperref[Fig.04]{Fig.~4(b,c)} for each of the two contributions given in Eq.~\eqref{Eq.16}. Specifically, for the $\mathcal{S}_{1,1}({\bf r};\omega;T)$ contribution, the zeroth-order current $\hat{\bf j}_0$ at one point of space $\boldsymbol{\rho}''$ generates first-order current densities $\hat{\bf j}_1$ at different positions $\boldsymbol{\rho}$ and $\boldsymbol{\rho}'$~[see~\hyperref[Fig.04]{Fig.~4b}]. Despite emerging from fluctuating currents at different positions, the non-locality induced by time modulation allows for nontrivial correlations and a nonzero contribution for the emission spectrum.

As shown in~\hyperref[Fig.04]{Fig.~4c}, the situation is slightly different for the $\mathcal{S}_{0,2}({\bf r};\omega;T)$ contribution. In this case, there are two hopping processes, so that an original current $\hat{\bf j}_0$, induces a first-order current $\hat{\bf j}_1$, which in turn induces a second-order current $\hat{\bf j}_2$. Then, the second-order and the original current generate fields with nontrivial correlations between them. As deliberately depicted in the representations, this second-order correlations are spatially non-local. This feature is a direct consequence of the temporal modulation of the susceptibility, and means that the correlations are associated to currents that may be spatially separated from each other, which ultimately translates into an enhanced spatial coherence.

\subsection{Releasing of ENZ thermal field fluctuations trapped in the substrate}
\label{Sect.V.D}

\begin{figure*}
	\centering
	\includegraphics[width=1\linewidth]{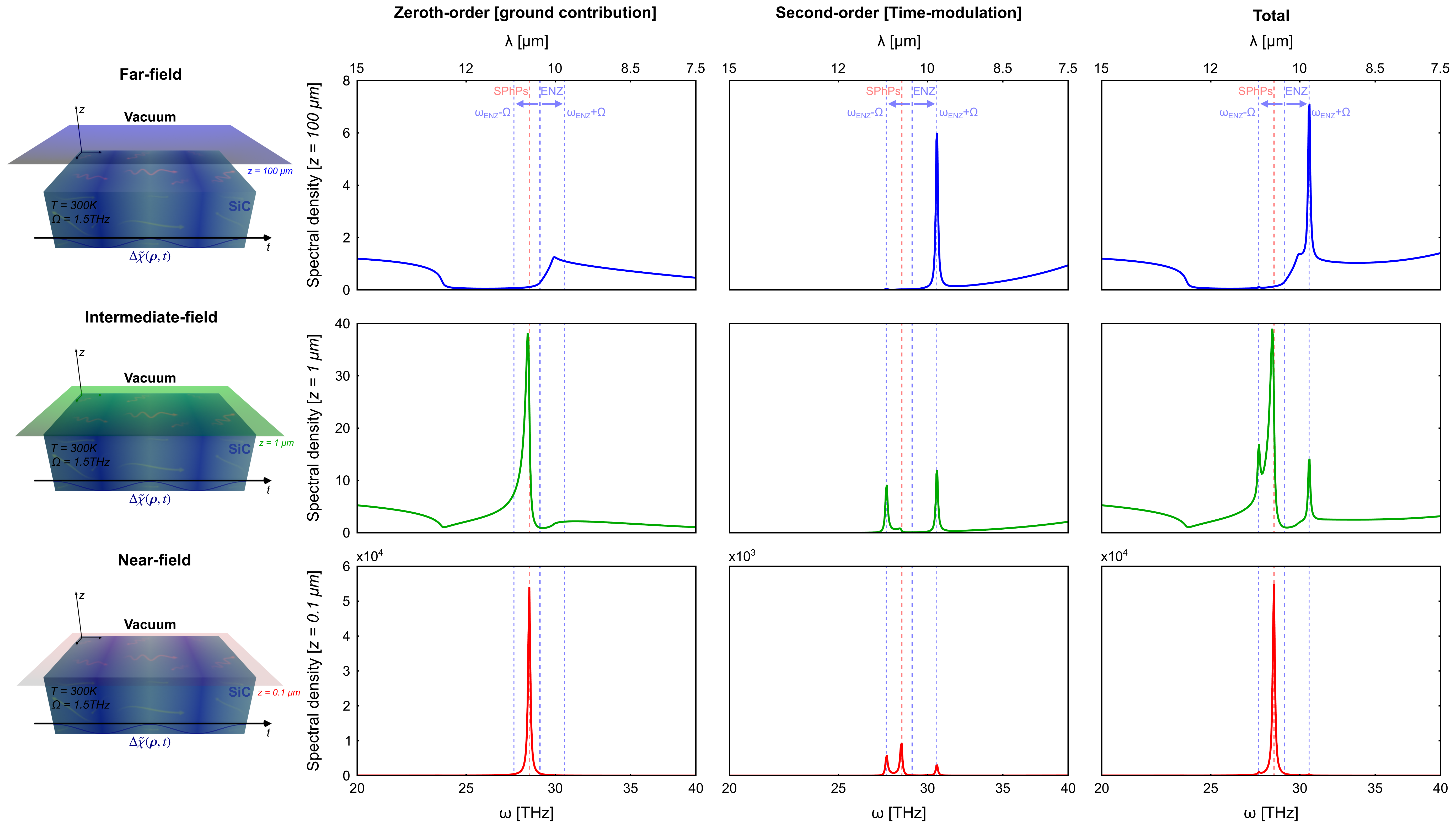}
	\caption{\textbf{Thermal emission spectra of a semi-infinite planar slab made of SiC at $\boldsymbol{\rm T=300}$ K with a time-varying susceptibility externally perturbed under a time-harmonic modulation ($\boldsymbol{\rm \Omega=1.5}$ THz and $\boldsymbol{\delta\chi=0.025}$).} The~thermal emission spectra display remarkable differences depending on the regime in which one perform the measurement: \textbf{a},~far-field ($100$ ${\rm \mu}$m); \textbf{b},~intermediate-field ($1$ ${\rm \mu}$m); \textbf{c},~near-field~($0.1$ ${\rm \mu}$m). Likewise, the time-modulation introduces new physical features, yielding an enhancement of the thermal emission spectra that appears in the epsilon-near-zero frequency~lines, noticing their shift due to the modulation. The emergence of these peaks are a direct consequence of the non-local character of the current density correlations in time-varying media, and it enables a mechanism for connecting near-to-far field radiation~effects.}
	\label{Fig.05}
\end{figure*}

Using the above theoretical apparatus, we obtain the emission spectra of the time-modulated SiC as a function of the distance $z$ above the interface. This is shown in~\hyperref[Fig.05]{Fig.~5}, where we plot separately each of the contributions associated to the zeroth and the second order, as well as the total spectrum at $z=100$~${\rm \mu}$m~(far-field), $z=1$~${\rm \mu}$m~(intermediate-field), and $z=0.1$~${\rm \mu}$m~(near-field). The zeroth-order contribution is associated to the free-evolving and non-perturbed part of the system, and, as anticipated, reproduces previous semiclassical results \cite{Shchegrov2000,Joulain2005}. In the far-zone, thermal emission is characterized by the black-body spectrum weighted over the emissivity of the substrate. Here, the main feature is a band of minimum emission associated with the Reststrahlen band of SiC~\cite{Caldwell2015}, where the permittivity is negative $\varepsilon<0$. In the near field, thermal radiation is dominated by the excitation of surface phonon polaritons (SPhPs), with a very narrow band spectrum centered at the frequency for which $\varepsilon\simeq -1$ (indicated as a dashed red line).

On the other hand, the second-order spectra are tied to the perturbed harmonic time-modulation of the susceptibility, where it has been assumed that $\Omega=1.5$~THz and $\delta\chi=0.025$. From the thermal emission spectra shown in~\hyperref[Fig.05]{Fig.~5}, we firstly highlight the appearance of a sharp peak in the far-field regime, centered at $\omega_{\rm ENZ}+\Omega$. That is, it appears at the ENZ frequency ($\omega_{\rm ENZ}$ for which $\varepsilon\simeq 0$), shifted by the temporal modulation frequency $\Omega$. Since this narrowband emission peak originates from a pure time-modulation of the substrate, it offers a greater flexibility and frequency agile capabilities than the nanofabrication engineering of thermal emitters. In addition, it constitutes a novel wave phenomenon with distinctive properties.

First, the existence of the emission peak is associated with strong thermal fluctuations trapped within the material body, which are released via the temporal modulation. This behavior is perfectly captured by our formalism. Indeed, inspecting the Green's function one~may realize that, within a same medium, $|{\bf G}({\bf r},\boldsymbol{\rho},\omega)|^{2}\to|\varepsilon''(\omega)|^{-2}$ when $\varepsilon(\omega)\to0$. Then, taking~into account that the source contribution $\braket{\hat{\bf j}_{\vphantom{^\dagger}0}\cdot\hat{\bf j}_0^\dagger}_{\rm th}\to\varepsilon''(\omega)$, it follows that the field correlation in a given medium is proportional to $|\varepsilon(\omega)|^{-1}$. This explains why thermal fields excited by fluctuating currents within~an~ENZ body are extremely strong in the limit of $\varepsilon(\omega)\to 0$~\cite{Liberal2018}. However, due to the extreme boundary conditions of ENZ media~\cite{Liberal2017A}, these fluctuations are trapped within the source's hosting, and particular resonance in the zeroth-order spectrum is found at the ENZ frequency, neither in the far nor in the near-field regimes. Again, the trapping effect of the boundary is mathematically reflected into the whole Green's function (now including the Fresnel's coefficients), which, in this case, i.e., for transitions between different media, adopts a form such that $|{\bf G}({\bf r},\boldsymbol{\rho},\omega)|^{2}\to|\varepsilon''(\omega)|^{0}$ for $\varepsilon(\omega)\to0$. Hence, when accounting also for the source's contribution, it results that the field correlations vanishes, thereby inhibiting the releasing of thermal radiation~\cite{Liberal2017B}. By contrast, when the substrate is time-modulated, the strong ENZ field fluctuations generate secondary currents at $\omega_{\rm ENZ}+\Omega$ and $\omega_{\rm ENZ}-\Omega$. For $\omega_{\rm ENZ}+\Omega$, the interface with the substrate no longer traps propagating thermal fields, and they can be observed as far-field radiation.

As aforementioned, this emission peak has some other characteristic features. For example, from the expression of the Green's function it can be drawn that the thermal fields within an ENZ body are enhanced for all $k_R$ wavenumbers. Therefore, when these fields are released via time-modulation, thermal emission is continuously enhanced from near to far-field regimes. Consequently, the thermal emission peak at $\omega_{\rm ENZ}+\Omega$ continuously exists in the near-field, intermediate-field, and far-field spectra. In addition, an additional peak appears at $\omega_{\rm ENZ}-\Omega$ for the intermediate-field and near field-spectra, since evanescent fields are allowed at that frequency. We emphasize that the significant persistence of the ENZ-induced emission peaks would allow for resonant thermal energy transfer from near to far-field.

It is also insightful to visualize this novel thermal emission effect as the dual of a spatial grating~\cite{Greffet2002}. For a spatial grating, a nanostructure fixes a given transversal wavenumber $k_R$. Then, thermal emission is observed for a continuum of frequencies supporting this wavenumber, typically scanning the direction of emission. On the contrary, the time-modulated system fixes a frequency of emission, and then narrowband thermal radiation is observed in a continuum of wavenumbers, from near to far-fields. Therefore, it represents a qualitatively different approach for engineering thermal emission.

\section{Conclusions and outlook}
\label{Sect.VI}

In this work we have elaborated a quantum theoretical formalism to address thermal emission processes in time-modulated materials. Consequentially, we have also conducted the corresponding extension of the FDT. Upon this basis, we have demonstrated the emergence of new physics associated to time-varying media, such as non-local correlations, and the stretching out of far-field thermal radiation beyond the black-body spectrum. Besides facilitating the control of thermal fluctuations, these properties increase the coherence of thermal fields, and, more importantly, give rise to novel thermal phenomena. Specifically, we have highlighted the role of the ENZ media as a genuine platform to release internal field fluctuations trapped within the boundary of a material. Further, we have underscored the permanence of such an ENZ-induced emission peak also in the near-field regime, representing a novel thermal emitter, dual to spatial gratings. While we have focused on pure time-varying media, the present approach could be extended to spatiotemporal metamaterials, likely uncovering additional thermal emission effects.

Finally, it should be noted that, although we have focused the application of the FDT on thermal emission processes, the scope of such a fundamental theorem, and hence its extension to time-varying media, is completely general, and it may concern to other non-classical phenomena, for example, non-contact friction forces, the dynamical Casimir effects, or other exciting mechanisms to amplify the quantum vacuum fluctuations. In this respect, we expect that our theoretical analysis may promote the further exploration of these features, unraveling other novel phenomena and unforeseen properties, as well as fostering the search of new venues to carry out its experimental realization.


\section{Acknowledgments}

This work was supported by ERC Starting Grant No.~ERC-2020-STG-948504-NZINATECH. 
I.L. further acknowledges support from Ram\'on y Cajal fellowship~RYC2018-024123-I and project RTI2018-093714-301J-I00 (MCIU/AEI/FEDER/UE).



\clearpage
\appendix
\onecolumngrid
\renewcommand\theequation{S\arabic{equation}}
\setcounter{equation}{0}
\renewcommand\thepage{S.\arabic{page}}
\setcounter{page}{1}
\thispagestyle{empty}
\begin{center}
	{\large \textbf{Supplementary Information:\\Incandescent temporal metamaterials}}
\end{center}

\begin{center}
	J. Enrique V\'azquez\,-Lozano$\blue{^*}$ and I\~nigo Liberal$\blue{^\dagger}$\\
	{\small\textit{
	Department of Electrical, Electronic and Communications Engineering,\\
	Institute of Smart Cities (ISC), Universidad P\'ublica de Navarra (UPNA), 31006 Pamplona, Spain}}
\end{center}
\vspace{0.25cm}
\begingroup
\par
\leftskip5.5em
\rightskip\leftskip
\small{This {\em Supplementary Information} provides a step-by-step derivation of the fluctuation-dissipation relations for time-varying quantum systems and bring out its practical application by analyzing a particular case of thermal emission. Specifically, we first elaborate on the equations of motion in the \textit{Heisenberg picture}, that naturally leads to the dynamic of the operators, and thus, to that of the fields, from which we derive a closed and analytical (though iterative) expression for the fluctuating electric field and currents. Then, just by evaluating their corresponding correlations, we find a connection to their dissipation features, thus putting forward a new form of the \textit{fluctuation-dissipation theorem for time-varying quantum systems}. Finally, we apply this formalism to the calculation of the thermal emission spectra, considering, as a particular case, a semi-infinite planar slab whose time-varying susceptibility is subjected to a harmonic modulation.}
\par
\endgroup
\vspace{-0.25cm}
\section{I. HAMILTONIAN OF A TIME\,-VARYING QUANTUM SYSTEM: A MACROSCOPIC QED APPROACH}
\label{SectSI.I}
\renewcommand\theequation{S.I.\arabic{equation}}
\setcounter{equation}{0}

We start by considering a time-varying quantum system whose dynamical behavior can be simply described by means of a perturbative Hamiltonian consisting in the sum of two contributions:
\begin{equation}
	\hat{\mathcal{H}}=\hat{\mathcal{H}}_0+\hat{\mathcal{H}}_{\rm T},
	\label{Eq.I.01}
\end{equation}
where
\setcounter{equation}{0}
\begin{subequations}
	\begin{align}
		\hat{\mathcal{H}}_0&\equiv\int{d^3{\bf r}\int_{0}^{+\infty}{d\omega_f \hbar \omega_f \hat{\bf f}^\dagger({\bf r},\omega_f;t)\cdot\hat{\bf f}({\bf r},\omega_f;t)}},
		\label{Eq.I.01a}\\
		\hat{\mathcal{H}}_{\rm T}&\equiv-\int{d^3{\bf r}\hat{\boldsymbol{\mathcal{P}}}({\bf r};t)\cdot\hat{\boldsymbol{\mathcal{E}}}({\bf r};t)}=-\int{d^3{\bf r}\left[\int_{0}^{t}{d\tau \Delta\chi({\bf r},t,\tau)\hat{\boldsymbol{\mathcal{E}}}({\bf r};\tau)}\right]\cdot\hat{\boldsymbol{\mathcal{E}}}({\bf r};t)},
		\label{Eq.I.01b}
	\end{align}
\end{subequations}
characterize, respectively, the photonic environment and the interaction term, this latter yielded by the polarization field induced by an external time-modulation~\cite{LoudonSI,BoydSI}. Notice that, for clarity, we are adopting a notation showing explicitly the time dependence of the creation/annihilation polaritonic operators (${\bf f}$ and ${\bf f}^\dagger$) \cite{VogelSI,Scheel2008SI,Delga2014SI,Liberal2019SI}. Furthermore, it is worth noticing the mathematical character of the electric and polarization vector fields, which, within the context of \textit{macroscopic~quantum~electrodynamics} (MQED) \cite{Scheel2008SI} are not functions, but quantum operators (indicated with a~hat), thus bearing well known properties such as their way to be applied over quantum states, or their, in general, non-commutative character. In this regard it should be noted that the electric field operator generally reads as:
\begin{equation}
	\hat{\boldsymbol{\mathcal{E}}}({\bf r};t)=\hat{\boldsymbol{\mathcal{E}}}^{(+)}({\bf r};t)+\hat{\boldsymbol{\mathcal{E}}}^{(-)}({\bf r};t),
	\label{Eq.I.02}
\end{equation}
with
\setcounter{equation}{1}
\begin{subequations}
	\begin{align}
		\hat{\boldsymbol{\mathcal{E}}}^{(+)}({\bf r};t)&=\int_{-\infty}^{+\infty}{d^3\boldsymbol{\rho}\int_{0}^{+\infty}{d\omega_f {\bf G}_{\rm E}({\bf r},\boldsymbol{\rho},\omega_f)\hat{\bf f}(\boldsymbol{\rho},\omega_f;t)}},
		\label{Eq.I.02a}\\
		\hat{\boldsymbol{\mathcal{E}}}^{(-)}({\bf r};t)&=\int_{-\infty}^{+\infty}{d^3\boldsymbol{\rho}\int_{0}^{+\infty}{d\omega_f {\bf G}_{\rm E}^*({\bf r},\boldsymbol{\rho},\omega_f)\hat{\bf f}^\dagger(\boldsymbol{\rho},\omega_f;t)}},
		\label{Eq.I.02b}
	\end{align}
\end{subequations}
being the positive- and negative-frequency components, and
\begin{equation}
	{\bf G}_{\rm E}({\bf r},\boldsymbol{\rho},\omega_f)\equiv i\sqrt{\frac{\hbar}{\pi\varepsilon_0}}\left(\frac{\omega_f}{c}\right)^2\sqrt{\text{Im}{\left[\varepsilon(\boldsymbol{\rho},\omega_f)\right]}}{\bf G}({\bf r},\boldsymbol{\rho},\omega_f),
	\label{Eq.I.03}
\end{equation}
the response function, depending in turn on the \textit{dyadic Green's function}, ${\bf G}({\bf r},\boldsymbol{\rho},\omega_f)$, characterizing the medium. Therefore, the non-commutativity of the electric field operator is underpinned by the \textit{equal-time commutation relations} of the polaritonic operators \cite{VogelSI,Scheel2008SI,LoudonSI}:
\begin{subequations}
	\begin{align}
		&\left[\hat{\bf f}({\bf r},\omega_f;t),\hat{\bf f}({\bf r}',\omega_f';t)\right]=\left[\hat{\bf f}^\dagger({\bf r},\omega_f;t),\hat{\bf f}^\dagger({\bf r}',\omega_f';t)\right]=0,
		\label{Eq.I.04a}\\
		&\left[\hat{\bf f}({\bf r},\omega_f;t),\hat{\bf f}^\dagger({\bf r}',\omega_f';t)\right]=\mathbb{I}\delta{\left[{\bf r}-{\bf r}'\right]}\delta{\left[\omega_f-\omega_f'\right]}.
		\label{Eq.I.04b}
	\end{align}
\end{subequations}

\section{II. HEISENBERG EQUATIONS OF MOTION OF THE POLARITONIC OPERATORS}
\label{SectSI.II}
\renewcommand\theequation{S.II.\arabic{equation}}
\setcounter{equation}{0}

As it has already been anticipated (by emphasizing the time dependence into the operators), throughout this work the analysis shall be carried out within the \textit{Heisenberg picture}. It should be noted that introducing the time dependence into the states (i.e., adopting the \textit{Schr\"odinger picture}) is not an easy task at all, since it would require for additional knowledge (or sharp assumptions) as for the dynamical regime of the system. On the contrary, under the Heisenberg picture, the dynamical behavior of the (operators characterizing the) system directly stems from the \textit{Heisenberg equation of motion}:
\begin{equation}
	i\hbar\partial_t\hat{\mathcal{O}}\equiv\left[\hat{\mathcal{O}},\hat{\mathcal{H}}\right].
	\label{Eq.II.01}
\end{equation}
Considering separately the two contributions of the Hamiltonian given in Eq.~\eqref{Eq.I.01}, it follows that:
\begin{equation}
	i\hbar\partial_t\hat{\bf f}({\bf r},\omega_f;t)=\underbracket{i\hbar\partial_t\hat{\bf f}_0({\bf r},\omega_f;t)\vphantom{\left[\hat{\bf f}({\bf r},\omega_f;t),\hat{\mathcal{H}}\right]}}_{\text{free-evolving}}+\underbracket{i\hbar\partial_t\hat{\bf f}_{\rm T}({\bf r},\omega_f;t)\vphantom{\left[\hat{\bf f}({\bf r},\omega_f;t),\hat{\mathcal{H}}\right]}}_{\text{time-modulated}}=\left[\hat{\bf f}({\bf r},\omega_f;t),\hat{\mathcal{H}}\right],
	\label{Eq.II.02}
\end{equation}
where
\begin{align*}
	&\left[\hat{\bf f}({\bf r},\omega_f;t),\hat{\mathcal{H}}_0\right]=\hat{\bf f}({\bf r},\omega_f;t)\hat{\mathcal{H}}_0-\hat{\mathcal{H}}_0\hat{\bf f}({\bf r},\omega_f;t)\\
	&=\hat{\bf f}({\bf r},\omega_f;t)\left\{\int{d^3{\bf r}'\int\limits_{0}^{+\infty}{d\omega_f' \hbar \omega_f' \hat{\bf f}^\dagger({\bf r}',\omega_f';t)\hat{\bf f}({\bf r}',\omega_f';t)}}\right\}-\left\{\int{d^3{\bf r}'\int\limits_{0}^{+\infty}{d\omega_f' \hbar \omega_f' \hat{\bf f}^\dagger({\bf r}',\omega_f';t)\hat{\bf f}({\bf r}',\omega_f';t)}}\right\}\hat{\bf f}({\bf r},\omega_f;t)\\
	&=\int{d^3{\bf r}'\int\limits_{0}^{+\infty}{d\omega_f' \hbar \omega_f' \left[\hat{\bf f}({\bf r},\omega_f;t)\hat{\bf f}^\dagger({\bf r}',\omega_f';t)\right]\hat{\bf f}({\bf r}',\omega_f';t)}}-\int{d^3{\bf r}'\int\limits_{0}^{+\infty}{d\omega_f' \hbar \omega_f' \left[\hat{\bf f}^\dagger({\bf r}',\omega_f';t)\hat{\bf f}({\bf r}',\omega_f';t)\right]\hat{\bf f}({\bf r},\omega_f;t)}}\\
	&=\hbar\omega_f\hat{\bf f}({\bf r},\omega_f;t)+\int{d^3{\bf r}'\int\limits_{0}^{+\infty}{d\omega_f' \hbar \omega_f' \left[\hat{\bf f}^\dagger({\bf r}',\omega_f';t)\hat{\bf f}({\bf r}',\omega_f';t)\right]\hat{\bf f}({\bf r},\omega_f;t)}}-\int{d^3{\bf r}'\int\limits_{0}^{+\infty}{d\omega_f' \hbar \omega_f' \left[\hat{\bf f}^\dagger({\bf r}',\omega_f';t)\hat{\bf f}({\bf r}',\omega_f';t)\right]\hat{\bf f}({\bf r},\omega_f;t)}},
\end{align*}
thus leading to:
\begin{tcolorbox}[sharp corners]
	\vspace{-0.35cm}
	\begin{equation}
		\left[\hat{\bf f}({\bf r},\omega_f;t),\hat{\mathcal{H}}_0\right]=\hat{\bf f}({\bf r},\omega_f;t)\hat{\mathcal{H}}_0-\hat{\mathcal{H}}_0\hat{\bf f}({\bf r},\omega_f;t)=\hbar\omega_f\hat{\bf f}({\bf r},\omega_f;t).
		\label{Eq.II.03}
	\end{equation}
\end{tcolorbox}
\noindent Proceeding similarly with the creation operator it follows that:
\begin{tcolorbox}[sharp corners]
	\vspace{-0.35cm}
	\begin{equation}
		\left[\hat{\bf f}^\dagger({\bf r},\omega_f;t),\hat{\mathcal{H}}_0\right]=\hat{\bf f}^\dagger({\bf r},\omega_f;t)\hat{\mathcal{H}}_0-\hat{\mathcal{H}}_0\hat{\bf f}^\dagger({\bf r},\omega_f;t)=-\hbar\omega_f\hat{\bf f}^\dagger({\bf r},\omega_f;t).
		\label{Eq.II.04}
	\end{equation}
\end{tcolorbox}
\noindent So, in absence of time-modulation, the corresponding Heisenberg equations of motion for the polaritonic operators~are:
\begin{tcolorbox}[sharp corners,colback=blue!5!white,colframe=blue!50!white]
	\vspace{-0.35cm}
	\begin{subequations}
		\begin{align}
			&\partial_t\hat{\bf f}_0({\bf r},\omega_f;t)=-i\omega_f\hat{\bf f}_0({\bf r},\omega_f;t),
			\label{Eq.II.05a}\\
			&\partial_t\hat{\bf f}_0^\dagger({\bf r},\omega_f;t)=i\omega_f\hat{\bf f}_0^\dagger({\bf r},\omega_f;t).
			\label{Eq.II.05b}
		\end{align}
	\end{subequations}
\end{tcolorbox}

On the other side, the dynamic of the operators due to the time-modulation is given by the interaction Hamiltonian:
\begin{align*}
	&\left[\hat{\bf f}({\bf r},\omega_f;t),\hat{\mathcal{H}}_{\rm T}\right]=\hat{\bf f}({\bf r},\omega_f;t)\hat{\mathcal{H}}_{\rm T}-\hat{\mathcal{H}}_{\rm T}\hat{\bf f}({\bf r},\omega_f;t)\\
	&=\left\{\int{d^3{\bf r}'\left[\int\limits_{0}^{t}{d\tau \Delta\chi({\bf r}',t,\tau)\hat{\boldsymbol{\mathcal{E}}}({\bf r}',\tau)}\right]\cdot\hat{\boldsymbol{\mathcal{E}}}({\bf r}',t)}\right\}\hat{\bf f}({\bf r},\omega_f;t)-\hat{\bf f}({\bf r},\omega_f;t)\left\{\int{d^3{\bf r}'\left[\int\limits_{0}^{t}{d\tau \Delta\chi({\bf r}',t,\tau)\hat{\boldsymbol{\mathcal{E}}}({\bf r}',\tau)}\right]\cdot\hat{\boldsymbol{\mathcal{E}}}({\bf r}',t)}\right\}\\
	&=\left\{\int{d^3{\bf r}'\int\limits_{0}^{t}{d\tau \Delta\chi({\bf r}',t,\tau)\hat{\boldsymbol{\mathcal{E}}}({\bf r}',\tau)}\cdot\hat{\boldsymbol{\mathcal{E}}}({\bf r}',t)}\right\}\hat{\bf f}({\bf r},\omega_f;t)-\hat{\bf f}({\bf r},\omega_f;t)\left\{\int{d^3{\bf r}'\int\limits_{0}^{t}{d\tau \Delta\chi({\bf r}',t,\tau)\hat{\boldsymbol{\mathcal{E}}}({\bf r}',\tau)}\cdot\hat{\boldsymbol{\mathcal{E}}}({\bf r}',t)}\right\}\\
\end{align*}
\begin{align*}
	&=\int{d^3{\bf r}'\int\limits_{0}^{t}{d\tau \Delta\chi({\bf r}',t,\tau)\left[\int{d^3\boldsymbol{\rho}'\int\limits_{0}^{+\infty}{d\omega_f'\left[{\bf G}_{\rm E}({\bf r}',\boldsymbol{\rho}',\omega_f')\hat{\bf f}(\boldsymbol{\rho}',\omega_f';\tau)+{\bf G}^*_{\rm E}({\bf r}',\boldsymbol{\rho}',\omega_f')\hat{\bf f}^\dagger(\boldsymbol{\rho}',\omega_f';\tau)\right]}}\right]}}\\
	&\qquad\times\left[\int{d^3\boldsymbol{\rho}''\int\limits_{0}^{+\infty}{d\omega_f''\left[{\bf G}_{\rm E}({\bf r}',\boldsymbol{\rho}'',\omega_f'')\hat{\bf f}(\boldsymbol{\rho}'',\omega_f'';t)+{\bf G}^*_{\rm E}({\bf r}',\boldsymbol{\rho}'',\omega_f'')\hat{\bf f}^\dagger(\boldsymbol{\rho}'',\omega_f'';t)\right]}}\right]\hat{\bf f}({\bf r},\omega_f;t)\\
	&\quad-\hat{\bf f}({\bf r},\omega_f;t)\int{d^3{\bf r}'\int\limits_{0}^{t}{d\tau \Delta\chi({\bf r}',t,\tau)\left[\int{d^3\boldsymbol{\rho}'\int\limits_{0}^{+\infty}{d\omega_f'\left[{\bf G}_{\rm E}({\bf r}',\boldsymbol{\rho}',\omega_f')\hat{\bf f}(\boldsymbol{\rho}',\omega_f';\tau)+{\bf G}^*_{\rm E}({\bf r}',\boldsymbol{\rho}',\omega_f')\hat{\bf f}^\dagger(\boldsymbol{\rho}',\omega_f';\tau)\right]}}\right]}}\\
	&\qquad\times\left[\int{d^3\boldsymbol{\rho}''\int\limits_{0}^{+\infty}{d\omega_f''\left[{\bf G}_{\rm E}({\bf r}',\boldsymbol{\rho}'',\omega_f'')\hat{\bf f}(\boldsymbol{\rho}'',\omega_f'';t)+{\bf G}^*_{\rm E}({\bf r}',\boldsymbol{\rho}'',\omega_f'')\hat{\bf f}^\dagger(\boldsymbol{\rho}'',\omega_f'';t)\right]}}\right]\\
	&=\iiint{d^3{\bf r}'d^3\boldsymbol{\rho}'d^3\boldsymbol{\rho}''\iint\limits_{0}^{\quad+\infty}{d\omega_f'd\omega_f''\int\limits_{0}^{t}{d\tau \Delta\chi({\bf r}',t,\tau)}}}\\
	&\!\!\!\left\{\left[{\bf G}_{\rm E}({\bf r}',\boldsymbol{\rho}',\omega_f'){\bf G}_{\rm E}({\bf r}',\boldsymbol{\rho}'',\omega_f'')\hat{\bf f}(\boldsymbol{\rho}',\omega_f';\tau)\hat{\bf f}(\boldsymbol{\rho}'',\omega_f'';t)\hat{\bf f}({\bf r},\omega_f;t)\!+\!{\bf G}^*_{\rm E}({\bf r}',\boldsymbol{\rho}',\omega_f'){\bf G}^*_{\rm E}({\bf r}',\boldsymbol{\rho}'',\omega_f'')\hat{\bf f}^\dagger(\boldsymbol{\rho}',\omega_f';\tau)\hat{\bf f}^\dagger(\boldsymbol{\rho}'',\omega_f'';t)\hat{\bf f}({\bf r},\omega_f;t)\right.\right.\\
	&\!\!+\!\left.{\bf G}_{\rm E}({\bf r}',\boldsymbol{\rho}',\omega_f'){\bf G}^*_{\rm E}({\bf r}',\boldsymbol{\rho}'',\omega_f'')\hat{\bf f}(\boldsymbol{\rho}',\omega_f';\tau)\hat{\bf f}^\dagger(\boldsymbol{\rho}'',\omega_f'';t)\hat{\bf f}({\bf r},\omega_f;t)\!+\!{\bf G}^*_{\rm E}({\bf r}',\boldsymbol{\rho}',\omega_f'){\bf G}_{\rm E}({\bf r}',\boldsymbol{\rho}'',\omega_f'')\hat{\bf f}^\dagger(\boldsymbol{\rho}',\omega_f';\tau)\hat{\bf f}(\boldsymbol{\rho}'',\omega_f'';t)\hat{\bf f}({\bf r},\omega_f;t)\right]\\
	&\!\!-\!\left[{\bf G}_{\rm E}({\bf r}',\boldsymbol{\rho}',\omega_f'){\bf G}_{\rm E}({\bf r}',\boldsymbol{\rho}'',\omega_f'')\hat{\bf f}({\bf r},\omega_f;t)\hat{\bf f}(\boldsymbol{\rho}',\omega_f';\tau)\hat{\bf f}(\boldsymbol{\rho}'',\omega_f'';t)\!+\!{\bf G}^*_{\rm E}({\bf r}',\boldsymbol{\rho}',\omega_f'){\bf G}^*_{\rm E}({\bf r}',\boldsymbol{\rho}'',\omega_f'')\hat{\bf f}({\bf r},\omega_f;t)\hat{\bf f}^\dagger(\boldsymbol{\rho}',\omega_f';\tau)\hat{\bf f}^\dagger(\boldsymbol{\rho}'',\omega_f'';t)\right.\\
	&\!\!+\!\left.\left.{\bf G}_{\rm E}({\bf r}',\boldsymbol{\rho}',\omega_f'){\bf G}^*_{\rm E}({\bf r}',\boldsymbol{\rho}'',\omega_f'')\hat{\bf f}({\bf r},\omega_f;t)\hat{\bf f}(\boldsymbol{\rho}',\omega_f';\tau)\hat{\bf f}^\dagger(\boldsymbol{\rho}'',\omega_f'';t)\!+\!{\bf G}^*_{\rm E}({\bf r}',\boldsymbol{\rho}',\omega_f'){\bf G}_{\rm E}({\bf r}',\boldsymbol{\rho}'',\omega_f'')\hat{\bf f}({\bf r},\omega_f;t)\hat{\bf f}^\dagger(\boldsymbol{\rho}',\omega_f';\tau)\hat{\bf f}(\boldsymbol{\rho}'',\omega_f'';t)\right]\right\}\!.
\end{align*}

At this point, to proceed further with the calculations avoiding the introduction of time-dependent commutation relations~\cite{VogelSI}, we simplify the mathematical treatment by assuming a specific form for the time-varying susceptibility, in such a manner that it is considered as a kind of \textit{local time-modulation}, namely, displaying the following form~\cite{Sloan2021SI}:
\begin{equation}
	\Delta\chi({\bf r},t,\tau)\equiv\Delta\tilde{\chi}({\bf r},t)\delta{\left[t-\tau\right]}.
	\label{Eq.II.06}
\end{equation}
Under this consideration, the above time integration is merely reduced to a simple substitution of $\tau\to t$, thus allowing the usage of the equal-time commutation relations given in Eqs.~\eqref{Eq.I.04a} and \eqref{Eq.I.04b}:
\begin{align*}
	&\left[\hat{\bf f}({\bf r},\omega_f;t),\hat{\mathcal{H}}_{\rm T}\right]=\iiint{d^3{\bf r}'d^3\boldsymbol{\rho}'d^3\boldsymbol{\rho}''\iint\limits_{0}^{\quad+\infty}{d\omega_f'd\omega_f''\Delta\tilde{\chi}({\bf r}',t)}}\\
	&\qquad\left\{{\bf G}_{\rm E}({\bf r}',\boldsymbol{\rho}',\omega_f'){\bf G}_{\rm E}({\bf r}',\boldsymbol{\rho}'',\omega_f'')\left[\hat{\bf f}(\boldsymbol{\rho}',\omega_f';t)\hat{\bf f}(\boldsymbol{\rho}'',\omega_f'';t)\hat{\bf f}({\bf r},\omega_f;t)-\hat{\bf f}({\bf r},\omega_f;t)\hat{\bf f}(\boldsymbol{\rho}',\omega_f';t)\hat{\bf f}(\boldsymbol{\rho}'',\omega_f'';t)\right]\right.\\
	&\qquad+{\bf G}^*_{\rm E}({\bf r}',\boldsymbol{\rho}',\omega_f'){\bf G}^*_{\rm E}({\bf r}',\boldsymbol{\rho}'',\omega_f'')\left[\hat{\bf f}^\dagger(\boldsymbol{\rho}',\omega_f';t)\hat{\bf f}^\dagger(\boldsymbol{\rho}'',\omega_f'';t)\hat{\bf f}({\bf r},\omega_f;t)-\hat{\bf f}({\bf r},\omega_f;t)\hat{\bf f}^\dagger(\boldsymbol{\rho}',\omega_f';t)\hat{\bf f}^\dagger(\boldsymbol{\rho}'',\omega_f'';t)\right]\\
	&\qquad+{\bf G}_{\rm E}({\bf r}',\boldsymbol{\rho}',\omega_f'){\bf G}^*_{\rm E}({\bf r}',\boldsymbol{\rho}'',\omega_f'')\left[\hat{\bf f}(\boldsymbol{\rho}',\omega_f';t)\hat{\bf f}^\dagger(\boldsymbol{\rho}'',\omega_f'';t)\hat{\bf f}({\bf r},\omega_f;t)-\hat{\bf f}({\bf r},\omega_f;t)\hat{\bf f}(\boldsymbol{\rho}',\omega_f';t)\hat{\bf f}^\dagger(\boldsymbol{\rho}'',\omega_f'';t)\right]\\
	&\qquad+\left.{\bf G}^*_{\rm E}({\bf r}',\boldsymbol{\rho}',\omega_f'){\bf G}_{\rm E}({\bf r}',\boldsymbol{\rho}'',\omega_f'')\left[\hat{\bf f}^\dagger(\boldsymbol{\rho}',\omega_f';t)\hat{\bf f}(\boldsymbol{\rho}'',\omega_f'';t)\hat{\bf f}({\bf r},\omega_f;t)-\hat{\bf f}({\bf r},\omega_f;t)\hat{\bf f}^\dagger(\boldsymbol{\rho}',\omega_f';t)\hat{\bf f}(\boldsymbol{\rho}'',\omega_f'';t)\right]\right\}\\
	&=-\iiint{d^3{\bf r}'d^3\boldsymbol{\rho}'d^3\boldsymbol{\rho}''\iint\limits_{0}^{\quad+\infty}{d\omega_f'd\omega_f''\Delta\tilde{\chi}({\bf r}',t)\mathbb{I}\delta{\left[{\bf r}-\boldsymbol{\rho}'\right]}\delta{\left[\omega_f-\omega_f'\right]}}}\left[{\bf G}^*_{\rm E}({\bf r}',\boldsymbol{\rho}',\omega_f'){\bf G}^*_{\rm E}({\bf r}',\boldsymbol{\rho}'',\omega_f'')\hat{\bf f}^\dagger(\boldsymbol{\rho}'',\omega_f'';t)\right]\\
	&\quad-\iiint{d^3{\bf r}'d^3\boldsymbol{\rho}'d^3\boldsymbol{\rho}''\iint\limits_{0}^{\quad+\infty}{d\omega_f'd\omega_f''\Delta\tilde{\chi}({\bf r}',t)\mathbb{I}\delta{\left[{\bf r}-\boldsymbol{\rho}'\right]}\delta{\left[\omega_f-\omega_f'\right]}}}\left[{\bf G}^*_{\rm E}({\bf r}',\boldsymbol{\rho}',\omega_f'){\bf G}_{\rm E}({\bf r}',\boldsymbol{\rho}'',\omega_f'')\hat{\bf f}(\boldsymbol{\rho}'',\omega_f'';t)\right]\\
	&\quad-\iiint{d^3{\bf r}'d^3\boldsymbol{\rho}'d^3\boldsymbol{\rho}''\iint\limits_{0}^{\quad+\infty}{d\omega_f'd\omega_f''\Delta\tilde{\chi}({\bf r}',t)\mathbb{I}\delta{\left[{\bf r}-\boldsymbol{\rho}''\right]}\delta{\left[\omega_f-\omega_f''\right]}}}\left[{\bf G}^*_{\rm E}({\bf r}',\boldsymbol{\rho}',\omega_f'){\bf G}^*_{\rm E}({\bf r}',\boldsymbol{\rho}'',\omega_f'')\hat{\bf f}^\dagger(\boldsymbol{\rho}',\omega_f';t)\right]\\
	&\quad-\iiint{d^3{\bf r}'d^3\boldsymbol{\rho}'d^3\boldsymbol{\rho}''\iint\limits_{0}^{\quad+\infty}{d\omega_f'd\omega_f''\Delta\tilde{\chi}({\bf r}',t)\mathbb{I}\delta{\left[{\bf r}-\boldsymbol{\rho}''\right]}\delta{\left[\omega_f-\omega_f''\right]}}}\left[{\bf G}_{\rm E}({\bf r}',\boldsymbol{\rho}',\omega_f'){\bf G}^*_{\rm E}({\bf r}',\boldsymbol{\rho}'',\omega_f'')\hat{\bf f}(\boldsymbol{\rho}',\omega_f';t)\right].
\end{align*}
\noindent Then, performing the integration over positions and frequencies and reordering the resulting expression leads to:
\begin{align*}
	\left[\hat{\bf f}({\bf r},\omega_f;t),\hat{\mathcal{H}}_{\rm T}\right]&=-\iint{d^3{\bf r}'d^3\boldsymbol{\rho}''\int\limits_{0}^{+\infty}{d\omega_f''\Delta\tilde{\chi}({\bf r}',t)}}\left\{{\bf G}^*_{\rm E}({\bf r}',{\bf r},\omega_f)\left[{\bf G}_{\rm E}({\bf r}',\boldsymbol{\rho}'',\omega_f'')\hat{\bf f}(\boldsymbol{\rho}'',\omega_f'';t)+{\bf G}^*_{\rm E}({\bf r}',\boldsymbol{\rho}'',\omega_f'')\hat{\bf f}^\dagger(\boldsymbol{\rho}'',\omega_f'';t)\right]\right\}\\
	&\quad-\iint{d^3{\bf r}'d^3\boldsymbol{\rho}'\int\limits_{0}^{+\infty}{d\omega_f'\Delta\tilde{\chi}({\bf r}',t)}}\left\{{\bf G}^*_{\rm E}({\bf r}',{\bf r},\omega_f)\left[{\bf G}_{\rm E}({\bf r}',\boldsymbol{\rho}',\omega_f')\hat{\bf f}(\boldsymbol{\rho}',\omega_f';t)+{\bf G}^*_{\rm E}({\bf r}',\boldsymbol{\rho}',\omega_f')\hat{\bf f}^\dagger(\boldsymbol{\rho}',\omega_f';t)\right]\right\}\\
	&=-2\iint{d^3{\bf r}'d^3\boldsymbol{\rho}'\int\limits_{0}^{+\infty}{d\omega_f'\Delta\tilde{\chi}({\bf r}',t)}}\left\{{\bf G}^*_{\rm E}({\bf r}',{\bf r},\omega_f)\left[{\bf G}_{\rm E}({\bf r}',\boldsymbol{\rho}',\omega_f')\hat{\bf f}(\boldsymbol{\rho}',\omega_f';t)+{\bf G}^*_{\rm E}({\bf r}',\boldsymbol{\rho}',\omega_f')\hat{\bf f}^\dagger(\boldsymbol{\rho}',\omega_f';t)\right]\right\}\!.
\end{align*}
Hence, from the expressions for the electric field operators [i.e., those given in Eqs.~\eqref{Eq.I.02a} and \eqref{Eq.I.02b}], it follows~that:
\begin{tcolorbox}[sharp corners]
	\vspace{-0.35cm}
	\begin{equation}
		\left[\hat{\bf f}({\bf r},\omega_f;t),\hat{\mathcal{H}}_{\rm T}\right]=-2\int{d^3\boldsymbol{\rho}\Delta\tilde{\chi}(\boldsymbol{\rho},t){\bf G}_{\rm E}^*(\boldsymbol{\rho},{\bf r},\omega_f)\hat{\boldsymbol{\mathcal{E}}}(\boldsymbol{\rho};t)}.
		\label{Eq.II.07}
	\end{equation}
\end{tcolorbox}
\noindent Proceeding similarly with the creation operator it can be shown that:
\begin{tcolorbox}[sharp corners]
	\vspace{-0.35cm}
	\begin{equation}
		\left[\hat{\bf f}^\dagger({\bf r},\omega_f;t),\hat{\mathcal{H}}_{\rm T}\right]=2\int{d^3\boldsymbol{\rho}\Delta\tilde{\chi}(\boldsymbol{\rho},t){\bf G}_{\rm E}(\boldsymbol{\rho},{\bf r},\omega_f)\hat{\boldsymbol{\mathcal{E}}}(\boldsymbol{\rho};t)}.
		\label{Eq.II.08}
	\end{equation}
\end{tcolorbox}
\noindent Therefore, the corresponding \textit{Heisenberg equations of motion for the time-modulated polaritonic operators} are:
\begin{tcolorbox}[sharp corners,colback=blue!5!white,colframe=blue!50!white]
	\vspace{-0.35cm}
	\begin{subequations}
		\begin{align}
			&\partial_t\hat{\bf f}_{\rm T}({\bf r},\omega_f;t)=\frac{2i}{\hbar}\int{d^3\boldsymbol{\rho}\Delta\tilde{\chi}(\boldsymbol{\rho},t){\bf G}^*_{\rm E}(\boldsymbol{\rho},{\bf r},\omega_f)\hat{\boldsymbol{\mathcal{E}}}(\boldsymbol{\rho};t)},
			\label{Eq.II.09a}\\
			&\partial_t\hat{\bf f}_{\rm T}^\dagger({\bf r},\omega_f;t)=\frac{-2i}{\hbar}\int{d^3\boldsymbol{\rho}\Delta\tilde{\chi}(\boldsymbol{\rho},t){\bf G}_{\rm E}(\boldsymbol{\rho},{\bf r},\omega_f)\hat{\boldsymbol{\mathcal{E}}}(\boldsymbol{\rho};t)}.
			\label{Eq.II.09b}
		\end{align}
	\end{subequations}
\end{tcolorbox}

From the results given in Eqs.~\eqref{Eq.II.05a} and \eqref{Eq.II.05b}, and Eqs.~\eqref{Eq.II.09a} and \eqref{Eq.II.09b}, along with the decomposition sketched out in Eq.~\eqref{Eq.II.02}, the dynamic of the operators brought about by both the free-evolving and the time-modulated contributions reads as,
\begin{subequations}
	\begin{align}
		\nonumber\partial_t\hat{\bf f}({\bf r},\omega_f;t)=\frac{1}{i\hbar}\left[\hat{\bf f}({\bf r},\omega_f;t),\hat{\mathcal{H}}\right]&=\partial_t\hat{\bf f}_0({\bf r},\omega_f;t)+\partial_t\hat{\bf f}_{\rm T}({\bf r},\omega_f;t)\\
		&=-i\omega_f\hat{\bf f}({\bf r},\omega_f;t)+\frac{2i}{\hbar}\int{d^3\boldsymbol{\rho}\Delta\tilde{\chi}(\boldsymbol{\rho},t){\bf G}^*_{\rm E}(\boldsymbol{\rho},{\bf r},\omega_f)\hat{\boldsymbol{\mathcal{E}}}(\boldsymbol{\rho};t)},
		\label{Eq.II.10a}\\
		\nonumber\partial_t\hat{\bf f}^\dagger({\bf r},\omega_f;t)=\frac{1}{i\hbar}\left[\hat{\bf f}^\dagger({\bf r},\omega_f;t),\hat{\mathcal{H}}\right]&=\partial_t\hat{\bf f}^\dagger_0({\bf r},\omega_f;t)+\partial_t\hat{\bf f}^\dagger_{\rm T}({\bf r},\omega_f;t)\\
		&=i\omega_f\hat{\bf f}^\dagger({\bf r},\omega_f;t)-\frac{2i}{\hbar}\int{d^3\boldsymbol{\rho}\Delta\tilde{\chi}(\boldsymbol{\rho},t){\bf G}_{\rm E}(\boldsymbol{\rho},{\bf r},\omega_f)\hat{\boldsymbol{\mathcal{E}}}(\boldsymbol{\rho};t)},
		\label{Eq.II.10b}
	\end{align}
\end{subequations}
thus yielding:
\begin{tcolorbox}[sharp corners,colback=red!5!white,colframe=red!50!white]
	\vspace{-0.35cm}
	\begin{subequations}
		\begin{align}
			\!\!\!\!\!\!\nonumber\hat{\bf f}({\bf r},\omega_f;t)&=\hat{\bf f}_0({\bf r},\omega_f;t)+\hat{\bf f}_{\rm T}({\bf r},\omega_f;t)\\
			&=\hat{\bf f}({\bf r},\omega_f;t=0)e^{-i\omega_f t}+\frac{2i}{\hbar}e^{-i\omega_ft}\left[\int_{0}^{t}{d\tau\int_{-\infty}^{+\infty}{d^3\boldsymbol{\rho}e^{i\omega_f\tau}\Delta\tilde{\chi}(\boldsymbol{\rho},\tau){\bf G}^*_{\rm E}(\boldsymbol{\rho},{\bf r},\omega_f)\hat{\boldsymbol{\mathcal{E}}}(\boldsymbol{\rho};\tau)}}\right],
			\label{Eq.II.11a}\\\nonumber\\
			\!\!\!\!\!\!\nonumber\hat{\bf f}^\dagger({\bf r},\omega_f;t)&=\hat{\bf f}^\dagger_0({\bf r},\omega_f;t)+\hat{\bf f}^\dagger_{\rm T}({\bf r},\omega_f;t)\\
			&=\hat{\bf f}^\dagger({\bf r},\omega_f;t=0)e^{+i\omega_f t}-\frac{2i}{\hbar}e^{+i\omega_ft}\left[\int_{0}^{t}{d\tau\int_{-\infty}^{+\infty}{d^3\boldsymbol{\rho}e^{-i\omega_f\tau}\Delta\tilde{\chi}(\boldsymbol{\rho},\tau){\bf G}_{\rm E}(\boldsymbol{\rho},{\bf r},\omega_f)\hat{\boldsymbol{\mathcal{E}}}(\boldsymbol{\rho};\tau)}}\right].
			\label{Eq.II.11b}
		\end{align}
	\end{subequations}
\end{tcolorbox}

\section{III. ELECTRIC FIELD OPERATOR: FREE-EVOLVING AND TIME-MODULATED CONTRIBUTIONS}
\label{SectSI.III}
\renewcommand\theequation{S.III.\arabic{equation}}
\setcounter{equation}{0}

Building on the above results, in this section we calculate the electric field operator making an explicit distinction between the free-evolving and the time-modulated terms. For the sake of completeness, as well as for convenience in the ensuing developments, we present both the real-valued fields in time domain, and the complex amplitudes in the frequency domain.

\subsection{A. Free-evolving electric field operator}
\label{SectSI.III.A}

The free-evolving positive-frequency electric field operator in time domain can be straightforwardly obtained just by performing the corresponding substitution of the (creation/annihilation) polaritonic operators as given in Eq.~\eqref{Eq.II.11a} (or Eq.~\eqref{Eq.II.11b} for the negative-frequency part), into the electric field operator given in Eq.~\eqref{Eq.I.02a} (or Eq.~\eqref{Eq.I.02b} for the negative-frequency part):
\begin{align}
	\nonumber\hat{\boldsymbol{\mathcal{E}}}^{(+)}_0({\bf r};t)&=\int_{-\infty}^{+\infty}{d^3\boldsymbol{\rho}\int_{0}^{+\infty}{d\omega_f{\bf G}_{\rm E}({\bf r},\boldsymbol{\rho},\omega_f)\hat{\bf f}_0(\boldsymbol{\rho},\omega_f;t)}}\\
	&=\int_{-\infty}^{+\infty}{d^3\boldsymbol{\rho}\int_{0}^{+\infty}{d\omega_f{\bf G}_{\rm E}({\bf r},\boldsymbol{\rho},\omega_f)\hat{\bf f}_0(\boldsymbol{\rho},\omega_f;t=0)e^{-i\omega_f t}}}.
	\label{Eq.III.01}
\end{align}
From the above expression, and noticing that we are only regarding the positive-frequency part of the field, the respective complex-like field operator in frequency domain can be directly obtained by means of the Laplace transform:
\begin{align}
	\nonumber\hat{\bf E}^{(+)}_0({\bf r};\omega)&=\int_{0}^{+\infty}{dt \hat{\boldsymbol{\mathcal{E}}}^{(+)}_0({\bf r};t)e^{i\omega t} }\\
	\nonumber&=\int_{0}^{+\infty}{dt \left[\int_{-\infty}^{+\infty}{d^3\boldsymbol{\rho} \int_{0}^{+\infty}{d\omega_f {\bf G}_{\rm E}({\bf r},\boldsymbol{\rho},\omega_f)\hat{\bf f}(\boldsymbol{\rho},\omega_f;t=0)e^{-i\omega_f t} }}\right] e^{i\omega t} }\\
	\nonumber&=\int_{-\infty}^{+\infty}{d^3\boldsymbol{\rho} \int_{0}^{+\infty}{d\omega_f {\bf G}_{\rm E}({\bf r},\boldsymbol{\rho},\omega_f)\hat{\bf f}(\boldsymbol{\rho},\omega_f;t=0)\left[\frac{i}{\omega-\omega_f+i0^+}\right] } }\\
	\nonumber&=\int_{-\infty}^{+\infty}{d^3\boldsymbol{\rho} \int_{0}^{+\infty}{d\omega_f {\bf G}_{\rm E}({\bf r},\boldsymbol{\rho},\omega_f)\hat{\bf f}(\boldsymbol{\rho},\omega_f;t=0)\left\{\pi\delta{[\omega-\omega_f]}+i\mathcal{P}{\left[\frac{1}{\omega-\omega_f}\right]}\right\} } }\\
	\nonumber&=\pi\int_{-\infty}^{+\infty}{d^3\boldsymbol{\rho} \int_{0}^{+\infty}{d\omega_f \delta{[\omega-\omega_f]}{\bf G}_{\rm E}({\bf r},\boldsymbol{\rho},\omega_f)\hat{\bf f}(\boldsymbol{\rho},\omega_f;t=0) } }+i\mathcal{P}{\left[\int_{-\infty}^{+\infty}{d^3\boldsymbol{\rho} \int_{0}^{+\infty}{d\omega_f \frac{{\bf G}_{\rm E}({\bf r},\boldsymbol{\rho},\omega_f)\hat{\bf f}(\boldsymbol{\rho},\omega_f;t=0)}{\omega-\omega_f} } }\right]}\\
	&=\pi\int_{-\infty}^{+\infty}{d^3\boldsymbol{\rho} {\bf G}_{\rm E}({\bf r},\boldsymbol{\rho},\omega)\hat{\bf f}(\boldsymbol{\rho},\omega;t=0)  }+i\pi\frac{1}{\pi}\mathcal{P}{\left[\int_{-\infty}^{+\infty}{\!\!d^3\boldsymbol{\rho} \int_{0}^{+\infty}{\!\!d\omega_f \frac{{\bf G}_{\rm E}({\bf r},\boldsymbol{\rho},\omega_f)\hat{\bf f}(\boldsymbol{\rho},\omega_f;t=0)}{\omega-\omega_f} } }\right]},
	\label{Eq.III.02}
\end{align}
where it has been used the identity,
\begin{equation}
	\int_{0}^{+\infty}{dt e^{i\Delta t}}=\frac{i}{\Delta+i0^+},
	\label{Eq.III.03}
\end{equation}
and the so-called \textit{Sokhotski-Plemelj theorem},
\begin{equation}
	\lim\limits_{\varepsilon\to0}{\frac{1}{\Delta\pm i\varepsilon^+}}\to\mp i\pi\delta{[\Delta]}+\mathcal{P}{\left[\frac{1}{\Delta}\right]},
	\label{Eq.III.04}
\end{equation}
where $\mathcal{P}$ stands for the \textit{Cauchy principal value}, thereby accounting for the poles of causal functions. Here it is worth stressing that in order to complete the Green's function, we would have to consider the extension of the lower limit of the frequency integral to $-\infty$. Such an approximation is commonly used (see, e.g., Refs.~\cite{VogelSI,LoudonSI}), and it is often claimed that leads to some missing terms resulting from the subsequent application of the \textit{rotating wave approximation}. However, in this case it should be noted that no rotating wave approximation is employed at all, so that there are no missing terms coming from such an assumption. Consequently, the second integral can be performed as follows:
\begin{align}
	\nonumber\frac{1}{\pi}\mathcal{P}{\left[\int_{-\infty}^{+\infty}{d^3\boldsymbol{\rho} \int_{0}^{+\infty}{d\omega_f \frac{{\bf G}_{\rm E}({\bf r},\boldsymbol{\rho},\omega_f)\hat{\bf f}(\boldsymbol{\rho},\omega_f;t=0)}{\omega-\omega_f} } }\right]}&=\int_{-\infty}^{+\infty}{d^3\boldsymbol{\rho} \frac{1}{\pi}\mathcal{P}{\left[\int_{0}^{+\infty}{d\omega_f\frac{\mathcal{G}({\bf r},\boldsymbol{\rho},\omega_f)}{\omega-\omega_f}}\right]}}\\
	&=\int_{-\infty}^{+\infty}{d^3\boldsymbol{\rho}\left[\text{Im}{\left[\mathcal{G}({\bf r},\boldsymbol{\rho},\omega)\right]}-i\text{Re}{\left[\mathcal{G}({\bf r},\boldsymbol{\rho},\omega)\right]}\right]},
	\label{Eq.III.05}
\end{align}
where the computation of the integral has been done on account of the \textit{Kramers-Kronig relations},
\begin{subequations}
	\begin{align}
		\text{Re}{\left[\mathcal{G}({\bf r},\boldsymbol{\rho},\omega)\right]}&=\frac{1}{\pi}\mathcal{P}{\left[\int_{-\infty}^{+\infty}{d\omega'} \frac{\text{Im}{\left[\mathcal{G}({\bf r},\boldsymbol{\rho},\omega)\right]}}{\omega'-\omega}\right]},
		\label{Eq.III.06a}\\
		\text{Im}{\left[\mathcal{G}({\bf r},\boldsymbol{\rho},\omega)\right]}&=\frac{1}{\pi}\mathcal{P}{\left[\int_{-\infty}^{+\infty}{d\omega'} \frac{\text{Re}{\left[\mathcal{G}({\bf r},\boldsymbol{\rho},\omega)\right]}}{\omega-\omega'}\right]},
		\label{Eq.III.06b}
	\end{align}
\end{subequations}
over the function $\mathcal{G}({\bf r},\boldsymbol{\rho},\omega_f)={\bf G}_{\rm E}({\bf r},\boldsymbol{\rho},\omega_f)\hat{\bf f}(\boldsymbol{\rho},\omega_f;t=0)$.
Therefore, putting it all together, the positive-frequency part of the free-evolving electric field operator in frequency domain reads as:
\begin{tcolorbox}[sharp corners, colback=blue!5!white,colframe=blue!50!white]
	\vspace{-0.35cm}
	\begin{equation}
		\hat{\bf E}^{(+)}_0({\bf r};\omega)=2\pi\int_{-\infty}^{+\infty}{d^3\boldsymbol{\rho} {\bf G}_{\rm E}({\bf r},\boldsymbol{\rho},\omega)\hat{\bf f}(\boldsymbol{\rho},\omega;t=0)}.
		\label{Eq.III.07}
	\end{equation}
\end{tcolorbox}

\subsection{B. Time-modulated electric field operator}
\label{SectSI.III.B}

Similarly, the time-modulated electric field operator in time domain can be simply obtained by inserting the corresponding polaritonic operator into the electric field operator. Once again, focusing only into the positive-frequency part it follows that,
\begin{align}
	\nonumber\hat{\boldsymbol{\mathcal{E}}}^{(+)}_{\rm T}({\bf r};t)&=\int_{-\infty}^{+\infty}{d^3\boldsymbol{\rho}\int_{0}^{+\infty}{d\omega_f {\bf G}_{\rm E}({\bf r},\boldsymbol{\rho},\omega_f)\hat{\bf f}_{\rm T}(\boldsymbol{\rho},\omega_f;t)}}\\
	\nonumber&=\frac{2i}{\hbar}\int_{-\infty}^{+\infty}{d^3\boldsymbol{\rho}\int_{0}^{+\infty}{d\omega_f {\bf G}_{\rm E}({\bf r},\boldsymbol{\rho},\omega_f)\left[e^{-i\omega_f t}\int_{0}^{t}{d\tau'\int_{-\infty}^{+\infty}{d^3\boldsymbol{\rho}' e^{i\omega_f\tau'}\Delta\tilde{\chi}(\boldsymbol{\rho}',\tau'){\bf G}^*_{\rm E}(\boldsymbol{\rho}',\boldsymbol{\rho},\omega_f)\hat{\boldsymbol{\mathcal{E}}}(\boldsymbol{\rho}';\tau') } }\right]}}\\
	&=\int_{-\infty}^{+\infty}{d^3\boldsymbol{\rho}'\int_{0}^{t}{d\tau' \left\{\frac{2i}{\pi\varepsilon_0c^2}\int_{0}^{+\infty}{d\omega_f \omega_f^2e^{-i\omega_f(t-\tau')}\text{Im}{\left[{\bf G}({\bf r},\boldsymbol{\rho}',\omega_f)\right]} } \right\}\Delta\tilde{\chi}(\boldsymbol{\rho}',\tau')\hat{\boldsymbol{\mathcal{E}}}(\boldsymbol{\rho}';\tau')}},
	\label{Eq.III.08}
\end{align}
where it has been used the \textit{completeness relation of the dyadic Green's function} \cite{VogelSI,Scheel2008SI}:
\begin{equation}
	\int_{-\infty}^{+\infty}{d^3\boldsymbol{\rho} \frac{\omega^2}{c^2}\text{Im}{\left[\varepsilon(\boldsymbol{\rho},\omega)\right]}{\bf G}({\bf r},\boldsymbol{\rho},\omega){\bf G}^*({\bf r}',\boldsymbol{\rho},\omega) }=\text{Im}{\left[{\bf G}({\bf r},{\bf r}',\omega)\right]},
	\label{Eq.III.09}
\end{equation}
which can be derived from the \textit{Schwarz reflection principle}, ${\bf G}^*({\bf r},{\bf r}',\omega)={\bf G}({\bf r},{\bf r}',-\omega^*)$, the \textit{Onsager reciprocity theorem},  ${\bf G}^{\rm T}({\bf r},{\bf r}',\omega)={\bf G}({\bf r}',{\bf r},\omega)$, requiring the condition that ${\bf G}({\bf r},{\bf r}',\omega)\to0$ at ${\bf r}\to \infty$ (namely, ensuring that there is no net energy transport, i.e., the time-averaged Poynting vector is to be zero for any point in space), and making use of the own definition of ${\bf G}({\bf r},{\bf r}',\omega)$:
\begin{equation}
	\nabla\times\nabla\times{\bf G}({\bf r},{\bf r}',\omega)-k^2{\bf G}({\bf r},{\bf r}',\omega)=\mathbb{I}\delta{[{\bf r}-{\bf r}']}.
	\label{Eq.III.10}
\end{equation}
In this way, the real-valued time-modulated field operator can be written as the spatial integral of a \textit{memory kernel}, acting on the polarization field operator:
\begin{tcolorbox}[sharp corners, colback=blue!5!white,colframe=blue!50!white]
	\vspace{-0.35cm}
	\begin{equation}
		\hat{\boldsymbol{\mathcal{E}}}^{(+)}_{\rm T}({\bf r};t)=\int_{-\infty}^{+\infty}{d^3\boldsymbol{\rho}\left\{\int_{0}^{t}{d\tau \boldsymbol{\mathcal{K}}^{(+)}_{\rm E}({\bf r},\boldsymbol{\rho},t-\tau)\hat{\boldsymbol{\mathcal{P}}}(\boldsymbol{\rho};\tau) }\right\}},
		\label{Eq.III.11}
	\end{equation}
	with $\hat{\boldsymbol{\mathcal{P}}}(\boldsymbol{\rho};\tau)\equiv\Delta\tilde{\chi}(\boldsymbol{\rho},\tau)\hat{\boldsymbol{\mathcal{E}}}(\boldsymbol{\rho};\tau)$, and
	\begin{equation}
		\boldsymbol{\mathcal{K}}^{(+)}_{\rm E}({\bf r},\boldsymbol{\rho},t-\tau)=\frac{2i}{\pi\varepsilon_0c^2}\int_{0}^{+\infty}{d\omega_f \omega_f^2\text{Im}{\left[{\bf G}({\bf r},\boldsymbol{\rho},\omega_f)\right]} e^{-i\omega_f(t-\tau)}}.
		\label{Eq.III.12}
	\end{equation}
\end{tcolorbox}

Just like the free-evolving contribution, the complex-like electric field operator in frequency domain can be calculated from the above expressions [Eqs.~\eqref{Eq.III.11} and \eqref{Eq.III.12}] taking the Laplace transform:
\begin{align}
	\nonumber\hat{\bf E}^{(+)}_{\rm T}({\bf r};\omega)&=\int_{0}^{+\infty}{dt\hat{\boldsymbol{\mathcal{E}}}^{(+)}_{\rm T}({\bf r};t)e^{i\omega t}}\\
	\nonumber&=\int_{0}^{+\infty}{dt\left[\int_{-\infty}^{+\infty}{d^3\boldsymbol{\rho}\left\{\int_{0}^{t}{d\tau \boldsymbol{\mathcal{K}}^{(+)}_{\rm E}({\bf r},\boldsymbol{\rho},t-\tau)\hat{\boldsymbol{\mathcal{P}}}(\boldsymbol{\rho};\tau) }\right\} }\right]e^{i\omega t}}\\
	\nonumber&=\int_{-\infty}^{+\infty}{d^3\boldsymbol{\rho}\left\{\int_{0}^{+\infty}{dt \left[\int_{0}^{t}{d\tau \boldsymbol{\mathcal{K}}^{(+)}_{\rm E}({\bf r},\boldsymbol{\rho},t-\tau)\Delta\tilde{\chi}(\boldsymbol{\rho},\tau)\hat{\boldsymbol{\mathcal{E}}}(\boldsymbol{\rho};\tau) }\right]e^{i\omega(t-\tau)}}e^{i\omega\tau}\right\}}\\
	\nonumber&=\int_{-\infty}^{+\infty}{d^3\boldsymbol{\rho}\left\{\int_{0}^{+\infty}{d\tau \left[\int_{\tau}^{+\infty}{dt \boldsymbol{\mathcal{K}}^{(+)}_{\rm E}({\bf r},\boldsymbol{\rho},t-\tau)\Delta\tilde{\chi}(\boldsymbol{\rho},\tau)\hat{\boldsymbol{\mathcal{E}}}(\boldsymbol{\rho};\tau) }\right]e^{i\omega(t-\tau)}}e^{i\omega\tau}\right\}}\\
	&=\int_{-\infty}^{+\infty}{d^3\boldsymbol{\rho}\left\{\left[\int_{0}^{+\infty}{d\tilde{t}\boldsymbol{\mathcal{K}}^{(+)}_{\rm E}({\bf r},\boldsymbol{\rho},\tilde{t})e^{i\omega\tilde{t}} }\right]\cdot\left[\int_{0}^{+\infty}{d\tau \Delta\tilde{\chi}(\boldsymbol{\rho},\tau)\hat{\boldsymbol{\mathcal{E}}}(\boldsymbol{\rho};\tau)e^{i\omega\tau} }\right] \right\}},
	\label{Eq.III.13}
\end{align}
where it has been made the change of variable $\tilde{t}=t-\tau$. So, the positive-frequency part of the time-modulated electric field operator in frequency domain reads as:
\begin{tcolorbox}[sharp corners, colback=blue!5!white,colframe=blue!50!white]
	\vspace{-0.35cm}
	\begin{equation}
		\hat{\bf E}^{(+)}_{\rm T}({\bf r};\omega)=\int_{-\infty}^{+\infty}{d^3\boldsymbol{\rho}{\bf K}^{(+)}_{\rm E}({\bf r},\boldsymbol{\rho},\omega)\mathcal{L}_\omega{[\Delta\tilde{\chi}(\boldsymbol{\rho},\tau)\hat{\boldsymbol{\mathcal{E}}}(\boldsymbol{\rho};\tau)]} },
		\label{Eq.III.14}
	\end{equation}
	where
	\begin{equation}
		\mathcal{L}_\omega{[\Delta\tilde{\chi}(\boldsymbol{\rho},\tau)\hat{\boldsymbol{\mathcal{E}}}(\boldsymbol{\rho};\tau)]}\equiv\int_{0}^{+\infty}{d\tau \Delta\tilde{\chi}(\boldsymbol{\rho},\tau)\hat{\boldsymbol{\mathcal{E}}}(\boldsymbol{\rho};\tau)e^{i\omega\tau}},
		\label{Eq.III.15}
	\end{equation}
	is the Laplace transform of the polarization field operator $\hat{\boldsymbol{\mathcal{P}}}(\boldsymbol{\rho};\tau)\equiv \Delta\tilde{\chi}(\boldsymbol{\rho},\tau)\hat{\boldsymbol{\mathcal{E}}}(\boldsymbol{\rho};\tau)$, and
	\begin{equation}
		{\bf K}^{(+)}_{\rm E}({\bf r},\boldsymbol{\rho},\omega)=\frac{2\omega^2}{\varepsilon_0c^2}{\bf G}({\bf r},\boldsymbol{\rho},\omega).
		\label{Eq.III.16}
	\end{equation}
\end{tcolorbox}

\subsection{C. Total electric field operator}
\label{SectSI.III.C}

The Eq.~\eqref{Eq.III.11}, and consequently Eq.~\eqref{Eq.III.14}, are self-contained expressions in which the time-modulated positive-frequency electric field operator, $\hat{\boldsymbol{\mathcal{E}}}^{(+)}_{\rm T}$, is directly related to the total field, $\hat{\boldsymbol{\mathcal{E}}}=\hat{\boldsymbol{\mathcal{E}}}^{(+)}_0+\hat{\boldsymbol{\mathcal{E}}}^{(-)}_0+\hat{\boldsymbol{\mathcal{E}}}^{(+)}_{\rm T}+\hat{\boldsymbol{\mathcal{E}}}^{(-)}_{\rm T}$. Namely, the annihilation polaritonic operator (associated to the positive-frequency component of the electric field) shall lead to a combination of the creation and the annihilation operators, being modified (actually shifted) by the time-modulation. Furthermore, in contrast to the case of a time-invariant system (e.g., that considering a quantum single emitter), there is a spatial integral that makes the kernel and the field operator tied to each other, i.e., they appear to be tightly coupled. In turn, as a consequence of the time-modulation of the system, the susceptibility function and the field operator are also coupled through the Laplace transform. At any rate, and in view of the above results, it could be inferred that, in general, for time-varying systems is impossible to express the frequency-dependent source-like (time-modulated) electric field as an algebraic product of a kernel function times a polarization field (as it actually occurs in the stationary case). This is essentially due to the coupling that exists between the medium and the external field in this kind of time-varying systems, that translates into the fact that the Laplace transform cannot be expressed in a factorized manner. Far from being a hurdle to overcome, this is indeed a distinctive feature of time-varying systems, increasing both the richness and the variety of the achievable physical effects; in particular those involving convoluted correlations that may arise from the own time-modulation of the medium. Therefore, Eq.~\eqref{Eq.III.14}, along with Eqs.~\eqref{Eq.III.15} and \eqref{Eq.III.16}, is arguably the most compact form in which one may express the relationship between the time-modulated electric field operator and the total field in the frequency domain. So, in order to give a more explicit result, we shall particularize the time-dependent susceptibility $\Delta\tilde{\chi}$ to a special case, so as to be able to get an explicit expression for its associated Laplace transform.\\

Be that as it may, the time-modulated electric field operator, $\hat{\boldsymbol{\mathcal{E}}}_{\rm T}$, appears both in the right- and left-hand sides of the latter expression. So, at least as a first approach, it could be useful to think on a iterative procedure in which one solves order-by-order the full expression of the time-modulated electric field operator. In this way, one could then obtain, recursively, the total electric field operator in the following way:
\begin{align*}
	\!\!\!\!\!\!\!\!\!\!\!\!\!\!\!\!\!\!\!\!\!\!\!\!\!\!\!\!\!\!\!\!\!\!\!\!\!\!\!\!\hat{\boldsymbol{\mathcal{E}}}^{(+)}({\bf r};t)&=\hat{\boldsymbol{\mathcal{E}}}^{(+)}_0({\bf r};t)+\hat{\boldsymbol{\mathcal{E}}}^{(+)}_{\rm T}({\bf r};t)\\
	\!\!\!\!\!\!\!\!\!\!\!\!\!\!\!\!\!\!\!\!\!\!\!\!\!\!\!\!\!\!\!\!\!\!\!\!\!\!\!\!&=\hat{\boldsymbol{\mathcal{E}}}^{(+)}_0({\bf r};t)+\int\limits_{-\infty}^{+\infty}{d^3\boldsymbol{\rho}'\left\{\int\limits_{0}^{t}{d\tau' \boldsymbol{\mathcal{K}}^{(+)}_{\rm E}({\bf r},\boldsymbol{\rho}',t-\tau')\Delta\tilde{\chi}(\boldsymbol{\rho}',\tau')\hat{\boldsymbol{\mathcal{E}}}(\boldsymbol{\rho}';\tau') }\right\}}\\
	\qquad\qquad\;&=\hat{\boldsymbol{\mathcal{E}}}^{(+)}_0({\bf r};t)+\int\limits_{-\infty}^{+\infty}{d^3\boldsymbol{\rho}'\left\{\int\limits_{0}^{t}{d\tau' \boldsymbol{\mathcal{K}}^{(+)}_{\rm E}({\bf r},\boldsymbol{\rho}',t-\tau')\Delta\tilde{\chi}(\boldsymbol{\rho}',\tau')\hat{\boldsymbol{\mathcal{E}}}_0(\boldsymbol{\rho}';\tau')}\right\}}\\
	&\quad+\int\limits_{-\infty}^{+\infty}{d^3\boldsymbol{\rho}'\left\{\int\limits_{0}^{t}{d\tau' \boldsymbol{\mathcal{K}}^{(+)}_{\rm E}({\bf r},\boldsymbol{\rho}',t-\tau')\Delta\tilde{\chi}(\boldsymbol{\rho}',\tau')\left[\int\limits_{-\infty}^{+\infty}{d^3\boldsymbol{\rho}'' \left\{\int\limits_{0}^{\tau'}{d\tau''\boldsymbol{\mathcal{K}}_{\rm E}(\boldsymbol{\rho}',\boldsymbol{\rho}'',\tau'-\tau'')\Delta\tilde{\chi}(\boldsymbol{\rho}'';\tau'')\hat{\boldsymbol{\mathcal{E}}}(\boldsymbol{\rho}'';\tau'') }\right\}}\right] }\right\}}\\
	&=\hat{\boldsymbol{\mathcal{E}}}^{(+)}_0({\bf r};t)+\int\limits_{-\infty}^{+\infty}{d^3\boldsymbol{\rho}'\left\{\int\limits_{0}^{t}{d\tau' \boldsymbol{\mathcal{K}}^{(+)}_{\rm E}({\bf r},\boldsymbol{\rho}',t-\tau')\Delta\tilde{\chi}(\boldsymbol{\rho}',\tau')\hat{\boldsymbol{\mathcal{E}}}_0(\boldsymbol{\rho}';\tau')}\right\}}\\
	&\quad+\int\limits_{-\infty}^{+\infty}{d^3\boldsymbol{\rho}'\left\{\int\limits_{0}^{t}{d\tau' \boldsymbol{\mathcal{K}}^{(+)}_{\rm E}({\bf r},\boldsymbol{\rho}',t-\tau')\Delta\tilde{\chi}(\boldsymbol{\rho}',\tau')\left[\int\limits_{-\infty}^{+\infty}{d^3\boldsymbol{\rho}'' \left\{\int\limits_{0}^{\tau'}{d\tau''\boldsymbol{\mathcal{K}}_{\rm E}(\boldsymbol{\rho}',\boldsymbol{\rho}'',\tau'-\tau'')\Delta\tilde{\chi}(\boldsymbol{\rho}'';\tau'')\hat{\boldsymbol{\mathcal{E}}}_0(\boldsymbol{\rho}'';\tau'') }\right\}}\right] }\right\}}\\
	&\quad+\int\limits_{-\infty}^{+\infty}{d^3\boldsymbol{\rho}'\left\{\int\limits_{0}^{t}{d\tau' \boldsymbol{\mathcal{K}}^{(+)}_{\rm E}({\bf r},\boldsymbol{\rho}',t-\tau')\Delta\tilde{\chi}(\boldsymbol{\rho}',\tau')\left[\int\limits_{-\infty}^{+\infty}{d^3\boldsymbol{\rho}'' \left\{\int\limits_{0}^{\tau'}{d\tau''\boldsymbol{\mathcal{K}}_{\rm E}(\boldsymbol{\rho}',\boldsymbol{\rho}'',\tau'-\tau'')\Delta\tilde{\chi}(\boldsymbol{\rho}'';\tau'')\hat{\boldsymbol{\mathcal{E}}}_{\rm T}(\boldsymbol{\rho}'';\tau'')}\right\}}\right] }\right\}},
\end{align*}
with $\hat{\boldsymbol{\mathcal{E}}}_{\rm 0/T}=\hat{\boldsymbol{\mathcal{E}}}^{(+)}_{\rm 0/T}+\hat{\boldsymbol{\mathcal{E}}}^{(-)}_{\rm 0/T}$ and $\boldsymbol{\mathcal{K}}_{\rm E}=\boldsymbol{\mathcal{K}}^{(+)}_{\rm E}+\boldsymbol{\mathcal{K}}^{(-)}_{\rm E}$. According to the above result, it is clear that we can express the electric field operator as a sort of series expansion of successive higher-order terms:
\begin{equation}
	\hat{\boldsymbol{\mathcal{E}}}^{(+)}({\bf r};t)=\sum_{i=0}^{\infty}{\hat{\boldsymbol{\mathcal{E}}}^{(+)}_i({\bf r};t)},
	\label{Eq.III.17}
\end{equation}
where:
\setcounter{equation}{16}
\begin{subequations}
	\begin{align}
		\hat{\boldsymbol{\mathcal{E}}}^{(+)}_0({\bf r};t)&=\hat{\boldsymbol{\mathcal{E}}}^{(+)}_0({\bf r};t),
		\label{Eq.III.17a}\\
		\hat{\boldsymbol{\mathcal{E}}}^{(+)}_1({\bf r};t)&=\int\limits_{-\infty}^{+\infty}{d^3\boldsymbol{\rho}'\int\limits_{0}^{t}{d\tau' \boldsymbol{\mathcal{K}}^{(+)}_{\rm E}({\bf r},\boldsymbol{\rho}',t-\tau')\Delta\tilde{\chi}(\boldsymbol{\rho}',\tau')\hat{\boldsymbol{\mathcal{E}}}_0(\boldsymbol{\rho}';\tau')}},
		\label{Eq.III.17b}\\
		\hat{\boldsymbol{\mathcal{E}}}^{(+)}_2({\bf r};t)&=\int\limits_{-\infty}^{+\infty}{d^3\boldsymbol{\rho}'\int\limits_{-\infty}^{+\infty}{d^3\boldsymbol{\rho}''\int\limits_{0}^{t}{d\tau'\int\limits_{0}^{\tau'}{d\tau''\boldsymbol{\mathcal{K}}^{(+)}_{\rm E}({\bf r},\boldsymbol{\rho}',t-\tau')\Delta\tilde{\chi}(\boldsymbol{\rho}',\tau')\boldsymbol{\mathcal{K}}_{\rm E}(\boldsymbol{\rho}',\boldsymbol{\rho}'',\tau'-\tau'')\Delta\tilde{\chi}(\boldsymbol{\rho}'',\tau'')\hat{\boldsymbol{\mathcal{E}}}_0(\boldsymbol{\rho}'';\tau'')}}}},
		\label{Eq.III.17c}\\
		\nonumber&\;\;\vdots\\
		\hat{\boldsymbol{\mathcal{E}}}^{(+)}_N({\bf r};t)&=\int\limits_{-\infty}^{+\infty}{d^3\boldsymbol{\rho}'\int\limits_{-\infty}^{+\infty}{d^3\boldsymbol{\rho}''\ldots\int\limits_{-\infty}^{+\infty}{d^3\boldsymbol{\rho}^{N}\int\limits_{0}^{t}{d\tau'\int\limits_{0}^{\tau'}{d\tau''\ldots\int\limits_{0}^{\tau^{N-1}}{d\tau^N\boldsymbol{\mathcal{K}}^{(+)}_{\rm E}({\bf r},\boldsymbol{\rho}',t-\tau')\tilde{\boldsymbol{\Xi}}^N(\boldsymbol{\rho}',\boldsymbol{\rho}'',\ldots,\boldsymbol{\rho}^{N},\tau',\tau'',\ldots,\tau^N)}}}}}},
		\label{Eq.III.17d}
	\end{align}
\end{subequations}
with
\begin{align}
	&\nonumber\tilde{\boldsymbol{\Xi}}^N(\boldsymbol{\rho}',\boldsymbol{\rho}'',\ldots,\boldsymbol{\rho}^{N},\tau',\tau'',\ldots,\tau^N)\equiv\Delta\tilde{\chi}(\boldsymbol{\rho}',\tau')\boldsymbol{\mathcal{K}}_{\rm E}(\boldsymbol{\rho}',\boldsymbol{\rho}'',\tau'-\tau'')\Delta\tilde{\chi}(\boldsymbol{\rho}'',\tau'')
	\boldsymbol{\mathcal{K}}_{\rm E}(\boldsymbol{\rho}'',\boldsymbol{\rho}''',\tau''-\tau''')\ldots\\
	&\qquad\qquad\qquad\qquad\qquad\qquad\qquad\;\;\ldots\Delta\tilde{\chi}(\boldsymbol{\rho}^{N-1},\tau^{N-1})
	\boldsymbol{\mathcal{K}}_{\rm E}(\boldsymbol{\rho}^{N-1},\boldsymbol{\rho}^N,\tau^{N-1}-\tau^{N})\Delta\tilde{\chi}(\boldsymbol{\rho}^{N},\tau^{N})
	\hat{\boldsymbol{\mathcal{E}}}_0(\boldsymbol{\rho}^{N};\tau^{N}).
	\label{Eq.III.18}
\end{align}

As indicated previously, to calculate the total contribution of the complex-like electric field operator in the frequency domain, we take the Laplace transform over every term given above, so that:
\begin{subequations}
	\begin{align}
		\hat{\bf E}^{(+)}_0({\bf r};\omega)&=2\pi\int\limits_{-\infty}^{+\infty}{d^3\boldsymbol{\rho} {\bf G}_{\rm E}({\bf r},\boldsymbol{\rho},\omega)\hat{\bf f}(\boldsymbol{\rho},\omega;t=0)},
		\label{Eq.III.19a}\\
		\hat{\bf E}^{(+)}_1({\bf r};\omega)&=\int\limits_{-\infty}^{+\infty}{d^3\boldsymbol{\rho}'{\bf K}^{(+)}_{\rm E}({\bf r},\boldsymbol{\rho}',\omega)\mathcal{L}_\omega{[\Delta\tilde{\chi}(\boldsymbol{\rho}',\tau')\hat{\boldsymbol{\mathcal{E}}}_0(\boldsymbol{\rho}';\tau')]}},
		\label{Eq.III.19b}\\
		\hat{\bf E}^{(+)}_2({\bf r};\omega)&=\iint\limits_{-\infty}^{\quad+\infty}{d^3\boldsymbol{\rho}'d^3\boldsymbol{\rho}''{\bf K}^{(+)}_{\rm E}({\bf r},\boldsymbol{\rho}',\omega)\left\{\!\int\limits_{0}^{+\infty}{\!d\tau''\int\limits_{0}^{+\infty}{\!d\tilde{\tau}'\Delta\tilde{\chi}(\boldsymbol{\rho}',\tilde{\tau}'+\tau'')\boldsymbol{\mathcal{K}}_{\rm E}(\boldsymbol{\rho}',\boldsymbol{\rho}'',\tilde{\tau}')\Delta\tilde{\chi}(\boldsymbol{\rho}'',\tau'')\hat{\boldsymbol{\mathcal{E}}}_0(\boldsymbol{\rho}'';\tau'')e^{i\omega\tilde{\tau}'}e^{i\omega\tau''}}}\right\}},
		\label{Eq.III.19c}
	\end{align}
\end{subequations}
where $\tilde{t}=t-\tau'$ and $\tilde{\tau}'=\tau'-\tau''$. From this latter expression it can be observed that the second-order term, $\hat{\bf E}^{(+)}_2({\bf r};\omega)$, gives rise to some mixing (or coupling) terms involving ancillary times appearing in the time-varying susceptibility function, $\Delta\tilde{\chi}(\boldsymbol{\rho}',\tilde{\tau}'+\tau'')$. This in turn would lead to a intertwining of Laplace transforms, hindering so the simplification of the expression to get a sort of nested (or concatenated) product of frequency-dependent functions. Still, one can take into consideration the following property of Laplace transforms relating products to convolutions:
\begin{equation}
	\mathcal{L}_\omega{[f(t)g(t)]}\equiv\int\limits_{0}^{+\infty}{dt f(t)g(t)e^{i\omega t}}=\frac{1}{2\pi}\lim\limits_{\xi\to\infty}{\int\limits_{-\xi+i\zeta}^{+\xi+i\zeta}{d\omega' F(\omega')G(\omega-\omega') } }=\frac{1}{2\pi}\int\limits_{-\infty+i\zeta}^{+\infty+i\zeta}{d\omega' F(\omega')G(\omega-\omega') },
	\label{Eq.III.20}
\end{equation}
where it has been used the following definition for the Laplace transform: $\mathcal{L}_{\omega}{[f(t)]}\equiv F(\omega)\equiv\int_{0}^{+\infty}{dtf(t)e^{i\omega t} }$. It is important to note that the convergence of $F(\omega')$ should be ensured along the horizontal line $\text{Im}{\left[\omega'\right]}=\zeta$, that could be set to $\zeta\to 0^+$. By proceeding in this way the computation merely reduces to only one integral (eventually requiring the application of \textit{Cauchy's residue theorem}), simplifying so considerably the mathematics. Then, the above results can be recast as:
\begin{subequations}
	\begin{align}
		\hat{\bf E}^{(+)}_0({\bf r};\omega)&=2\pi\int\limits_{-\infty}^{+\infty}{d^3\boldsymbol{\rho} {\bf G}_{\rm E}({\bf r},\boldsymbol{\rho},\omega)\hat{\bf f}(\boldsymbol{\rho},\omega;t=0)},
		\label{Eq.III.21a}\\
		\hat{\bf E}^{(+)}_1({\bf r};\omega)&=\int\limits_{-\infty}^{+\infty}{d^3\boldsymbol{\rho}'{\bf K}^{(+)}_{\rm E}({\bf r},\boldsymbol{\rho}',\omega)\frac{1}{2\pi}\int\limits_{-\infty+i0^+}^{+\infty+i0^+}{d\omega' \Delta\tilde{\chi}(\boldsymbol{\rho}',\omega-\omega')\hat{\bf{E}}_0(\boldsymbol{\rho}';\omega') } },
		\label{Eq.III.21b}\\
		\hat{\bf E}^{(+)}_2({\bf r};\omega)&=\iint\limits_{-\infty}^{\quad+\infty}{d^3\boldsymbol{\rho}'d^3\boldsymbol{\rho}'' {\bf K}^{(+)}_{\rm E}({\bf r},\boldsymbol{\rho}',\omega)\left[\frac{1}{2\pi }\right]^2\!\!\iint\limits_{-\infty+i0^+}^{\quad+\infty+i0^+}{\!\!\!\!d\omega'd\omega''\Delta\tilde{\chi}(\boldsymbol{\rho}',\omega-\omega'){\bf K}_{\rm E}(\boldsymbol{\rho}',\boldsymbol{\rho}'',\omega')\Delta\tilde{\chi}(\boldsymbol{\rho}'',\omega'-\omega'')\hat{\bf E}_0(\boldsymbol{\rho}'';\omega'') } },
		\label{Eq.III.21c}\\
		\nonumber&\;\;\vdots\\
		\hat{\bf E}^{(+)}_N({\bf r};\omega)&=\int\limits_{-\infty}^{+\infty}{d^3\boldsymbol{\rho}'\int\limits_{-\infty}^{+\infty}{d^3\boldsymbol{\rho}''\ldots\int\limits_{-\infty}^{+\infty}{d^3\boldsymbol{\rho}^N {\bf K}^{(+)}_{\rm E}({\bf r},\boldsymbol{\rho}',\omega)\int\limits_{-\infty+i0^+}^{+\infty+i0^+}{d\omega'\Delta\tilde{\chi}(\boldsymbol{\rho}',\omega-\omega')\boldsymbol{\Xi}^N(\boldsymbol{\rho}',\boldsymbol{\rho}'',\ldots,\boldsymbol{\rho}^{N},\omega',\omega'',\ldots,\omega^N) } } } },
		\label{Eq.III.21d}
	\end{align}
\end{subequations}
with
\begin{align}
	&\nonumber\boldsymbol{\Xi}^N(\boldsymbol{\rho}',\boldsymbol{\rho}'',\ldots,\boldsymbol{\rho}^{N},\omega',\omega'',\ldots,\omega^N)=\left[\frac{1}{2\pi}\right]^N\int\limits_{-\infty+i0^+}^{+\infty+i0^+}{d\omega'\Delta\tilde{\chi}(\boldsymbol{\rho}',\omega-\omega'){\bf K}_{\rm E}(\boldsymbol{\rho}',\boldsymbol{\rho}'',\omega')}\\
	&\qquad\qquad\int\limits_{-\infty+i0^+}^{+\infty+i0^+}{d\omega''\Delta\tilde{\chi}(\boldsymbol{\rho}'',\omega'-\omega''){\bf K}_{\rm E}(\boldsymbol{\rho}'',\boldsymbol{\rho}''',\omega'')}
	\quad\ldots\int\limits_{-\infty+i0^+}^{+\infty+i0^+}{d\omega^N\Delta\tilde{\chi}(\boldsymbol{\rho}^N,\omega^{N-1}-\omega^N)\hat{\bf E}(\boldsymbol{\rho}^{N};\omega^N)}.
	\label{Eq.III.22}
\end{align}

\section{IV. CURRENT DENSITY OPERATORS: DEFINITIONS AND CORRELATIONS}
\label{SectSI.IV}
\renewcommand\theequation{S.IV.\arabic{equation}}
\setcounter{equation}{0}

In this section we calculate the current density operator associated to the fields. Such a calculation can be made just by a simple comparison of the previous expressions with the following definition for the electric field:
\begin{align}
	\hat{\bf E}({\bf r};\omega)&=i\omega\mu_0\left[\int\limits_{-\infty}^{+\infty}{d^3\boldsymbol{\rho}{\bf G}({\bf r},\boldsymbol{\rho},\omega) \cdot \hat{\bf j}(\boldsymbol{\rho};\omega)}+\int\limits_{-\infty}^{+\infty}{d^3\boldsymbol{\rho}{\bf G}^*({\bf r},\boldsymbol{\rho},\omega) \cdot \hat{\bf j}^\dagger(\boldsymbol{\rho};\omega)}\right]=\hat{\bf E}^{(+)}({\bf r};\omega)+\hat{\bf E}^{(-)}({\bf r};\omega).
	\label{Eq.IV.01}
\end{align}
Then, since
\begin{equation}
	\nonumber\hat{\bf E}^{(+)}_0({\bf r};\omega)=2\pi\int\limits_{-\infty}^{+\infty}{d^3\boldsymbol{\rho}{\bf G}_{\rm E}({\bf r},\boldsymbol{\rho},\omega)\hat{\bf f}(\boldsymbol{\rho},\omega;t=0)}=2\pi i\sqrt{\frac{\hbar}{\pi\varepsilon_0}}\left(\frac{\omega}{c}\right)^2\int\limits_{-\infty}^{+\infty}{d^3\boldsymbol{\rho}\sqrt{\text{Im}{\left[\varepsilon(\boldsymbol{\rho},\omega)\right]}}{\bf G}({\bf r},\boldsymbol{\rho},\omega)\hat{\bf f}(\boldsymbol{\rho},\omega;t=0)},
\end{equation}
it follows that:
\begin{tcolorbox}[sharp corners, colback=blue!5!white,colframe=blue!50!white]
	\vspace{-0.35cm}
	\begin{equation}
		\hat{\bf j}_0({\bf r};\omega)=\omega \sqrt{4\pi\varepsilon_0\hbar}\sqrt{\text{Im}{\left[\varepsilon({\bf r},\omega)\right]}}\hat{\bf f}({\bf r},\omega;t=0).
		\label{Eq.IV.02}
	\end{equation}
\end{tcolorbox}
\noindent Similarly, for the first-order electric field,
\begin{equation*}
	\hat{\bf E}^{(+)}_1({\bf r};\omega)=\frac{2\omega^2}{\varepsilon_0c^2}\int\limits_{-\infty}^{+\infty}{d^3\boldsymbol{\rho}' {\bf G}({\bf r},\boldsymbol{\rho}',\omega)\frac{1}{2\pi}\int\limits_{-\infty+i0^+}^{+\infty+i0^+}{d\omega'\Delta\tilde{\chi}(\boldsymbol{\rho}',\omega-\omega')\hat{\bf E}_0(\boldsymbol{\rho}';\omega')}},
\end{equation*}
the associated current density operator reads as:
\begin{tcolorbox}[sharp corners, colback=blue!5!white,colframe=blue!50!white]
	\vspace{-0.35cm}
	\begin{equation}
		\hat{\bf j}_1({\bf r};\omega)=\frac{\omega\mu_0}{\pi}\int\limits_{-\infty}^{+\infty}{d^3\boldsymbol{\rho}'\int\limits_{-\infty+i0^+}^{+\infty+i0^+}{d\omega' \omega'\Delta\tilde{\chi}({\bf r},\omega-\omega')\left[{\bf G}({\bf r},\boldsymbol{\rho}',\omega')\hat{\bf j}_0(\boldsymbol{\rho}';\omega')+{\bf G}^*({\bf r},\boldsymbol{\rho}',\omega')\hat{\bf j}^\dagger_0(\boldsymbol{\rho}';\omega')\right]}}.
		\label{Eq.IV.03}
	\end{equation}
\end{tcolorbox}
\noindent Likewise, for the second-order field,
\begin{equation*}
	\hat{\bf E}^{(+)}_2({\bf r};\omega)=\frac{2\omega^2}{\varepsilon_0c^2}\iint\limits_{-\infty}^{\quad+\infty}{d^3\boldsymbol{\rho}'d^3\boldsymbol{\rho}'' {\bf G}({\bf r},\boldsymbol{\rho}',\omega)\left[\frac{1}{2\pi }\right]^2\!\!\iint\limits_{-\infty+i0^+}^{\quad+\infty+i0^+}{\!\!\!\!d\omega'd\omega''\Delta\tilde{\chi}(\boldsymbol{\rho}',\omega-\omega'){\bf K}_{\rm E}(\boldsymbol{\rho}',\boldsymbol{\rho}'',\omega')\Delta\tilde{\chi}(\boldsymbol{\rho}'',\omega'-\omega'')\hat{\bf E}_0(\boldsymbol{\rho}'';\omega'') } },
\end{equation*}
and the corresponding current density operator is:
\begin{tcolorbox}[sharp corners, colback=blue!5!white,colframe=blue!50!white]
	\vspace{-0.35cm}
	\begin{equation}
		\hat{\bf j}_2({\bf r};\omega)=\frac{\omega\mu_0}{\pi}\int\limits_{-\infty}^{+\infty}{d^3\boldsymbol{\rho}' \int\limits_{-\infty+i0^+}^{+\infty+i0^+}{d\omega'\omega'\Delta\tilde{\chi}({\bf r},\omega-\omega'){\bf G}({\bf r},\boldsymbol{\rho}',\omega')\hat{\bf j}_1(\boldsymbol{\rho}';\omega') } }.
		\label{Eq.IV.04}
	\end{equation}
\end{tcolorbox}

These expressions allow us to perform the calculation of the corresponding $n$th-order correlations. To do that, we evaluate $\braket{\hat{\bf j}^\dagger_l({\bf r};\omega)\cdot\hat{\bf j}_m({\bf r';\omega'})}_{\rm th}$, where $l+m=n$, and the subscript ${\rm th}$ stands for thermal average, namely, $\braket{\ldots}_{\rm th}\equiv\text{Tr}{[\ldots\hat{\rho}_{\rm th}]}$, with $\hat{\rho}_{\rm th}$ being the \textit{canonical density operator} of electromagnetic fields in thermal equilibrium at temperature $T$ (i.e., a \textit{thermal field}). Then, the \textbf{zeroth-order correlation} of the current density reads as:
\begin{tcolorbox}[sharp corners, colback=red!5!white,colframe=red!50!white]
	\vspace{-0.35cm}
	\begin{align}
		\nonumber\braket{\hat{\bf j}^\dagger_0({\bf r};\omega)\cdot\hat{\bf j}_0({\bf r';\omega'})}_{\rm th}&=4\pi\varepsilon_0\hbar\omega\omega'\sqrt{\text{Im}{\left[\varepsilon({\bf r},\omega)\right]}}\sqrt{\text{Im}{\left[\varepsilon({\bf r}',\omega')\right]}}\braket{\hat{\bf f}^\dagger({\bf r},\omega)\hat{\bf f}({\bf r}',\omega')}_{\rm th}\\
		&=\omega'\mathcal{S}({\bf r};\omega;T)\delta{[{\bf r}-{\bf r}']}\delta{[\omega-\omega']},
		\label{Eq.IV.05}
	\end{align}
	where
	\begin{equation}
		\mathcal{S}({\bf r};\omega;T)\equiv 4\pi\varepsilon_0\text{Im}{\left[\varepsilon({\bf r},\omega)\right]}\hbar\omega\Theta(\omega,T),
		\label{Eq.IV.06}
	\end{equation}
	and $\Theta(\omega,T)=[e^{\hbar \omega/(k_BT)}-1]^{-1}$ stands for the average number of thermal photons, with $\hbar$ and $k_B$ being, respectively, the reduced Planck's and the Boltzmann constants.
\end{tcolorbox}
\noindent Equation~\eqref{Eq.IV.05} is a very well-known result that provides the correlation of fluctuating currents in absence of any kind of time-modulation. Next, and from this result, one can straightforwardly calculate the \textbf{first-order correlation}:
\begin{align*}
	\braket{\hat{\bf j}^\dagger_0({\bf r};\omega)\cdot\hat{\bf j}_1({\bf r}';\omega')}_{\rm th}&=\frac{\omega'\mu_0}{\pi}\int\limits_{-\infty}^{+\infty}{d^3\boldsymbol{\rho}''\left[\int\limits_{-\infty+i0^+}^{+\infty+i0^+}{d\omega'' \omega''\Delta\tilde{\chi}({\bf r}',\omega'-\omega''){\bf G}({\bf r}',\boldsymbol{\rho}'',\omega'')\braket{\hat{\bf j}^\dagger_0({\bf r};\omega)\cdot\hat{\bf j}_0(\boldsymbol{\rho}'';\omega'') }_{\rm th}}\right]}\\
	&=\omega'4\pi\varepsilon_0\text{Im}{\left[\varepsilon({\bf r},\omega)\right]}\hbar\omega\Theta(\omega,T)\left[\frac{\mu_0}{\pi}\int\limits_{-\infty+i0^+}^{+\infty+i0^+}{d\omega'' (\omega'')^2\Delta\tilde{\chi}({\bf r}',\omega'-\omega''){\bf G}({\bf r}',{\bf r},\omega'')\delta{[\omega-\omega'']}}\right]\\
	&=\omega'\mathcal{S}({\bf r};\omega;T)\left[\frac{\mu_0\omega^2}{\pi}\Delta\tilde{\chi}({\bf r}',\omega'-\omega){\bf G}({\bf r}',{\bf r},\omega)\right],\\
	\braket{\hat{\bf j}^\dagger_1({\bf r};\omega)\cdot\hat{\bf j}_0({\bf r}';\omega')}_{\rm th}&=\frac{\omega\mu_0}{\pi}\int\limits_{-\infty}^{+\infty}{d^3\boldsymbol{\rho}''\left[\int\limits_{-\infty+i0^+}^{+\infty+i0^+}{d\omega''\omega''\Delta\tilde{\chi}^*({\bf r},\omega-\omega''){\bf G}^*({\bf r},\boldsymbol{\rho}'',\omega'')\braket{\hat{\bf j}^\dagger_0(\boldsymbol{\rho}'';\omega'')\cdot\hat{\bf j}_0({\bf r}';\omega')}_{\rm th} }\right]}\\
	&=\omega 4\pi\varepsilon_0\text{Im}{\left[\varepsilon({\bf r}',\omega')\right]}\hbar\omega'\Theta(\omega',T)\left[\frac{\mu_0}{\pi}\int\limits_{-\infty+i0^+}^{+\infty+i0^+}{d\omega''(\omega'')^2\Delta\tilde{\chi}^*({\bf r},\omega-\omega''){\bf G}^*({\bf r},{\bf r}',\omega'')\delta{[\omega''-\omega']} }\right]\\
	&=\omega\mathcal{S}({\bf r}';\omega';T)\left[\frac{\mu_0\omega'^2}{\pi}\Delta\tilde{\chi}^*({\bf r},\omega-\omega'){\bf G}^*({\bf r},{\bf r}',\omega') \right],
\end{align*}
so that:
\begin{tcolorbox}[sharp corners, colback=red!5!white,colframe=red!50!white]
	\vspace{-0.35cm}
	\begin{align}
		\nonumber\braket{\hat{\bf j}^\dagger_0({\bf r};\omega)\hat{\bf j}_1({\bf r}';\omega')+\hat{\bf j}^\dagger_1({\bf r};\omega)\hat{\bf j}_0({\bf r';\omega'})}_{\rm th}&=\frac{\omega\omega'\mu_0}{\pi}\left[\omega\mathcal{S}({\bf r};\omega;T)\Delta\tilde{\chi}({\bf r}',\omega'-\omega){\bf G}({\bf r}',{\bf r},\omega)\right]\\
		&\qquad\qquad+\frac{\omega\omega'\mu_0}{\pi}\left[\omega'\mathcal{S}({\bf r}',\omega';T)\Delta\tilde{\chi}^*({\bf r},\omega-\omega'){\bf G}^*({\bf r},{\bf r}',\omega')\right].
		\label{Eq.IV.08}
	\end{align}
\end{tcolorbox}
\noindent Similarly, from the above result one could calculate the \textbf{second-order correlation}:
\begin{align*}
	&\braket{\hat{\bf j}^\dagger_0({\bf r};\omega)\cdot\hat{\bf j}_2({\bf r}';\omega')}_{\rm th}=\frac{\omega'\mu_0}{\pi}\int\limits_{-\infty}^{+\infty}{d^3\boldsymbol{\rho}'' \left[\int\limits_{-\infty+i0^+}^{+\infty+i0^+}{d\omega'' \omega''\Delta\tilde{\chi}({\bf r}',\omega'-\omega''){\bf G}({\bf r}',\boldsymbol{\rho}'',\omega'')\braket{\hat{\bf j}^\dagger_0({\bf r};\omega)\cdot\hat{\bf j}_1(\boldsymbol{\rho}'';\omega'')}_{\rm th} }\right]}\\
	&\qquad=\omega'\mathcal{S}({\bf r};\omega;T)\left[\frac{\mu_0\omega}{\pi}\right]^2\int\limits_{-\infty}^{+\infty}{d^3\boldsymbol{\rho}''\left[\int\limits_{-\infty+i0^+}^{+\infty+i0^+}{d\omega''(\omega'')^2\Delta\tilde{\chi}({\bf r}',\omega'-\omega''){\bf G}({\bf r}',\boldsymbol{\rho}'',\omega'')\Delta\tilde{\chi}(\boldsymbol{\rho}'',\omega''-\omega){\bf G}(\boldsymbol{\rho}'',{\bf r},\omega)  }\right]},\\
	&\braket{\hat{\bf j}^\dagger_2({\bf r};\omega)\cdot\hat{\bf j}_0({\bf r}';\omega')}_{\rm th}=\frac{\omega\mu_0}{\pi}\int\limits_{-\infty}^{+\infty}{d^3\boldsymbol{\rho}'' \left[\int\limits_{-\infty+i0^+}^{+\infty+i0^+}{d\omega'' \omega''\Delta\tilde{\chi}^*({\bf r},\omega-\omega''){\bf G}^*({\bf r},\boldsymbol{\rho}'',\omega'')\braket{\hat{\bf j}^\dagger_1(\boldsymbol{\rho}'';\omega'')\cdot\hat{\bf j}_0({\bf r}';\omega')}_{\rm th} }\right]}\\
	&\qquad=\omega\mathcal{S}({\bf r}';\omega';T)\left[\frac{\mu_0\omega'}{\pi}\right]^2\int\limits_{-\infty}^{+\infty}{d^3\boldsymbol{\rho}''\left[\int\limits_{-\infty+i0^+}^{+\infty+i0^+}{d\omega''(\omega'')^2\Delta\tilde{\chi}^*({\bf r},\omega-\omega''){\bf G}^*({\bf r},\boldsymbol{\rho}'',\omega'')\Delta\tilde{\chi}^*(\boldsymbol{\rho}'',\omega''-\omega'){\bf G}^*(\boldsymbol{\rho}'',{\bf r}',\omega') }\right]},
\end{align*}
\begin{align*}
	&\braket{\hat{\bf j}^\dagger_1({\bf r};\omega)\cdot\hat{\bf j}_1({\bf r}';\omega')}_{\rm th}=\omega\omega'\left[\frac{\mu_0}{\pi}\right]^2\iint\limits_{-\infty}^{\quad+\infty}{d^3\boldsymbol{\rho}d^3\boldsymbol{\rho}'\iint\limits_{-\infty+i0^+}^{\quad+\infty+i0^+}{d\tilde{\omega}d\tilde{\omega}'\tilde{\omega}\tilde{\omega}'\Delta\tilde{\chi}^*({\bf r},\omega-\tilde{\omega})\Delta\tilde{\chi}({\bf r}',\omega'-\tilde{\omega}')}}\\
	&\qquad\qquad\qquad\qquad\qquad\qquad \left[{\bf G}^*({\bf r},\boldsymbol{\rho},\tilde{\omega}){\bf G}({\bf r}',\boldsymbol{\rho}',\tilde{\omega}')\braket{\hat{\bf j}^\dagger_0(\boldsymbol{\rho};\tilde{\omega})\cdot\hat{\bf j}_0(\boldsymbol{\rho}';\tilde{\omega}')}_{\rm th}+{\bf G}({\bf r},\boldsymbol{\rho},\tilde{\omega}){\bf G}^*({\bf r}',\boldsymbol{\rho}',\tilde{\omega}')\braket{\hat{\bf j}_0(\boldsymbol{\rho};\tilde{\omega})\cdot\hat{\bf j}^\dagger_0(\boldsymbol{\rho}';\tilde{\omega}')}_{\rm th}\right]\\
	&\qquad=4\hbar\omega\omega'\left[\frac{\mu_0}{\pi}\right]\int\limits_{-\infty+i0^+}^{+\infty+i0^+}{d\tilde{\omega}(\tilde{\omega})^2\Delta\tilde{\chi}^*({\bf r},\omega-\tilde{\omega})\Delta\tilde{\chi}({\bf r}',\omega'-\tilde{\omega})\left\{\text{Im}{\left[{\bf G}({\bf r},{\bf r}',\tilde{\omega})\right]}\left[1+\Theta(\tilde{\omega},T)\right]+\Theta(\tilde{\omega},T)\text{Im}{\left[{\bf G}({\bf r}',{\bf r},\tilde{\omega})\right]}\right\}}.
\end{align*}

\section{V. ELECTRIC-FIELD CORRELATIONS}
\label{SectSI.V}
\renewcommand\theequation{S.V.\arabic{equation}}
\setcounter{equation}{0}

In this section, and taking into account the expressions given above, we shall proceed with the calculation electric-field correlations of successive orders, a result that directly will lead us to the of thermal emission spectra.
\begin{itemize}
	\item \textbf{Zeroth-order of E-field correlations:}
\end{itemize}
\begin{align*}
	S_{0,0}({\bf r};\omega;T)=\braket{(\hat{\bf E}^{(+)}_0({\bf r};\omega))^\dagger\cdot\hat{\bf E}^{(+)}_0({\bf r};\omega)}_{\rm th}&=\omega^2\mu_0^2\iint\limits_{-\infty}^{\quad+\infty}{d^3\boldsymbol{\rho}d^3\boldsymbol{\rho}'{\bf G}^*({\bf r},\boldsymbol{\rho},\omega){\bf G}({\bf r},\boldsymbol{\rho}',\omega)\braket{\hat{\bf j}^\dagger_0(\boldsymbol{\rho};\omega)\cdot\hat{\bf j}_0(\boldsymbol{\rho}';\omega)}_{\rm th}}\\
	&=4\pi\omega\mu_0\hbar\omega\Theta(\omega,T)\int\limits_{-\infty}^{+\infty}{d^3\boldsymbol{\rho}\frac{\omega^2}{c^2}\text{Im}{\left[\varepsilon(\boldsymbol{\rho},\omega)\right]}{\bf G}^*({\bf r},\boldsymbol{\rho},\omega){\bf G}({\bf r},\boldsymbol{\rho},\omega)},
\end{align*}
\begin{tcolorbox}[sharp corners, colback=red!5!white,colframe=red!50!white]
	\vspace{-0.35cm}
	\begin{equation}
		S_{0}({\bf r};\omega;T)=S_{0,0}({\bf r};\omega;T)=4\pi\omega\mu_0\hbar\omega\Theta(\omega,T)\text{Im}{\left[{\bf G}({\bf r},{\bf r},\omega)\right]}.
		\label{Eq.V.01}
	\end{equation}
\end{tcolorbox}
\begin{itemize}
	\item \textbf{First-order of E-field correlations:}
\end{itemize}
\begin{align*}
	S_{0,1}({\bf r};\omega;T)=\braket{(\hat{\bf E}^{(+)}_0({\bf r};\omega))^\dagger\cdot\hat{\bf E}^{(+)}_1({\bf r};\omega)}_{\rm th}&=\omega^2\mu_0^2\iint\limits_{-\infty}^{\quad+\infty}{d^3\boldsymbol{\rho}d^3\boldsymbol{\rho}'{\bf G}^*({\bf r},\boldsymbol{\rho},\omega){\bf G}({\bf r},\boldsymbol{\rho}',\omega)\braket{\hat{\bf j}^\dagger_0(\boldsymbol{\rho};\omega)\cdot\hat{\bf j}_1(\boldsymbol{\rho}';\omega)}_{\rm th}}\\
	&=4\pi\omega\mu_0 \hbar\omega\Theta(\omega,T) \omega^2\frac{\mu_0}{\pi}\int\limits_{-\infty}^{+\infty}{d^3\boldsymbol{\rho}\text{Im}{\left[{\bf G}(\boldsymbol{\rho},{\bf r},\omega)\right]}{\bf G}({\bf r},\boldsymbol{\rho},\omega)\Delta\tilde{\chi}(\boldsymbol{\rho},0) },
\end{align*}

\begin{tcolorbox}[sharp corners, colback=red!5!white,colframe=red!50!white]
	\vspace{-0.35cm}
	\begin{equation}
		S_{1}({\bf r};\omega;T)=2\text{Re}{\left[S_{0,1}({\bf r};\omega;T)\right]}=8\omega^3\mu_0^2\hbar\omega\Theta(\omega,T)\int\limits_{-\infty}^{+\infty}{d^3\boldsymbol{\rho}\text{Im}{\left[{\bf G}({\bf r},\boldsymbol{\rho},\omega)\right]}\text{Re}{\left[{\bf G}({\bf r},\boldsymbol{\rho},\omega)\Delta\tilde{\chi}(\boldsymbol{\rho},0)\right]} },
		\label{Eq.V.02}
	\end{equation}
	where it should be noted that $S_{1,0}({\bf r};\omega;T)=[S_{0,1}({\bf r};\omega;T)]^\dagger$.
\end{tcolorbox}
\begin{itemize}
	\item \textbf{Second-order of E-field correlations:}
\end{itemize}
\begin{align*}
	S_{0,2}({\bf r};\omega;T)&=\braket{(\hat{\bf E}^{(+)}_0({\bf r};\omega))^\dagger\cdot\hat{\bf E}^{(+)}_2({\bf r};\omega)}_{\rm th}=\omega^2\mu_0^2\iint\limits_{-\infty}^{\quad+\infty}{d^3\boldsymbol{\rho}d^3\boldsymbol{\rho}'{\bf G}^*({\bf r},\boldsymbol{\rho},\omega){\bf G}({\bf r},\boldsymbol{\rho}',\omega)\braket{\hat{\bf j}^\dagger_0(\boldsymbol{\rho};\omega)\cdot\hat{\bf j}_2(\boldsymbol{\rho}';\omega)}_{\rm th}}\\
	&\!\!\!\!\!\!\!\!\!\!\!\!\!\!\!\!\!\!\!\!\!\!\!\!\!\!\!\!\!\!\!\!=4\pi\omega\mu_0\hbar\omega\Theta(\omega,T)\frac{\omega^2\mu_0^2}{\pi^2}\iint\limits_{-\infty}^{\quad+\infty}{d^3\boldsymbol{\rho}d^3\boldsymbol{\rho}'{\bf G}({\bf r},\boldsymbol{\rho},\omega)\text{Im}{\left[{\bf G}(\boldsymbol{\rho}',{\bf r},\omega)\right]}\int\limits_{-\infty+i0^+}^{+\infty+i0^+}{d\tilde{\omega}(\tilde{\omega})^2\Delta\tilde{\chi}(\boldsymbol{\rho},\omega-\tilde{\omega})\Delta\tilde{\chi}(\boldsymbol{\rho}',\tilde{\omega}-\omega){\bf G}(\boldsymbol{\rho},\boldsymbol{\rho}',\tilde{\omega})  } },\\
	S_{1,1}({\bf r};\omega;T)&=\braket{(\hat{\bf E}^{(+)}_1({\bf r};\omega))^\dagger\cdot\hat{\bf E}^{(+)}_1({\bf r};\omega)}_{\rm th}=\omega^2\mu_0^2\iint\limits_{-\infty}^{\quad+\infty}{d^3\boldsymbol{\rho}d^3\boldsymbol{\rho}'{\bf G}^*({\bf r},\boldsymbol{\rho},\omega){\bf G}({\bf r},\boldsymbol{\rho}',\omega)\braket{\hat{\bf j}^\dagger_1(\boldsymbol{\rho};\omega)\cdot\hat{\bf j}_1(\boldsymbol{\rho}';\omega)}_{\rm th}}\\
	&\!\!\!\!\!\!\!\!\!\!\!\!\!\!\!\!\!\!\!\!\!\!\!\!\!\!\!\!\!\!\!\!=4\hbar\omega^4\frac{\mu_0^3}{\pi}\iint\limits_{-\infty}^{\quad+\infty}{d^3\boldsymbol{\rho}d^3\boldsymbol{\rho}'{\bf G}^*({\bf r},\boldsymbol{\rho},\omega){\bf G}({\bf r},\boldsymbol{\rho}',\omega)\!\!\int\limits_{-\infty+i0^+}^{+\infty+i0^+}{\!\!d\tilde{\omega}\left[1+\Theta(\tilde{\omega},T)\right]\tilde{\omega}^2\Delta\tilde{\chi}^*(\boldsymbol{\rho},\omega-\tilde{\omega})\Delta\tilde{\chi}(\boldsymbol{\rho}',\omega-\tilde{\omega})\text{Im}{\left[{\bf G}(\boldsymbol{\rho},\boldsymbol{\rho}',\tilde{\omega})\right]} } }\\
	&\!\!\!\!\!\!\!\!\!\!\!\!\!\!\!\!\!\!\!\!\!\!\!\!\!\!\!\!\!\!\!\!+4\hbar\omega^4\frac{\mu_0^3}{\pi}\iint\limits_{-\infty}^{\quad+\infty}{d^3\boldsymbol{\rho}d^3\boldsymbol{\rho}'{\bf G}^*({\bf r},\boldsymbol{\rho},\omega){\bf G}({\bf r},\boldsymbol{\rho}',\omega)\!\!\int\limits_{-\infty+i0^+}^{+\infty+i0^+}{\!\!d\tilde{\omega}\Theta(\tilde{\omega},T)\tilde{\omega}^2\Delta\tilde{\chi}^*(\boldsymbol{\rho},\omega-\tilde{\omega})\Delta\tilde{\chi}(\boldsymbol{\rho}',\omega-\tilde{\omega})\text{Im}{\left[{\bf G}(\boldsymbol{\rho}',\boldsymbol{\rho},\tilde{\omega})\right]} } }.
\end{align*}

\section{VI. THERMAL EMISSION OF A TIME-MODULATED SEMI-INFINITE HOMOGENEOUS MEDIUM}
\label{SectSI.VI}
\renewcommand\theequation{S.VI.\arabic{equation}}
\setcounter{equation}{0}

\subsection{Angular spectrum representation of the dyadic Green's function}
\label{SectSI.VI.A}

In order to demonstrate the usefulness of the above theoretical formulation on the fluctuation-dissipation theorem for time-varying quantum systems, here below we make use of it to tackle on a practical case in which we analyze the phenomenon of thermal emission. Specifically, we consider a flat interface separating a semi-infinite homogeneous and nonmagnetic medium, held in local thermodynamic equilibrium at a uniform temperature $T$ ($z\leq0$), from a vacuum~($z>\nolinebreak0$). For such a system, and in general, for any translationally symmetric structure exhibiting a planar geometry (namely, slabs, interfaces, or layered media, including thin-films), one can take advantage of a powerful mathematical formalism commonly referred to as the \textit{angular spectrum representation}. Such an approach allows one to express the dyadic Green's function of a system as a superposition of elementary plane waves \cite{NovotnySI,MandelSI}:
\begin{tcolorbox}[sharp corners]
	\vspace{-0.35cm}
	\begin{equation}
		{\bf G}({\bf r},\boldsymbol{\rho},\omega)=\frac{k_0^2}{4\pi^2}\iint\limits_{-\infty}^{\quad+\infty}{d^2\boldsymbol{\kappa}_\parallel \hat{\bf G}(\kappa_x,\kappa_y;\omega|z,\rho_z)e^{ik_0\left[\kappa_x\left(x-\rho_x\right)+\kappa_y\left(y-\rho_y\right)\right]}}.
		\label{Eq.VI.01}
	\end{equation}
\end{tcolorbox}
\noindent Then, assuming a homogeneous, isotropic and linear medium~\cite{footnote01SI}, it follows that \cite{Joulain2005SI}:
\begin{tcolorbox}[sharp corners]
	\vspace{-0.35cm}
	\begin{equation*}
		{\bf k}_i=k_0(\kappa_x,\kappa_y,\tilde{k}_{z,i})=k_0(\kappa_x,\kappa_y,\sqrt{\varepsilon_i(\omega)}\kappa_{z,i}) \; \Rightarrow \; {\bf k}_i^2=k_0^2\left[\kappa_x^2+\kappa_y^2+\tilde{k}_{z,i}^2\right]=k_0^2\left[\kappa_x^2+\kappa_y^2+\varepsilon_i(\omega)\kappa_{z,i}^2\right]=\varepsilon_i(\omega)k_0^2,
	\end{equation*}
	\begin{equation}
		\kappa_R^2+\tilde{k}_{z,i}^2=\kappa_R^2+\varepsilon_i(\omega)\kappa_{z,i}^2=\varepsilon_i(\omega)\quad\Longrightarrow\quad\kappa_{z,i}=\sqrt{1-\frac{\kappa_R^2}{\varepsilon_i(\omega)}},
		\label{Eq.VI.02}
	\end{equation}
	with $\tilde{k}_{z,i}=k_{z,i}/k_0=\sqrt{\varepsilon_i(\omega)}\kappa_{z,i}$, and $\kappa_R^2\equiv\kappa_x^2+\kappa_y^2$.
\end{tcolorbox}
\noindent Notice that, solely in the case of lossless media (i.e., those in which $\varepsilon_i(\omega)$ is real), it is possible to establish the following usual correspondence:
\begin{tcolorbox}[sharp corners]
	\vspace{-0.35cm}
	\begin{equation}
		\kappa_{z,i}=\left\{\begin{matrix}
			\displaystyle \sqrt{1-\kappa_R^2/\varepsilon_i(\omega)},	&\text{if} &\kappa_R^2\leq \varepsilon_i(\omega)& \quad \Longrightarrow \quad & \text{propagating modes (far-field)},\\\\
			\displaystyle i\sqrt{\kappa_R^2/\varepsilon_i(\omega)-1},	&\text{if} &\kappa_R^2>\varepsilon_i(\omega)	&\quad \Longrightarrow \quad & \text{evanescent modes (near-field)},
		\end{matrix}\right.
		\label{Eq.VI.03}
	\end{equation}
	which for the vacuum leads to the most standard form:
	\begin{equation}
		\kappa_{z,vac}=\left\{\begin{matrix}
			\displaystyle \sqrt{1-\kappa_R^2},	&\text{if} &\kappa_R^2\leq 1& \quad \Longrightarrow \quad & \text{propagating modes (far-field)},\\\\
			\displaystyle i\sqrt{\kappa_R^2-1},	&\text{if} &\kappa_R^2>1	&\quad \Longrightarrow \quad & \text{evanescent modes (near-field)}.
		\end{matrix}\right.
		\label{Eq.VI.04}
	\end{equation}
\end{tcolorbox}
\noindent Regarding this general framework, henceforth we shall only deal with two different media:
\begin{itemize}
	\item \textbf{medium} ${\bf 1}$, that will be the vacuum, so that $\varepsilon_1(\omega)\equiv1$, and thus having a longitudinal wave vector such that $k_{z,1}=k_0\tilde{k}_{z,1}=k_0\kappa_{z,1}$, with $\kappa_{z,1}=\sqrt{1-\kappa_R^2}$, which will be denoted by means of $(x,y,z)$-coordinates;
	\item \textbf{medium} ${\bf 2}$, that will be a dispersive material, so that $\varepsilon_2(\omega)\equiv\varepsilon(\omega)$, which, in general will be a complex-like function, and thus having a longitudinal wave vector $k_{z,2}=k_0\tilde{k}_{z,2}=k_0\sqrt{\varepsilon(\omega)}\kappa_{z,2}$, with $\kappa_{z,2}=\sqrt{1-\kappa_R^2/\varepsilon(\omega)}$ (that, due to the complex-like character of $\varepsilon(\omega)$, it will not directly fulfill the above correspondence with propagating and evanescent modes). In this case, we shall use $(\rho_x,\rho_y,\rho_z)$-coordinates to denote this medium~\cite{footnote02SI}.
\end{itemize}

Likewise, and on this basis, it is worth setting down a general formulation to deal with the dyadic notation of the Green's tensors (cf. Refs.~\cite{NovotnySI,MandelSI,Joulain2005SI,Sipe1987SI} for further details on it). Specifically, considering a plane interface system as mentioned above, the Green's tensor relating currents in the lower half-space (medium $j$; with coordinates $\rho_z\leq0$) to fields in  the upper half-space (medium $i$; with coordinates $z>0$), can be generally expressed as:
\begin{tcolorbox}[sharp corners]
	\vspace{-0.35cm}
	\begin{equation}
		\hat{\bf G}^{(\updownarrow/\updownarrow)}_{i\leftarrow j}(\kappa_R,\kappa_\varphi;\omega|z,\rho_z)=\frac{i}{2}\frac{1}{k_{z,j}}\left[t^{\rm(s)}_{i\leftarrow j}\hat{\bf g}^{\rm (s)}_{i\leftarrow j}+t^{\rm(p)}_{i\leftarrow j}\hat{\bf g}^{\rm(p)(\updownarrow/\updownarrow)}_{i\leftarrow j}\right]\Gamma_{i\leftarrow j},
		\label{Eq.VI.05}
	\end{equation}
\end{tcolorbox}
\noindent where
\begin{tcolorbox}[sharp corners]
	\vspace{-0.35cm}
	\begin{subequations}
		\begin{align}
			&\hat{\bf g}^{\rm (s)}_{i\leftarrow j}\equiv \hat{\bf s}\otimes\hat{\bf s} =\begin{bmatrix}
				\sin^2{\kappa_\varphi} & -\cos{\kappa_\varphi} \sin{\kappa_\varphi} & 0 \\
				-\cos{\kappa_\varphi} \sin{\kappa_\varphi} & \cos^2{\kappa_\varphi} & 0 \\
				0 & 0 & 0
			\end{bmatrix},
			\label{Eq.VI.06a}\\
			&\hat{\bf g}^{\rm(p)(\updownarrow/\updownarrow)}_{i\leftarrow j}\equiv \hat{\bf p}^{\updownarrow}_i\otimes\hat{\bf p}^{\updownarrow}_j =\begin{bmatrix}
				\kappa_{z,i}\kappa_{z,j}\cos^2{\kappa_\varphi} & \kappa_{z,i}\kappa_{z,j}\cos{\kappa_\varphi} \sin{\kappa_\varphi} & \frac{\kappa_{z,i}\kappa_R\cos{\kappa_\varphi}}{(\mp)\sqrt{\varepsilon_j(\omega)}}\\
				\kappa_{z,i}\kappa_{z,j}\cos{\kappa_\varphi} \sin{\kappa_\varphi} & \kappa_{z,i}\kappa_{z,j}\sin^2{\kappa_\varphi} & \frac{\kappa_{z,i}\kappa_R\sin{\kappa_\varphi}}{(\mp)\sqrt{\varepsilon_j(\omega)}} \\
				\frac{\kappa_{z,j}\kappa_R\cos{\kappa_\varphi}}{(\mp)\sqrt{\varepsilon_i(\omega)}} & \frac{\kappa_{z,j}\kappa_R\sin{\kappa_\varphi}}{(\mp)\sqrt{\varepsilon_i(\omega)}} & \frac{\kappa_R^2}{(\pm)\sqrt{\varepsilon_i(\omega)}(\pm)\sqrt{\varepsilon_j(\omega)}}
			\end{bmatrix},
			\label{Eq.VI.06b}
		\end{align}
	\end{subequations}
	result from the dyadic product of the corresponding \textit{polarization-vector basis},
	\begin{subequations}
		\begin{align}
			&\hat{\bf s}\equiv\begin{pmatrix}\sin{\kappa_\varphi},-\cos{\kappa_\varphi},0\end{pmatrix},
			\label{Eq.VI.07a}\\
			&\hat{\bf p}_i^{\uparrow/\downarrow}\equiv\displaystyle\begin{pmatrix}- \kappa_{z,i}\cos{\kappa_\varphi},- \kappa_{z,i}\sin{\kappa_\varphi},\pm\kappa_R/\sqrt{\varepsilon_i(\omega)}\end{pmatrix},
			\label{Eq.VI.07b}
		\end{align}
	\end{subequations}
	with the signs $+$ and $-$ being associated to modes propagating upward ($\uparrow$) and downward ($\downarrow$). Furthermore,
	\begin{subequations}
		\begin{align}
			t^{\rm(s)}_{i\leftarrow j}&=\frac{2\sqrt{\varepsilon_j(\omega)}\kappa_{z,j}}{\sqrt{\varepsilon_i(\omega)}\kappa_{z,i}+\sqrt{\varepsilon_j(\omega)}\kappa_{z,j}},
			\label{Eq.VI.08a}\\
			t^{\rm(p)}_{i\leftarrow j}&=\frac{2\sqrt{\varepsilon_i(\omega)}\varepsilon_j(\omega)\kappa_{z,j}}{\varepsilon_j(\omega)\sqrt{\varepsilon_i(\omega)}\kappa_{z,i}+\varepsilon_i(\omega)\sqrt{\varepsilon_j(\omega)}\kappa_{z,j}},
			\label{Eq.VI.08b}
		\end{align}
	\end{subequations}
	stand for the \textit{Fresnel transmission coefficients}, and
	\begin{equation}
		\Gamma_{i\leftarrow j}=e^{ik_0\left[\sqrt{\varepsilon_i(\omega)}\kappa_{z,i}z-\sqrt{\varepsilon_j(\omega)}\kappa_{z,j}\rho_z\right]},
		\label{Eq.VI.09}
	\end{equation}
	is the \textit{field propagator}, where it has been implicitly assumed that the medium $j$ is underneath the medium $i$.
\end{tcolorbox}
On the other hand, in the case of two points lying in the same medium, e.g., in the lower medium $j$, the above Green's tensor has to be recast as:
\begin{tcolorbox}[sharp corners]
	\vspace{-0.35cm}
	\begin{align}
		\nonumber\hat{\bf G}^{(\updownarrow/\updownarrow)}_{i\leftarrow j}(\kappa_R,\kappa_\varphi;\omega|\rho_z,\rho_z')&=\frac{i}{2}\frac{1}{k_{z,j}}\left\{\left[r^{\rm(s)}_{j\leftarrow i}\hat{\bf g}^{\rm (s)}_{j\leftarrow j'}+r^{\rm(p)}_{j\leftarrow i}\hat{\bf g}^{\rm(p)(\downarrow/\uparrow)}_{j\leftarrow j'}\right]e^{-ik_0\sqrt{\varepsilon_j(\omega)}\kappa_{z,j}(\rho_z+\rho_z')}\right\}\\
		&+\frac{i}{2}\frac{1}{k_{z,j}}\left\{\left[\hat{\bf g}^{\rm (s)}_{j\leftarrow j'}+\hat{\bf g}^{\rm(p)(\updownarrow/\updownarrow)}_{j\leftarrow j'}\right]e^{ik_0\sqrt{\varepsilon_j(\omega)}\kappa_{z,j}\left|\rho_z-\rho_z'\right|}\right\}-\frac{\delta{[\rho_z-\rho_z']}}{k_0^2\varepsilon_j}[\hat{\bf z}\otimes\hat{\bf z}],
		\label{Eq.VI.10}
	\end{align}
	where
	\begin{subequations}
		\begin{align}
			r^{\rm(s)}_{j\leftarrow i}&=\frac{\sqrt{\varepsilon_j(\omega)}\kappa_{z,j}-\sqrt{\varepsilon_i(\omega)}\kappa_{z,i}}{\sqrt{\varepsilon_i(\omega)}\kappa_{z,i}+\sqrt{\varepsilon_j(\omega)}\kappa_{z,j}},
			\label{Eq.VI.11a}\\
			r^{\rm(p)}_{j\leftarrow i}&=\frac{\varepsilon_i(\omega)\sqrt{\varepsilon_j(\omega)}\kappa_{z,j}-\varepsilon_j(\omega)\sqrt{\varepsilon_i(\omega)}\kappa_{z,i}}{\varepsilon_j(\omega)\sqrt{\varepsilon_i(\omega)}\kappa_{z,i}+\varepsilon_i(\omega)\sqrt{\varepsilon_j(\omega)}\kappa_{z,j}},
			\label{Eq.VI.11b}
		\end{align}
	\end{subequations}
	are the \textit{Fresnel reflection coefficients}.
\end{tcolorbox}
\noindent Finally, regarding this dyadic notation it is also worth highlighting the following relations:
\begin{subequations}
	\begin{align}
		&\hat{\bf g}^{\rm (s)}_{i\leftarrow j}\cdot \hat{\bf g}^{\rm (s)}_{m\leftarrow n}=\left[\hat{\bf s}\otimes\hat{\bf s}\right]\cdot \left[\hat{\bf s}\otimes\hat{\bf s}\right]=\hat{\bf s}\left[\hat{\bf s}\cdot\hat{\bf s}\right]\hat{\bf s}=\hat{\bf s}\otimes\hat{\bf s}=\hat{\bf g}^{\rm (s)}_{i\leftarrow n}=\hat{\bf g}^{\rm (s)},
		\label{Eq.VI.12a}\\
		&\hat{\bf g}^{\rm (s)}_{i\leftarrow j}\cdot \hat{\bf g}^{\rm(p)(\updownarrow/\updownarrow)}_{m\leftarrow n}=\left[\hat{\bf s}\otimes\hat{\bf s}\right]\cdot \left[\hat{\bf p}^{\updownarrow}_{m}\otimes\hat{\bf p}^{\updownarrow}_{n}\right]=\hat{\bf s}\left[\hat{\bf s}\cdot\hat{\bf p}^{\updownarrow}_{m}\right]\hat{\bf p}^{\updownarrow}_{n}=\textbf{0},
		\label{Eq.VI.12b}
	\end{align}
\end{subequations}
\begin{subequations}
	\begin{align}
		&\hat{\bf g}^{\rm(p)(\updownarrow/\updownarrow)}_{i\leftarrow j}\cdot\hat{\bf g}^{\rm (s)}_{m\leftarrow n}=\left[\hat{\bf p}^{\updownarrow}_{i}\otimes\hat{\bf p}^{\updownarrow}_{j}\right]\cdot\left[\hat{\bf s}\otimes\hat{\bf s}\right]=\hat{\bf p}^{\updownarrow}_{i}\left[\hat{\bf p}^{\updownarrow}_{j}\cdot\hat{\bf s}\right]\hat{\bf s}=\textbf{0},
		\label{Eq.VI.12c}
		\tag{S.VI.12c}\\
		&\hat{\bf g}^{\rm(p)(\updownarrow/\updownarrow)}_{i\leftarrow j}\cdot\hat{\bf g}^{\rm(p)(\updownarrow/\updownarrow)}_{m\leftarrow n}=\left[\hat{\bf p}^{\updownarrow}_{i}\otimes\hat{\bf p}^{\updownarrow}_{j}\right]\cdot\left[\hat{\bf p}^{\updownarrow}_{m}\otimes\hat{\bf p}^{\updownarrow}_{n}\right]=\hat{\bf p}^{\updownarrow}_{i}\left[\hat{\bf p}^{\updownarrow}_{j}\cdot\hat{\bf p}^{\updownarrow}_{m}\right]\hat{\bf p}^{\updownarrow}_{n}=\left[\hat{\bf p}^{\updownarrow}_{j}\cdot\hat{\bf p}^{\updownarrow}_{m}\right]\hat{\bf g}^{\rm(p)(\updownarrow/\updownarrow)}_{i\leftarrow n},
		\label{Eq.VI.12d}
		\tag{S.VI.12d}
	\end{align}
\end{subequations}
with
\setcounter{equation}{12}
\begin{equation}
	\left[\hat{\bf p}^{\updownarrow}_{j}\cdot\hat{\bf p}^{\updownarrow}_{m}\right]=\kappa_{z,j}\kappa_{z,m}+\frac{\kappa_R^2}{(\pm)\sqrt{\varepsilon_j(\omega)}(\pm)\sqrt{\varepsilon_m(\omega)}}.
	\label{Eq.VI.13}
\end{equation}

\subsection{Zeroth-order of thermal emission}
\label{SectSI.VI.B}

Taking into account the previous section, the zeroth-order electric field correlation can be recast as follows:
\begin{align*}
	\mathcal{S}_{0,0}({\bf r};\omega;T)=\braket{[\hat{\bf E}^{(+)}_0({\bf r};\omega)]^\dagger\cdot\hat{\bf E}^{(+)}_0({\bf r};\omega)}_{\rm th}&=\omega^2\mu_0^2\iint\limits_{\mathcal{V}}{d^3\boldsymbol{\rho}d^3\boldsymbol{\rho}'\text{Tr}{[{\bf G}^*({\bf r},\boldsymbol{\rho},\omega){\bf G}({\bf r},\boldsymbol{\rho}',\omega)]}\braket{\hat{\bf j}^\dagger_0(\boldsymbol{\rho};\omega)\cdot\hat{\bf j}_0(\boldsymbol{\rho}';\omega)}_{\rm th}}\\
	&=\frac{4\pi}{\varepsilon_0}\frac{\omega^3}{c^4}\text{Im}{\left[\varepsilon(\omega)\right]}\hbar\omega\Theta(\omega,T)\left[\int\limits_{\mathcal{V}}{d^3\boldsymbol{\rho}\text{Tr}{[{\bf G}^*({\bf r},\boldsymbol{\rho},\omega){\bf G}({\bf r},\boldsymbol{\rho},\omega)]}}\right].
\end{align*}
Then:
\begin{tcolorbox}[sharp corners,colback=blue!5!white,colframe=blue!50!white]
	\vspace{-0.35cm}
	\begin{equation}
		\mathcal{S}_{0}({\bf r};\omega;T)=\mathcal{S}_{0,0}({\bf r};\omega;T)=\frac{4\pi}{\varepsilon_0}\frac{\omega^3}{c^4}\text{Im}{\left[\varepsilon(\omega)\right]}\hbar\omega\Theta(\omega,T)\mathcal{G}_{0,0}({\bf r},\omega),
		\label{Eq.VI.14}
	\end{equation}
	where
	\begin{equation}
		\mathcal{G}_{0,0}({\bf r},\omega)=\int\limits_{\mathcal{V}}{d^3\boldsymbol{\rho}\text{Tr}{[{\bf G}^*({\bf r},\boldsymbol{\rho},\omega){\bf G}({\bf r},\boldsymbol{\rho},\omega)]}},
		\label{Eq.VI.15}
	\end{equation}
	with $\mathcal{V}$ being the volume of the material system (i.e., medium $2$), occupying the half-space $z\leq 0$.
\end{tcolorbox}

\subsubsection{Calculation of $\mathcal{G}_{0,0}{\rm :}$}
\label{SectSI.VI.B.I}

We now proceed with the explicit calculation of $\mathcal{G}_{0,0}$. To do that, we start by substituting Eq.~\eqref{Eq.VI.01} into~\eqref{Eq.VI.15}:
\begin{align*}
	\mathcal{G}_{0,0}({\bf r},\omega)&=\int\limits_{\mathcal{V}}{d^3\boldsymbol{\rho}\text{Tr}{[{\bf G}^*({\bf r},\boldsymbol{\rho},\omega){\bf G}({\bf r},\boldsymbol{\rho},\omega)]}}\\
	&=\iiint\limits_{\mathcal{V}}{d\rho_xd\rho_yd\rho_z\left(\frac{k_0^2}{4\pi^2}\right)^2\iiiint\limits_{-\infty}^{\quad+\infty}{d^2\boldsymbol{\kappa}_\parallel d^2\boldsymbol{\kappa}'_\parallel\text{Tr}{[\hat{\bf G}^*(\kappa_x,\kappa_y;\omega|z,\rho_z)\hat{\bf G}(\kappa_x',\kappa_y';\omega|z,\rho_z)]}e^{ik_0\left[(\kappa_x'-\kappa_x)(x-\rho_x)+(\kappa_y'-\kappa_x')(y-\rho_y)\right]}}}\\
	&=\frac{k_0^2}{4\pi^2}\int\limits_{-\infty}^{0}{d\rho_z\iint\limits_{-\infty}^{\quad+\infty}{d^2\boldsymbol{\kappa}_\parallel\text{Tr}{[\hat{\bf G}^*(\kappa_x,\kappa_y;\omega|z,\rho_z)\hat{\bf G}(\kappa_x,\kappa_y;\omega|z,\rho_z)]}}}\\
	&=\frac{k_0^2}{4\pi^2}\int\limits_{-\infty}^{0}{d\rho_z\iint\limits_{-\infty}^{\quad+\infty}{d\kappa_xd\kappa_y\norm{\hat{\bf G}(\kappa_x,\kappa_y;\omega|z,\rho_z)}^2_{\mathcal{F}}}},
\end{align*}
where the subscript $\mathcal{F}$ stands for the so-called \textit{Frobenius norm}, defined as:
\begin{equation}
	||\hat{\bf A}||_{\mathcal{F}}=\sqrt{\sum_{i,j}{\left|a_{ij}\right|^2}}=\sqrt{\text{Tr}{[\hat{\bf A}^{\!^{\dagger}}\cdot\hat{\bf A}]}}.
	\label{Eq.VI.16}
\end{equation}
\noindent Thus:
\begin{tcolorbox}[sharp corners,colback=blue!5!white,colframe=blue!50!white]
	\vspace{-0.35cm}
	\begin{equation}
		\mathcal{G}_{0,0}({\bf r},\omega)=\frac{k_0^2}{2\pi}\int\limits_{-\infty}^{0}{d\rho_z\int\limits_{0}^{+\infty}{d\kappa_R\kappa_R\norm{\hat{\bf G}^{(\uparrow/\uparrow)}_{z \leftarrow \rho_z}(\kappa_R,\kappa_\varphi;\omega|z,\rho_z)}^2_{\mathcal{F}}}},
		\label{Eq.VI.17}
	\end{equation}
\end{tcolorbox}
\noindent where, in accordance with the aforementioned notation, we have set the correspondence $\hat{\bf G}(\kappa_x,\kappa_y;\omega|z,\rho_z)\to2\pi\hat{\bf G}^{(\uparrow/\uparrow)}_{z \leftarrow \rho_z}(\kappa_R,\kappa_\varphi;\omega|z,\rho_z)$, just by making a change of variables from Cartesian to cylindrical coordinates, thus leading to the factor $2\pi$ which results from the integration over the angles $\kappa_\varphi$~\cite{footnote03SI}. We have also included the superscript ``$(\uparrow/\uparrow)$'' as well as the subscript ``$z\leftarrow \rho_z$'' to indicate the polarization as well as the coordinates associated to the involved media. In this way, the explicit expression of the dyadic Green's function in the angular spectrum representation for this case is given by (cf. Refs.~\cite{Carminati1999SI,Shchegrov2000SI}):
\begin{equation}
	\!\!\!\!\!\!\hat{\bf G}^{(\uparrow/\uparrow)}_{z\leftarrow \rho_z}(\kappa_R,\kappa_\varphi;\omega|z,\rho_z)\!=\!\frac{i}{2}\frac{1}{k_{z,2}}\!\left\{t^{\rm (s)}_{z\leftarrow \rho_z}(\kappa_R,\kappa_\varphi;\omega)\!\left[\hat{\bf s}\otimes\hat{\bf s}\right]\!+\!t^{\rm (p)}_{z\leftarrow \rho_z}(\kappa_R,\kappa_\varphi;\omega)\!\left[\hat{\bf p}^{\uparrow}_{z}\otimes\hat{\bf p}^{\uparrow}_{\rho_z}\right]\!\right\}\!e^{ik_0\left[\tilde{k}_{z,1}z-\tilde{k}_{z,2}\rho_z\right]},\!\!\!\!\!\!
	\label{Eq.VI.18}
\end{equation}
with $k_{z,2}=k_0\tilde{k}_{z,2}=k_0\sqrt{\varepsilon(\omega)}\kappa_{z,2}$, $\hat{\bf s}$ and $\hat{\bf p}^{\uparrow}_{i}$ (with $i=\left\{z(=1),\rho_z(=2)\right\}$) being the corresponding polarization-vector basis [see. Eqs.~\eqref{Eq.VI.07a} and \eqref{Eq.VI.07b}]:
\begin{tcolorbox}[sharp corners]
	\vspace{-0.35cm}
	\begin{subequations}
		\begin{align}
			&\hat{\bf s}\equiv\begin{pmatrix}\sin{\kappa_\varphi},-\cos{\kappa_\varphi},0\end{pmatrix},
			\label{Eq.VI.19a}\\
			&\hat{\bf p}_i^{\uparrow}\equiv\displaystyle\begin{pmatrix}- \kappa_{z,i}\cos{\kappa_\varphi},- \kappa_{z,i}\sin{\kappa_\varphi},\kappa_R/\sqrt{\varepsilon_i(\omega)}\end{pmatrix},
			\label{Eq.VI.19b}
		\end{align}
	\end{subequations}
\end{tcolorbox}
\noindent and $t^{\rm (s)}_{z\leftarrow \rho_z}(\kappa_R,\kappa_\varphi;\omega)$ and $t^{\rm (p)}_{z\leftarrow \rho_z}(\kappa_R,\kappa_\varphi;\omega)$, the \textit{Fresnel transmission coefficients}:
\begin{tcolorbox}[sharp corners]
	\vspace{-0.35cm}
	\begin{subequations}
		\begin{align}
			t^{\rm (s)}_{z\leftarrow \rho_z}(\kappa_R,\kappa_\varphi;\omega)&\equiv\frac{2\varepsilon(\omega)\kappa_{z,2}}{\sqrt{\varepsilon(\omega)}\kappa_{z,1}+\varepsilon(\omega)\kappa_{z,2}},
			\label{Eq.VI.20a}\\
			t^{\rm (p)}_{z\leftarrow \rho_z}(\kappa_R,\kappa_\varphi;\omega)&\equiv\frac{2\varepsilon(\omega)\kappa_{z,2}}{\varepsilon(\omega)\kappa_{z,1}+\sqrt{\varepsilon(\omega)}\kappa_{z,2}}.
			\label{Eq.VI.20b}
		\end{align}
	\end{subequations}
\end{tcolorbox}
\noindent Then, inserting Eqs.~\eqref{Eq.VI.19a}--\eqref{Eq.VI.20b} into \eqref{Eq.VI.18}, the dyadic Green's function reads as:
\begin{tcolorbox}[sharp corners,colback=blue!5!white,colframe=blue!50!white]
	\vspace{-0.35cm}
	\begin{equation}
		\hat{\bf G}^{(\uparrow/\uparrow)}_{z\leftarrow \rho_z}(\kappa_R,\kappa_\varphi;\omega|z,\rho_z)=\frac{i}{2}\frac{1}{k_0\tilde{k}_{z,2}}\left[t^{\rm (s)}_{z\leftarrow \rho_z} \hat{\bf g}^{\rm (s)}_{z\leftarrow \rho_z}+t^{\rm(p)}_{z\leftarrow \rho_z}\hat{\bf g}^{\rm(p)(\uparrow/\uparrow)}_{z\leftarrow \rho_z} \right]\Gamma_{z\leftarrow \rho_z},
		\label{Eq.VI.21}
	\end{equation}
	with
	\begin{subequations}
		\begin{align}
			&\hat{\bf g}^{\rm (s)}_{z\leftarrow \rho_z}\equiv \hat{\bf s}\otimes\hat{\bf s} =\begin{bmatrix}
				\sin^2{\kappa_\varphi} & -\cos{\kappa_\varphi} \sin{\kappa_\varphi} & 0 \\
				-\cos{\kappa_\varphi} \sin{\kappa_\varphi} & \cos^2{\kappa_\varphi} & 0 \\
				0 & 0 & 0
			\end{bmatrix},
			\label{Eq.VI.22a}\\
			&\hat{\bf g}^{\rm(p)(\uparrow/\uparrow)}_{z\leftarrow \rho_z}\equiv \hat{\bf p}^{\uparrow}_{z}\otimes\hat{\bf p}^{\uparrow}_{\rho_z} =\begin{bmatrix}
				\kappa_{z,1}\kappa_{z,2}\cos^2{\kappa_\varphi} & \kappa_{z,1}\kappa_{z,2}\cos{\kappa_\varphi} \sin{\kappa_\varphi} & -\frac{\kappa_{z,1}\kappa_R\cos{\kappa_\varphi}}{\sqrt{\varepsilon(\omega)}}\\
				\kappa_{z,1}\kappa_{z,2}\cos{\kappa_\varphi} \sin{\kappa_\varphi} & \kappa_{z,1}\kappa_{z,2}\sin^2{\kappa_\varphi} & -\frac{\kappa_{z,1}\kappa_R\sin{\kappa_\varphi}}{\sqrt{\varepsilon(\omega)}} \\
				-\kappa_{z,2}\kappa_R\cos{\kappa_\varphi} & -\kappa_{z,2}\kappa_R\sin{\kappa_\varphi} & \frac{\kappa_R^2}{\sqrt{\varepsilon(\omega)}}
			\end{bmatrix},
			\label{Eq.VI.22b}\\
			&\Gamma_{z\leftarrow \rho_z}=e^{ik_0\left[\tilde{k}_{z,1}z-\tilde{k}_{z,2}\rho_z\right]}.
			\label{Eq.VI.22c}
		\end{align}
	\end{subequations}
\end{tcolorbox}
\noindent Thus, putting it all together it follows that:
\begin{equation}
	\norm{\hat{\bf G}^{(\uparrow/\uparrow)}_{z\leftarrow \rho_z}(\kappa_R,\kappa_\varphi;\omega|z,\rho_z)}^2_{\mathcal{F}}=\frac{1}{4}\frac{1}{k_0^2|\tilde{k}_{z,2}|^2}\left[\left|t^{\rm (s)}_{z\leftarrow \rho_z}\right|^2 \norm{\hat{\bf g}^{\rm (s)}_{z\leftarrow \rho_z}}^2_{\mathcal{F}}+\left|t^{\rm (p)}_{z\leftarrow \rho_z}\right|^2 \norm{\hat{\bf g}^{\rm(p)(\uparrow/\uparrow)}_{z\leftarrow \rho_z}}^2_{\mathcal{F}}\right]\left|\Gamma_{z\leftarrow \rho_z}\right|^2,
	\label{Eq.VI.23}
\end{equation}
with
\begin{align*}
	&\norm{\hat{\bf g}^{\rm (s)}_{z\leftarrow \rho_z}}^2_{\mathcal{F}}=\left(\hat{\bf s}\cdot\hat{\bf s}^*\right)\left(\hat{\bf s}\cdot\hat{\bf s}^*\right)=1,\\
	&\norm{\hat{\bf g}^{\rm(p)(\uparrow/\uparrow)}_{z\leftarrow \rho_z}}^2_{\mathcal{F}}=\left(\hat{\bf p}^{\uparrow}_{z}\cdot[\hat{\bf p}_{z}^{\uparrow}]^*\right)\left(\hat{\bf p}^{\uparrow}_{\rho_z}\cdot[\hat{\bf p}_{\rho_z}^{\uparrow}]^*\right)=\frac{1}{\left|\varepsilon(\omega)\right|}\left\{|\tilde{k}_{z,1}|^2|\tilde{k}_{z,2}|^2+\kappa_R^2\left[|\tilde{k}_{z,2}|^2+|\tilde{k}_{z,1}|^2+\kappa_R^2\right]\right\},\\
	&\left|\Gamma_{z\leftarrow \rho_z}\right|^2=\Gamma_{z\leftarrow \rho_z}^*\Gamma_{z\leftarrow \rho_z}=e^{-ik_0\left[\tilde{k}_{z,1}^*z-\tilde{k}_{z,2}^*\rho_z\right]}e^{ik_0\left[\tilde{k}_{z,1}z-\tilde{k}_{z,2}\rho_z\right]}=e^{-2k_0\text{Im}{[\tilde{k}_{z,1}]z}}e^{2k_0\text{Im}{[\tilde{k}_{z,2}]}\rho_z},
\end{align*}
so that:
\begin{tcolorbox}[sharp corners,colback=red!5!white,colframe=red!50!white]
	\vspace{-0.35cm}
	\begin{align}
		\!\!\!\!\!\!\!\!\!\!\nonumber \mathcal{G}_{0,0}({\bf r},\omega)&\!=\!\frac{k_0^2}{2\pi}\int\limits_{-\infty}^{0}{d\rho_z\int\limits_{0}^{+\infty}{d\kappa_R\kappa_R \left\{\frac{1}{4}\frac{1}{k_0^2|\tilde{k}_{z,2}|^2}\left[\left|t^{\rm (s)}_{z\leftarrow \rho_z}\right|^2 \norm{\hat{\bf g}^{\rm (s)}_{z\leftarrow \rho_z}}^2_{\mathcal{F}}+\left|t^{\rm (p)}_{z\leftarrow \rho_z}\right|^2 \norm{\hat{\bf g}^{\rm(p)(\uparrow/\uparrow)}_{z\leftarrow \rho_z}}^2_{\mathcal{F}}\right]\left|\Gamma_{z\leftarrow \rho_z}\right|^2\right\}}}\\
		&\!=\!\frac{1}{16\pi k_0}\int\limits_{0}^{+\infty}{d\kappa_R \frac{\kappa_R}{|\tilde{k}_{z,2}|^2}\left[\left|t^{\rm (s)}_{z\leftarrow \rho_z}\right|^2 \norm{\hat{\bf g}^{\rm (s)}_{z\leftarrow \rho_z}}^2_{\mathcal{F}}+\left|t^{\rm (p)}_{z\leftarrow \rho_z}\right|^2 \norm{\hat{\bf g}^{\rm(p)(\uparrow/\uparrow)}_{z\leftarrow \rho_z}}^2_{\mathcal{F}}\right]\frac{e^{-2k_0\text{Im}{[\tilde{k}_{z,1}]z}}}{\text{Im}{[\tilde{k}_{z,2}]}}}.
		\label{Eq.VI.24}
	\end{align}
\end{tcolorbox}
\noindent Finally:
\begin{tcolorbox}[sharp corners,colback=red!5!white,colframe=red!50!white]
	\vspace{-0.35cm}
	\begin{equation}
		\mathcal{S}_0({\bf r};\omega;T)=\frac{1}{4\varepsilon_0}\frac{\omega^2}{c^3}\text{Im}{\left[\varepsilon(\omega)\right]}\hbar\omega\Theta(\omega,T)\left[\int\limits_{0}^{+\infty}{d\kappa_R \frac{\kappa_R}{|\tilde{k}_{z,2}|^2}\left[\norm{\hat{\mathcal{G}}_{0,0}^{\rm (s)}}^2_{\mathcal{F}}+\norm{\hat{\mathcal{G}}_{0,0}^{\rm (p)}}^2_{\mathcal{F}}\right]\frac{e^{-2k_0\text{Im}{[\tilde{k}_{z,1}]z}}}{\text{Im}{[\tilde{k}_{z,2}]}}}\right],
		\label{Eq.VI.25}
	\end{equation}
	with
	\begin{align*}
		&\norm{\hat{\mathcal{G}}_{0,0}^{\rm (s)}}^2_{\mathcal{F}}\equiv\left|t^{\rm (s)}_{z\leftarrow \rho_z}\right|^2 \norm{\hat{\bf g}^{\rm (s)}_{z\leftarrow \rho_z}}^2_{\mathcal{F}}=\left|t^{\rm (s)}\right|^2,\\
		&\norm{\hat{\mathcal{G}}_{0,0}^{\rm (p)}}^2_{\mathcal{F}}\equiv \left|t^{\rm (p)}_{z\leftarrow \rho_z}\right|^2 \norm{\hat{\bf g}^{\rm(p)(\uparrow/\uparrow)}_{z\leftarrow \rho_z}}^2_{\mathcal{F}}=\left|t^{\rm (p)}\right|^2\frac{1}{\left|\varepsilon(\omega)\right|}\left\{|\tilde{k}_{z,1}|^2|\tilde{k}_{z,2}|^2+\kappa_R^2\left[|\tilde{k}_{z,2}|^2+|\tilde{k}_{z,1}|^2+\kappa_R^2\right]\right\}.
	\end{align*}
\end{tcolorbox}

\subsection{Second-order of thermal emission}
\label{SectSI.VI.C}

As shown above, the zeroth-order correlation, $\mathcal{S}_0({\bf r};\omega;T)$, leads to the ground spectra in which the time-modulation does not enter into play. In this sense, the first-order correlation $\mathcal{S}_1({\bf r};\omega;T)$ should provide for the first insights about the role of time-modulation into the system, e.g., showing up some non-local features described by means of the spatial integral. Notwithstanding, it can be observed that the time-modulation, characterized by means of the Laplace transform of the time-dependent susceptibility, does not account for effective frequency correlations. Essentially, this is because, in the calculation of thermal emission spectra, one is actually interested in correlations evaluated at a given frequency. For the first-order, only frequencies that are separated by the material's modulating frequency can correlate, so that, according to the result shown in Eq.~\eqref{Eq.V.02}, currents at the same frequency would not contribute to the first-order spectrum. Hence, the first term in the development yielding a non-null contribution due to the time-modulation would be the second one. As pointed out previously, this term can be expressed as a sum of three contributions: $\mathcal{S}_{2}({\bf r};\omega;T)=\mathcal{S}_{2,0}({\bf r};\omega;T)+\mathcal{S}_{0,2}({\bf r};\omega;T)+\mathcal{S}_{1,1}({\bf r};\omega;T)$, each of them displaying different behaviors depending on the specific currents involved. In this regard, correlations $\mathcal{S}_{2,0}({\bf r};\omega;T)+\mathcal{S}_{0,2}({\bf r};\omega;T)$ will connect different frequencies appearing at different locations of the system, thus introducing a kind of non-local behavior that preserves the distribution (and so, the average energy) of the oscillators. On the other hand, the term $\mathcal{S}_{1,1}({\bf r};\omega;T)$ owns the additional feature of affecting to the photon distribution, producing a frequency-shift related to the frequency of modulation. In the following, we shall explicitly look into each of these contributions for the particular case pointed~out above.

Looking at the results given in the previous section, the second-order correlations of the electric-field are given by:
\begin{subequations}
	\begin{align}
		\nonumber\mathcal{S}_{0,2}({\bf r};\omega;T)&=\braket{[\hat{\bf E}^{(+)}_0({\bf r};\omega)]^\dagger\cdot\hat{\bf E}^{(+)}_2({\bf r};\omega)}_{\rm th}\\
		&=\omega^2\mu_0^2\iint\limits_{\mathcal{V}}{d^3\boldsymbol{\rho}d^3\boldsymbol{\rho}'\text{Tr}{[{\bf G}^*({\bf r},\boldsymbol{\rho},\omega){\bf G}({\bf r},\boldsymbol{\rho}',\omega)]}\braket{\hat{\bf j}^\dagger_0(\boldsymbol{\rho};\omega)\cdot\hat{\bf j}_2(\boldsymbol{\rho}';\omega)}_{\rm th}},
		\label{Eq.VI.26a}\\\nonumber\\
		\nonumber\mathcal{S}_{1,1}({\bf r};\omega;T)&=\braket{[\hat{\bf E}^{(+)}_1({\bf r};\omega)]^\dagger\cdot\hat{\bf E}^{(+)}_1({\bf r};\omega)}_{\rm th}\\
		&=\omega^2\mu_0^2\iint\limits_{\mathcal{V}}{d^3\boldsymbol{\rho}d^3\boldsymbol{\rho}'\text{Tr}{[{\bf G}^*({\bf r},\boldsymbol{\rho},\omega){\bf G}({\bf r},\boldsymbol{\rho}',\omega)]}\braket{\hat{\bf j}^\dagger_1(\boldsymbol{\rho};\omega)\cdot\hat{\bf j}_1(\boldsymbol{\rho}';\omega)}_{\rm th}},
		\label{Eq.VI.26b}
	\end{align}
\end{subequations}
noticing that $\mathcal{S}_{2,0}({\bf r};\omega;T)=\braket{[\hat{\bf E}^{(+)}_2({\bf r};\omega)]^\dagger\cdot\hat{\bf E}^{(+)}_0({\bf r};\omega)}_{\rm th}=[\braket{[\hat{\bf E}^{(+)}_0({\bf r};\omega)]^\dagger\cdot\hat{\bf E}^{(+)}_2({\bf r};\omega)}_{\rm th}]^\dagger=[\mathcal{S}_{0,2}({\bf r};\omega;T)]^\dagger$. Likewise, the current density correlations involved in the above expressions are given by:
\begin{align*}
	&\braket{\hat{\bf j}^\dagger_0(\boldsymbol{\rho};\omega)\cdot\hat{\bf j}_2(\boldsymbol{\rho}';\omega)}_{\rm th}=\frac{\omega\mu_0}{\pi}\int\limits_{\mathcal{V}}{d^3\boldsymbol{\rho}''\left[\int\limits_{-\infty}^{+\infty}{d\omega'' \omega''\Delta\tilde{\chi}(\boldsymbol{\rho}',\omega-\omega'')\text{Tr}{[{\bf G}(\boldsymbol{\rho}',\boldsymbol{\rho}'',\omega'')]}\braket{\hat{\bf j}_0^\dagger(\boldsymbol{\rho};\omega)\cdot\hat{\bf j}_1(\boldsymbol{\rho}'';\omega'')}_{\rm th}}\right]},\\
	&\braket{\hat{\bf j}^\dagger_0(\boldsymbol{\rho};\omega)\cdot\hat{\bf j}_1(\boldsymbol{\rho}'';\omega'')}_{\rm th}=\frac{\omega''\mu_0}{\pi}\int\limits_{\mathcal{V}}{d^3\boldsymbol{\rho}'''\left[\int\limits_{-\infty}^{+\infty}{d\omega'''\omega'''\Delta\tilde{\chi}(\boldsymbol{\rho}'',\omega''-\omega''')\text{Tr}{[{\bf G}(\boldsymbol{\rho}'',\boldsymbol{\rho}''',\omega''')]}\braket{\hat{\bf j}_0^\dagger(\boldsymbol{\rho};\omega)\cdot\hat{\bf j}_0(\boldsymbol{\rho}''';\omega''')}_{\rm th}}\right]},\\
	&\braket{\hat{\bf j}^\dagger_1(\boldsymbol{\rho};\omega)\cdot\hat{\bf j}_1(\boldsymbol{\rho}';\omega)}_{\rm th}=\frac{\omega^2\mu_0^2}{\pi^2}\iint\limits_{\mathcal{V}}{d^3\tilde{\boldsymbol{\rho}}d^3\tilde{\boldsymbol{\rho}}' \iint\limits_{-\infty}^{\quad +\infty}{d\tilde{\omega}d\tilde{\omega}'\tilde{\omega}\tilde{\omega}'\Delta\tilde{\chi}^*(\boldsymbol{\rho},\omega-\tilde{\omega})\Delta\tilde{\chi}(\boldsymbol{\rho}',\omega-\tilde{\omega}')}}\\
	&\qquad\qquad\quad \left[\text{Tr}{[{\bf G}^*(\boldsymbol{\rho},\tilde{\boldsymbol{\rho}},\tilde{\omega}){\bf G}(\boldsymbol{\rho}',\tilde{\boldsymbol{\rho}}',\tilde{\omega}')]}\braket{\hat{\bf j}_0^\dagger(\tilde{\boldsymbol{\rho}};\tilde{\omega})\cdot\hat{\bf j}_0(\tilde{\boldsymbol{\rho}}';\tilde{\omega}')}_{\rm th}+\text{Tr}{[{\bf G}(\boldsymbol{\rho},\tilde{\boldsymbol{\rho}},\tilde{\omega}){\bf G}^*(\boldsymbol{\rho}',\tilde{\boldsymbol{\rho}}',\tilde{\omega}')]}\braket{\hat{\bf j}_0(\tilde{\boldsymbol{\rho}};\tilde{\omega})\cdot\hat{\bf j}_0^\dagger(\tilde{\boldsymbol{\rho}}';\tilde{\omega}')}_{\rm th}\right].
\end{align*}
For simplicity, as well as convenience, henceforth we shall particularize to time-harmonic modulations of the form:
\begin{equation}
	\Delta\tilde{\chi}({\bf r},t)=\varepsilon_0\delta\chi\Delta\tilde{\chi}(t)=\varepsilon_0\delta\chi\sin{(\Omega t)}=\frac{\varepsilon_0\delta\chi}{2i}\left[e^{i\Omega t}-e^{-i\Omega t}\vphantom{\frac{\varepsilon_0^2}{2}}\right],
	\label{Eq.VI.27}
\end{equation}
and whose Laplace transform is simply given by:
\begin{align}
	\nonumber\Delta\tilde{\chi}({\bf r},\omega)=\varepsilon_0\delta\chi\Delta\tilde{\chi}(\omega)=\varepsilon_0\delta\chi\mathcal{L}_{\omega}{\left[\Delta\tilde{\chi}(t)\right]}&=\frac{\varepsilon_0\delta\chi}{2i}\left\{\mathcal{L}_{\omega}{\left[e^{i\Omega t}\right]}-\mathcal{L}_{\omega}{\left[e^{-i\Omega t}\right]}\right\}\\
	\nonumber&=\frac{\varepsilon_0\delta\chi}{2i}\left\{\frac{i}{\omega+\Omega+i0^+}-\frac{i}{\omega-\Omega+i0^+}\right\}\\
	\nonumber&=\frac{\varepsilon_0\delta\chi\pi}{2i}\left(\delta{[\omega+\Omega]}-\delta{[\omega-\Omega]}+i\left\{\frac{1}{\pi}\mathcal{P}{\left[\frac{1}{\omega+\Omega}\right]}-\frac{1}{\pi}\mathcal{P}{\left[\frac{1}{\omega-\Omega}\right]}\right\}\right)\\
	&\to\frac{\varepsilon_0\delta\chi\pi}{i}\left(\delta{[\omega+\Omega]}-\delta{[\omega-\Omega]}\right).
	\label{Eq.VI.28}
\end{align}
Therefore,
\begin{align*}
	\braket{\hat{\bf j}^\dagger_0(\boldsymbol{\rho};\omega)\cdot\hat{\bf j}_1(\boldsymbol{\rho}'';\omega'')}_{\rm th}&=\frac{\delta\chi\omega''}{ic^2}\int\limits_{\mathcal{V}}{d^3\boldsymbol{\rho}'''(\omega''+\Omega)\text{Tr}{[{\bf G}(\boldsymbol{\rho}'',\boldsymbol{\rho}''',\omega''+\Omega)]}\braket{\hat{\bf j}_0^\dagger(\boldsymbol{\rho};\omega)\cdot\hat{\bf j}_0(\boldsymbol{\rho}''';\omega''+\Omega)}_{\rm th}}\\
	&-\frac{\delta\chi\omega''}{ic^2}\int\limits_{\mathcal{V}}{d^3\boldsymbol{\rho}'''(\omega''-\Omega)\text{Tr}{[{\bf G}(\boldsymbol{\rho}'',\boldsymbol{\rho}''',\omega''-\Omega)]}\braket{\hat{\bf j}_0^\dagger(\boldsymbol{\rho};\omega)\cdot\hat{\bf j}_0(\boldsymbol{\rho}''';\omega''-\Omega)}_{\rm th}}\\
	&=\frac{4\pi\varepsilon_0\delta\chi}{i}\text{Im}{\left[\varepsilon(\omega)\right]}\hbar\omega\Theta(\omega,T)\frac{\omega''}{c^2}(\omega''+\Omega)^2\text{Tr}{[{\bf G}(\boldsymbol{\rho}'',\boldsymbol{\rho},\omega''+\Omega)]}\delta{[\omega-(\omega''+\Omega)]} \\
	&-\frac{4\pi\varepsilon_0\delta\chi}{i}\text{Im}{\left[\varepsilon(\omega)\right]}\hbar\omega\Theta(\omega,T)\frac{\omega''}{c^2}(\omega''-\Omega)^2\text{Tr}{[{\bf G}(\boldsymbol{\rho}'',\boldsymbol{\rho},\omega''-\Omega)]}\delta{[\omega-(\omega''-\Omega)]},
\end{align*}
\begin{align*}
	\braket{\hat{\bf j}^\dagger_0(\boldsymbol{\rho};\omega)\cdot\hat{\bf j}_2(\boldsymbol{\rho}';\omega)}_{\rm th}&=\frac{\delta\chi^2\omega}{ic^2}\int\limits_{\mathcal{V}}{d^3\boldsymbol{\rho}'' (\omega+\Omega)\text{Tr}{[{\bf G}(\boldsymbol{\rho}',\boldsymbol{\rho}'',\omega+\Omega)]}\braket{\hat{\bf j}_0^\dagger(\boldsymbol{\rho};\omega)\cdot\hat{\bf j}_1(\boldsymbol{\rho}'';\omega+\Omega)}_{\rm th}}\\
	&-\frac{\delta\chi^2\omega}{ic^2}\int\limits_{\mathcal{V}}{d^3\boldsymbol{\rho}'' (\omega-\Omega)\text{Tr}{[{\bf G}(\boldsymbol{\rho}',\boldsymbol{\rho}'',\omega-\Omega)]}\braket{\hat{\bf j}_0^\dagger(\boldsymbol{\rho};\omega)\cdot\hat{\bf j}_1(\boldsymbol{\rho}'';\omega-\Omega)}_{\rm th}}\\
	&=4\pi\varepsilon_0\delta\chi^2\text{Im}{\left[\varepsilon(\omega)\right]}\hbar\omega\Theta(\omega,T)\frac{(\omega+\Omega)^2\omega^3}{c^4}\left[\int\limits_{\mathcal{V}}{d^3\boldsymbol{\rho}'' \text{Tr}{[{\bf G}(\boldsymbol{\rho}',\boldsymbol{\rho}'',\omega+\Omega){\bf G}(\boldsymbol{\rho}'',\boldsymbol{\rho},\omega)]}}\right]\\
	&+4\pi\varepsilon_0\delta\chi^2\text{Im}{\left[\varepsilon(\omega)\right]}\hbar\omega\Theta(\omega,T)\frac{(\omega-\Omega)^2\omega^3}{c^4}\left[\int\limits_{\mathcal{V}}{d^3\boldsymbol{\rho}'' \text{Tr}{[{\bf G}(\boldsymbol{\rho}',\boldsymbol{\rho}'',\omega-\Omega){\bf G}(\boldsymbol{\rho}'',\boldsymbol{\rho},\omega)]}}\right],
\end{align*}
thereby leading to:
\begin{tcolorbox}[sharp corners,colback=blue!5!white,colframe=blue!50!white]
	\vspace{-0.35cm}
	\begin{equation}
		\mathcal{S}_{0,2}({\bf r};\omega;T)=\frac{4\pi\delta\chi^2}{\varepsilon_0}\frac{\omega^5}{c^8}\text{Im}{\left[\varepsilon(\omega)\right]}\hbar\omega\Theta(\omega,T)\left\{\left(\omega^+\right)^2\bar{\bar{\mathcal{G}}}_{0,2}^{(\omega^+)}+\left(\omega^-\right)^2\bar{\bar{\mathcal{G}}}_{0,2}^{(\omega^-)}\right\},
		\label{Eq.VI.29}
	\end{equation}
	with
	\begin{equation}
		\bar{\bar{\mathcal{G}}}_{0,2}^{(\omega^\pm)}\equiv\iiint\limits_{\mathcal{V}}{d^3\boldsymbol{\rho}d^3\boldsymbol{\rho}'d^3\boldsymbol{\rho}''\text{Tr}{[{\bf G}^*({\bf r},\boldsymbol{\rho},\omega){\bf G}({\bf r},\boldsymbol{\rho}',\omega){\bf G}(\boldsymbol{\rho}',\boldsymbol{\rho}'',\omega^\pm){\bf G}(\boldsymbol{\rho}'',\boldsymbol{\rho},\omega)]}},
		\label{Eq.VI.30}
	\end{equation}
	and where it has been defined $\omega^\pm\equiv\omega\pm\Omega$.
\end{tcolorbox}
\noindent Similarly,
\begin{align*}
	\braket{\hat{\bf j}^\dagger_1(\boldsymbol{\rho};\omega)\cdot\hat{\bf j}_1(\boldsymbol{\rho}';\omega)}_{\rm th}
	&=4\pi\varepsilon_0\delta\chi^2\frac{\omega^2}{c^4}\left\{(\omega^+)^3\text{Im}{\left[\varepsilon(\omega^+)\right]}\hbar\omega^+\Theta(\omega^+,T)\int\limits_{\mathcal{V}}{d^3\boldsymbol{\rho}''\text{Tr}{[{\bf G}^*(\boldsymbol{\rho},\boldsymbol{\rho}'',\omega^+){\bf G}(\boldsymbol{\rho}',\boldsymbol{\rho}'',\omega^+)]}}\right\}\\
	&+4\pi\varepsilon_0\delta\chi^2\frac{\omega^2}{c^4}\left\{(\omega^+)^3\text{Im}{\left[\varepsilon(\omega^+)\right]}\hbar\omega^+\left[\Theta(\omega^+,T)+1\right]\int\limits_{\mathcal{V}}{d^3\boldsymbol{\rho}''\text{Tr}{[{\bf G}(\boldsymbol{\rho},\boldsymbol{\rho}'',\omega^+){\bf G}^*(\boldsymbol{\rho}',\boldsymbol{\rho}'',\omega^+)]}}\right\}\\
	&+4\pi\varepsilon_0\delta\chi^2\frac{\omega^2}{c^4}\left\{(\omega^-)^3\text{Im}{\left[\varepsilon(\omega^-)\right]}\hbar\omega^-\Theta(\omega^-,T)\int\limits_{\mathcal{V}}{d^3\boldsymbol{\rho}''\text{Tr}{[{\bf G}^*(\boldsymbol{\rho},\boldsymbol{\rho}'',\omega^-){\bf G}(\boldsymbol{\rho}',\boldsymbol{\rho}'',\omega^-)]}}\right\}\\
	&+4\pi\varepsilon_0\delta\chi^2\frac{\omega^2}{c^4}\left\{(\omega^-)^3\text{Im}{\left[\varepsilon(\omega^-)\right]}\hbar\omega^-\left[\Theta(\omega^-,T)+1\right]\int\limits_{\mathcal{V}}{d^3\boldsymbol{\rho}''\text{Tr}{[{\bf G}(\boldsymbol{\rho},\boldsymbol{\rho}'',\omega^-){\bf G}^*(\boldsymbol{\rho}',\boldsymbol{\rho}'',\omega^-)]}}\right\},
\end{align*}
so that,
\begin{tcolorbox}[sharp corners,colback=blue!5!white,colframe=blue!50!white]
	\vspace{-0.4cm}
	\begin{equation}
		\mathcal{S}_{1,1}({\bf r};\omega;T)=\mathcal{S}_{1,1}^{(\omega^+)}({\bf r};\omega;T)+\mathcal{S}_{1,1}^{(\omega^-)}({\bf r};\omega;T),
		\label{Eq.VI.31}
	\end{equation}
	where
	\begin{equation}
		\mathcal{S}_{1,1}^{(\omega^\pm)}({\bf r};\omega;T)=\frac{4\pi\delta\chi^2}{\varepsilon_0}\frac{\omega^5}{c^8}\text{Im}{\left[\varepsilon(\omega^\pm)\right]}\hbar\omega^\pm(\omega^\pm)^2(1+\tilde{\Omega})\left\{\Theta(\omega^\pm,T)\bar{\mathcal{G}}_{1,1}^{(\omega^\pm)}+\left[\Theta(\omega^\pm,T)+1\right]\tilde{\mathcal{G}}_{1,1}^{(\omega^\pm)}\right\},
		\label{Eq.VI.32}
	\end{equation}
	with $\tilde{\Omega}\equiv\Omega/\omega$, and
	\begin{subequations}
		\begin{align}
			\bar{\mathcal{G}}_{1,1}^{(\omega^\pm)}&\equiv\iiint\limits_{\mathcal{V}}{d^3\boldsymbol{\rho}d^3\boldsymbol{\rho}'d^3\boldsymbol{\rho}''\text{Tr}{[{\bf G}^*({\bf r},\boldsymbol{\rho},\omega){\bf G}^*(\boldsymbol{\rho},\boldsymbol{\rho}'',\omega^\pm){\bf G}({\bf r},\boldsymbol{\rho}',\omega){\bf G}(\boldsymbol{\rho}',\boldsymbol{\rho}'',\omega^\pm)]}},
			\label{Eq.VI.33a}\\
			\tilde{\mathcal{G}}_{1,1}^{(\omega^\pm)}&\equiv\iiint\limits_{\mathcal{V}}{d^3\boldsymbol{\rho}d^3\boldsymbol{\rho}'d^3\boldsymbol{\rho}''\text{Tr}{[{\bf G}^*({\bf r},\boldsymbol{\rho},\omega){\bf G}(\boldsymbol{\rho},\boldsymbol{\rho}'',\omega^\pm){\bf G}({\bf r},\boldsymbol{\rho}',\omega){\bf G}^*(\boldsymbol{\rho}',\boldsymbol{\rho}'',\omega^\pm)]}}.
			\label{Eq.VI.33b}
		\end{align}
	\end{subequations}
\end{tcolorbox}

\subsubsection{Calculation of $\bar{\mathcal{G}}_{1,1}^{(\omega^\pm)}{\rm :}$}
\label{SectSI.VI.C.I}

To carry on with the calculation of $\bar{\mathcal{G}}_{1,1}^{(\omega^\pm)}$, first of all it is important to ensure that all of the matrices involved are properly arranged:
\begin{equation*}
	[\hat{E}^{(+)}_{1,\alpha}]^\dagger\cdot\hat{E}^{(+)}_{1,\alpha}\propto G^*_{\alpha,\gamma}\hat{j}^\dagger_{1,\gamma} G_{\alpha,\delta}\hat{j}_{1,\delta}\propto G^*_{\alpha,\gamma}G^*_{\gamma,\beta}\hat{j}^\dagger_{0,\beta} G_{\alpha,\delta}G_{\delta,\zeta}\hat{j}_{0,\zeta}\propto G^*_{\alpha,\gamma}G^*_{\gamma,\beta} G_{\alpha,\delta}G_{\delta,\zeta}\delta_{\beta,\zeta}\propto G^*_{\alpha,\gamma}G^*_{\gamma,\beta} G_{\alpha,\delta}G_{\delta,\beta}.
\end{equation*}
In that way it is easy to see that the four matrices can be grouped as:
\begin{align*}
	{\bf G}^*({\bf r},\boldsymbol{\rho},\omega){\bf G}^*(\boldsymbol{\rho},\boldsymbol{\rho}'',\omega^\pm){\bf G}({\bf r},\boldsymbol{\rho}',\omega){\bf G}(\boldsymbol{\rho}',\boldsymbol{\rho}'',\omega^\pm)&=G^*_{\alpha,\gamma}({\bf r},\boldsymbol{\rho},\omega)G^*_{\gamma,\beta}(\boldsymbol{\rho},\boldsymbol{\rho}'',\omega^\pm)G_{\alpha,\delta}({\bf r},\boldsymbol{\rho}',\omega)G_{\delta,\beta}(\boldsymbol{\rho}',\boldsymbol{\rho}'',\omega^\pm)\\
	&=\bar{G}^*_{\alpha,\beta}({\bf r},\boldsymbol{\rho},\boldsymbol{\rho}'',\omega,\omega^\pm)\bar{G}_{\alpha,\beta}({\bf r},\boldsymbol{\rho}',\boldsymbol{\rho}'',\omega,\omega^\pm),
\end{align*}
where $\bar{G}^*_{\alpha,\beta}({\bf r},\boldsymbol{\rho},\boldsymbol{\rho}'',\omega,\omega^\pm)\equiv G^*_{\alpha,\gamma}({\bf r},\boldsymbol{\rho},\omega)G^*_{\gamma,\beta}(\boldsymbol{\rho},\boldsymbol{\rho}'',\omega^\pm)$ and $\bar{G}_{\alpha,\beta}({\bf r},\boldsymbol{\rho}',\boldsymbol{\rho}'',\omega,\omega^\pm)\equiv G_{\alpha,\delta}({\bf r},\boldsymbol{\rho}',\omega)G_{\delta,\beta}(\boldsymbol{\rho}',\boldsymbol{\rho}'',\omega^\pm)$, noticing that the summation over $\gamma$ and $\delta$ actually refers to the customary matrix product. Inserting this latter expression into \eqref{Eq.VI.33a}, and making use of the expansion given in Eq.~\eqref{Eq.VI.01}, it follows that:
\begin{align*}
	&\bar{\mathcal{G}}_{1,1}^{(\omega^\pm)}=\iiint\limits_{\mathcal{V}}{d^3\boldsymbol{\rho}d^3\boldsymbol{\rho}'d^3\boldsymbol{\rho}''\text{Tr}{[{\bf G}^*({\bf r},\boldsymbol{\rho},\omega){\bf G}^*(\boldsymbol{\rho},\boldsymbol{\rho}'',\omega^\pm){\bf G}({\bf r},\boldsymbol{\rho}',\omega){\bf G}(\boldsymbol{\rho}',\boldsymbol{\rho}'',\omega^\pm)]}}\\
	&=\frac{k_0^2}{4\pi^2}\iiint\limits_{-\infty}^{\quad 0}{d\rho_zd\rho_z'd\rho_z'' \iint\limits_{-\infty}^{\quad +\infty}{d^2\boldsymbol{\kappa}_\parallel\text{Tr}{[\hat{\bf G}^*(\kappa_x,\kappa_y;\omega|z,\rho_z)\hat{\bf G}^*(\kappa_x^\pm,\kappa_y^\pm;\omega^\pm|\rho_z,\rho_z'')\hat{\bf G}(\kappa_x,\kappa_y;\omega|z,\rho_z')\hat{\bf G}(\kappa_x^\pm,\kappa_y^\pm;\omega^\pm|\rho_z',\rho_z'')]}}},
\end{align*}
where $\kappa_i^\pm\equiv\kappa_i/(1\pm\tilde{\Omega})$, with $i=\left\{x,y\right\}$~\cite{footnote04SI}. Hence:
\begin{tcolorbox}[sharp corners,colback=blue!5!white,colframe=blue!50!white]
	\vspace{-0.35cm}
	\begin{equation}
		\bar{\mathcal{G}}_{1,1}^{(\omega^\pm)}=\frac{k_0^2}{4\pi^2}\iiint\limits_{-\infty}^{\quad 0}{d\rho_zd\rho_z'd\rho_z'' \int\limits_{0}^{+\infty}{d\kappa_R \kappa_R \int\limits_{0}^{2\pi}{d\kappa_\varphi \norm{\left[\hat{\bar{{\bf G}}}_{z\leftarrow \rho_z\leftarrow \rho_z''}^{(\omega^\pm)[(\uparrow/\uparrow)(\updownarrow/\updownarrow)]}\right]^* \hat{\bar{{\bf G}}}_{z\leftarrow \rho_z'\leftarrow \rho_z''}^{(\omega^\pm)[(\uparrow/\uparrow)(\updownarrow/\updownarrow)]}}_{\mathcal{F}}}}},
		\label{Eq.VI.34}
	\end{equation}
	where
	\begin{subequations}
		\begin{align}
			&\left[\hat{\bar{{\bf G}}}_{z\leftarrow \rho_z\leftarrow \rho_z''}^{(\omega^\pm)[(\uparrow/\uparrow)(\updownarrow/\updownarrow)]}\right]^*\equiv\left[\hat{\bar{{\bf G}}}_{z\leftarrow \rho_z\leftarrow \rho_z''}^{(\omega^\pm)[(\updownarrow/\updownarrow)]}\right]^*=\left[\hat{\bf G}^{(\uparrow/\uparrow)}(\kappa_x,\kappa_y;\omega|z,\rho_z)\right]^*\cdot\left[\hat{\bf G}^{(\updownarrow/\updownarrow)}(\kappa_x^\pm,\kappa_y^\pm;\omega^\pm|\rho_z,\rho_z'')\right]^*,
			\label{Eq.VI.35a}\\
			&\hat{\bar{{\bf G}}}_{z\leftarrow \rho_z'\leftarrow \rho_z''}^{(\omega^\pm)[(\uparrow/\uparrow)(\updownarrow/\updownarrow)]}\equiv\hat{\bar{{\bf G}}}_{z\leftarrow \rho_z'\leftarrow \rho_z''}^{(\omega^\pm)[(\updownarrow/\updownarrow)]}=\hat{\bf G}^{(\uparrow/\uparrow)}(\kappa_x,\kappa_y;\omega|z,\rho_z')\cdot\hat{\bf G}^{(\updownarrow/\updownarrow)}(\kappa_x^\pm,\kappa_y^\pm;\omega^\pm|\rho_z',\rho_z'').
			\label{Eq.VI.35b}
		\end{align}
	\end{subequations}
\end{tcolorbox}
\noindent There are some similarities between these results and those obtained for the zeroth-order contribution [compare Eq.~\eqref{Eq.VI.34} with Eq.~\eqref{Eq.VI.17}]. However, in this case the matrices in the product of Eq.~\eqref{Eq.VI.34} are not the same, thus resulting not in the \textit{Frobenius norm}, but in the so-called \textit{Frobenius inner product}. Furthermore, in each of the products given in Eqs.~\eqref{Eq.VI.35a} and \eqref{Eq.VI.35b}, the second matrix involves variables lying in the same medium. So, whilst the first matrix will be of the form of Eq.~\eqref{Eq.VI.05}, the second one will be like that given in Eq.~\eqref{Eq.VI.10}. At any rate, one can proceed just by following the notation sketched out in Eqs.~\eqref{Eq.VI.05}--\eqref{Eq.VI.12d}:
\begin{align*}
	\left[\hat{\bar{{\bf G}}}_{z\leftarrow \rho_z\leftarrow \rho_z''}^{(\omega^\pm)[\blue{(\updownarrow/\updownarrow)}]}\right]^*=&-\frac{1}{4}\frac{1}{k_{z,2}^*k_{z,2}^{\pm*}}\left[t^{\rm (s)}\hat{\bf g}^{\rm (s)}+t^{\rm (p)}\left(\hat{\bf p}_{\rho_z}^\uparrow\cdot \hat{\bf p}_{\rho_z}^{(\omega^\pm)\blue{\updownarrow}}\right)\left(\hat{\bf p}_{z}^{\uparrow}\otimes\hat{\bf p}_{\rho_z''}^{\rm (\omega^\pm)\blue{\updownarrow}}\right)\right]^*e^{-i\left[k_{z,1}^*z-k_{z,2}^*\rho_z\right]}e^{-ik_{z,2}^{\pm*}\left|\rho_z-\rho_z''\right|}\\
	&-\frac{1}{4}\frac{1}{k_{z,2}^*k_{z,2}^{\pm*}}\left[t^{\rm (s)}r^{\rm (s)(\omega^\pm)} \hat{\bf g}^{\rm (s)}+t^{\rm (p)}r^{\rm (p)(\omega^\pm)} \left(\hat{\bf p}_{\rho_z}^\uparrow\cdot \hat{\bf p}_{\rho_z}^{(\omega^\pm)\blue{\downarrow}}\right)\left(\hat{\bf p}_{z}^{\uparrow}\otimes\hat{\bf p}_{\rho_z''}^{\rm (\omega^\pm)\blue{\uparrow}}\right)\right]^*e^{-i\left[k_{z,1}^*z-k_{z,2}^*\rho_z\right]}e^{ik_{z,2}^{\pm*}\left[\rho_z+\rho_z''\right]}\\
	&+\frac{i}{2}\frac{\delta{[\rho_z-\rho_z'']}}{k_{z,2}^*k_0^2(1\pm\tilde{\Omega})^2\varepsilon^*(\omega^\pm)}t^{\rm (p)*}\left( \left[\hat{\bf p}_{\rho_z}^\uparrow\right]^*\cdot{\bf z}\right)\left(\left[\hat{\bf p}_{z}^\uparrow\right]^*\otimes\hat{\bf z}\right)e^{-i\left[k_{z,1}^*z-k_{z,2}^*\rho_z\right]}\\
	=&\;\left[\hat{\bar{{\bf T}}}_{z\leftarrow \rho_z\leftarrow \rho_z''}^{(\omega^\pm)[\blue{(\updownarrow/\updownarrow)}]}\right]^*+\left[\hat{\bar{{\bf R}}}_{z\leftarrow \rho_z\leftarrow \rho_z''}^{(\omega^\pm)[\blue{(\downarrow/\uparrow)}]}\right]^*+\left[\hat{\bar{{\bf Z}}}_{z\leftarrow \rho_z\leftarrow \rho_z''}^{(\omega^\pm)}\right]^*,
\end{align*}
\begin{align*}
	\hat{\bar{{\bf G}}}_{z\leftarrow \rho_z'\leftarrow \rho_z''}^{(\omega^\pm)[\red{(\updownarrow/\updownarrow)}]}
	=&-\frac{1}{4}\frac{1}{k_{z,2}k_{z,2}^{\pm}}\left[t^{\rm (s)} \hat{\bf g}^{\rm (s)}+t^{\rm (p)} \left(\hat{\bf p}_{\rho_z'}^\uparrow\cdot \hat{\bf p}_{\rho_z'}^{(\omega^\pm)\red{\updownarrow}}\right)\left(\hat{\bf p}_{z}^{\uparrow}\otimes\hat{\bf p}_{\rho_z''}^{\rm (\omega^\pm)\red{\updownarrow}}\right)\right]e^{i\left[k_{z,1}z-k_{z,2}\rho_z'\right]}e^{ik_{z,2}^\pm\left|\rho_z'-\rho_z''\right|}\\
	&-\frac{1}{4}\frac{1}{k_{z,2}k_{z,2}^{\pm}}\left[t^{\rm (s)}r^{\rm (s)(\omega^\pm)} \hat{\bf g}^{\rm (s)}+t^{\rm (p)}r^{\rm (p)(\omega^\pm)} \left(\hat{\bf p}_{\rho_z'}^\uparrow\cdot \hat{\bf p}_{\rho_z'}^{(\omega^\pm)\red{\downarrow}}\right)\left(\hat{\bf p}_{z}^{\uparrow}\otimes\hat{\bf p}_{\rho_z''}^{\rm (\omega^\pm)\red{\uparrow}}\right)\right]e^{i\left[k_{z,1}z-k_{z,2}\rho_z'\right]}e^{-ik_{z,2}^\pm\left[\rho_z'+\rho_z''\right]}\\
	&-\frac{i}{2}\frac{\delta{[\rho_z'-\rho_z'']}}{k_{z,2}k_0^2(1\pm\tilde{\Omega})^2\varepsilon(\omega^\pm)}t^{\rm (p)}\left(\hat{\bf p}_{\rho_z'}^\uparrow\cdot\hat{\bf z}\right)\left(\hat{\bf p}_{z}^\uparrow\otimes\hat{\bf z}\vphantom{\left(\hat{\bf p}_{\rho_z'}^\uparrow\cdot\hat{\bf z}\right)}\right)e^{i\left[k_{z,1}z-k_{z,2}\rho_z'\right]}\\
	=&\;\hat{\bar{{\bf T}}}_{z\leftarrow \rho_z'\leftarrow \rho_z''}^{(\omega^\pm)[\red{(\updownarrow/\updownarrow)}]}+\hat{\bar{{\bf R}}}_{z\leftarrow \rho_z'\leftarrow \rho_z''}^{(\omega^\pm)[\red{(\downarrow/\uparrow)}]}+\hat{\bar{{\bf Z}}}_{z\leftarrow \rho_z'\leftarrow \rho_z''}^{(\omega^\pm)}.
\end{align*}
Then, according to this latter expressions, Eq.~\eqref{Eq.VI.34} can be recast as:
\begin{tcolorbox}[sharp corners,colback=red!5!white,colframe=red!50!white]
	\vspace{-0.35cm}
	\begin{equation}
		\bar{\mathcal{G}}_{1,1}^{(\omega^\pm)}=\bar{\mathcal{G}}_{\rm T-T\,(1,1)}^{(\omega^\pm)}+\bar{\mathcal{G}}_{\rm R-R\,(1,1)}^{(\omega^\pm)}+2\text{Re}{\left[\bar{\mathcal{G}}_{\rm T-R\,(1,1)}^{(\omega^\pm)}\right]}+\bar{\mathcal{G}}_{\rm Z-Z\,(1,1)}^{(\omega^\pm)}+2\text{Re}{\left[\bar{\mathcal{G}}_{\rm T-Z\,(1,1)}^{(\omega^\pm)}\right]}+2\text{Re}{\left[\bar{\mathcal{G}}_{\rm R-Z\,(1,1)}^{(\omega^\pm)}\right]}.
		\label{Eq.VI.36}
	\end{equation}
\end{tcolorbox}

In the following we will proceed with the calculation of Eq.~\eqref{Eq.VI.36} by analyzing separately each of the above contributions:
\begin{tcolorbox}[sharp corners,colback=blue!5!white,colframe=blue!50!white]
	\vspace{-0.35cm}
	\begin{equation}
		\norm{\left[\hat{\bar{{\bf T}}}_{z\leftarrow \rho_z\leftarrow \rho_z''}^{(\omega^\pm)\blue{(\updownarrow/\updownarrow)}}\right]^*\hat{\bar{{\bf T}}}_{z\leftarrow \rho_z'\leftarrow \rho_z''}^{(\omega^\pm)\red{(\updownarrow/\updownarrow)}}}_{\mathcal{F}}=\frac{1}{16}\frac{1}{\left|k_{z,2}\right|^2\left|k_{z,2}^{\pm}\right|^2}\sum_{i={\rm s,p}}{\norm{\left[\hat{\bar{{\bf t}}}_{z\leftarrow \rho_z\leftarrow \rho_z''}^{(\omega^\pm)\blue{(\updownarrow/\updownarrow)}}\right]^*\hat{\bar{{\bf t}}}_{z\leftarrow \rho_z'\leftarrow \rho_z''}^{(\omega^\pm)\red{(\updownarrow/\updownarrow)}}}^{\rm (i)}_{\mathcal{F}}}\bar{\Gamma}_{\rm T-T\,(1,1)}^{(\omega^\pm)(\blue{\updownarrow}/\red{\updownarrow})},
		\label{Eq.VI.37}
	\end{equation}
	with (cf. \hyperref[SectSI.Appendix.VI.A]{\!Appendix VI.A} for the demonstration of the $p$-like contribution),
	\begin{align*}
		\norm{\left[\hat{\bar{{\bf t}}}_{z\leftarrow \rho_z\leftarrow \rho_z''}^{(\omega^\pm)\blue{(\updownarrow/\updownarrow)}}\right]^*\hat{\bar{{\bf t}}}_{z\leftarrow \rho_z'\leftarrow \rho_z''}^{(\omega^\pm)\red{(\updownarrow/\updownarrow)}}}^{\rm (s)}_{\mathcal{F}}&=\left|t^{\rm (s)}\right|^2\norm{\hat{\bf g}^{\rm (s)}}^2_{\mathcal{F}},\\\\
		\norm{\left[\hat{\bar{{\bf t}}}_{z\leftarrow \rho_z\leftarrow \rho_z''}^{(\omega^\pm)\blue{(\updownarrow/\updownarrow)}}\right]^*\hat{\bar{{\bf t}}}_{z\leftarrow \rho_z'\leftarrow \rho_z''}^{(\omega^\pm)\red{(\updownarrow/\updownarrow)}}}^{\rm (p)}_{\mathcal{F}}&=\left|t^{\rm (p)}\right|^2\left(\hat{\bf p}_{\rho_z}^\uparrow\cdot\hat{\bf p}_{\rho_z}^{(\omega^\pm)\blue{\updownarrow}}\right)^*\left(\hat{\bf p}_{\rho_z'}^\uparrow\cdot \hat{\bf p}_{\rho_z'}^{(\omega^\pm)\red{\updownarrow}}\right)\left(\left[\hat{\bf p}_z^{\uparrow}\right]^*\cdot\hat{\bf p}_{z}^{\uparrow}\right)\left(\left[\hat{\bf p}_{\rho_z''}^{(\omega^\pm)\blue{\updownarrow}}\right]^*\cdot\hat{\bf p}_{\rho_z''}^{(\omega^\pm)\red{\updownarrow}}\right).
	\end{align*}
	Likewise, regarding the spatial integrals,
	\begin{align}
		\nonumber\bar{\Gamma}_{\rm T-T\,(1,1)}^{(\omega^\pm)(\blue{\updownarrow}/\red{\updownarrow})}(z,\rho_z,\rho_z',\rho_z'')&=e^{-i\left[k_{z,1}^*z-k_{z,2}^*\rho_z\right]}e^{-ik_{z,2}^{\pm*}\blue{\left|\rho_z-\rho_z''\right|}}e^{i\left[k_{z,1}z-k_{z,2}\rho_z'\right]}e^{ik_{z,2}^{\pm}\red{\left|\rho_z'-\rho_z''\right|}}\\
		&=e^{-2\text{Im}{\left[k_{z,1}\right]}z}e^{ik_{z,2}^*\rho_z}e^{-ik_{z,2}\rho_z'}e^{-ik_{z,2}^{\pm*}\blue{\left|\rho_z-\rho_z''\right|}}e^{ik_{z,2}^{\pm}\red{\left|\rho_z'-\rho_z''\right|}}.
		\label{Eq.VI.38}
	\end{align}
\end{tcolorbox}
\noindent According to Eq.~\eqref{Eq.VI.34}, this latter expression has to be integrated over $\rho_z$, $\rho_z'$, and $\rho_z''$. To do so, it should be noted the presence of the absolute value in some of the exponents so as to distinguish between the four possible cases:
\begin{itemize}
	\item \textbf{Case ($\uparrow/\uparrow$)}: $\quad\rho_z>\rho_z''$ and $\rho_z'>\rho_z''$ \quad $\Longleftrightarrow$ \quad $\left(\rho_z,\rho_z'\right)>\rho_z''$;
	\item \textbf{Case ($\uparrow/\downarrow$)}: $\quad\rho_z>\rho_z''$ and $\rho_z'<\rho_z''$ \quad $\Longleftrightarrow$ \quad $\rho_z>\rho_z''>\rho_z'$;
	\item \textbf{Case ($\downarrow/\uparrow$)}: $\quad\rho_z<\rho_z''$ and $\rho_z'>\rho_z''$ \quad $\Longleftrightarrow$ \quad $\rho_z'>\rho_z''>\rho_z$;
	\item \textbf{Case ($\downarrow/\downarrow$)}: $\quad\rho_z<\rho_z''$ and $\rho_z'<\rho_z''$ \quad $\Longleftrightarrow$ \quad $\left(\rho_z,\rho_z'\right)<\rho_z''$.
\end{itemize}
Therefore, the spatial integral in Eq.~\eqref{Eq.VI.34} can be recast as:
\begin{equation}
	e^{-2\text{Im}{\left[k_{z,1}\right]}z}\int\limits_{-\infty}^{0}{d\rho_z''\left[\underbracket{\iint\limits_{\rho_z''}^{\quad0}{d\rho_z'd\rho_z}\vphantom{\int\limits_{-\infty}^{\rho_z''}{d\rho_z'}\int\limits_{\rho_z''}^{0}{d\rho_z}}}_{\text{Case ($\uparrow/\uparrow$)}}+\underbracket{\int\limits_{-\infty}^{\rho_z''}{d\rho_z'}\int\limits_{\rho_z''}^{0}{d\rho_z}\vphantom{\int\limits_{-\infty}^{\rho_z''}{d\rho_z'}\int\limits_{\rho_z''}^{0}{d\rho_z}}}_{\text{Case ($\uparrow/\downarrow$)}}+\underbracket{\int\limits_{\rho_z''}^{0}{d\rho_z'}\int\limits_{-\infty}^{\rho_z''}{d\rho_z}\vphantom{\int\limits_{-\infty}^{\rho_z''}{d\rho_z'}\int\limits_{\rho_z''}^{0}{d\rho_z}}}_{\text{Case ($\downarrow/\uparrow$)}}+\underbracket{\iint\limits_{-\infty}^{\quad\rho_z''}{d\rho_z'd\rho_z}\vphantom{\int\limits_{-\infty}^{\rho_z''}{d\rho_z'}\int\limits_{\rho_z''}^{0}{d\rho_z}}}_{\text{Case ($\downarrow/\downarrow$)}}\right]},
	\label{Eq.VI.39}
\end{equation}
thereby covering the whole time-modulated medium. Obviously, all of these integrals are to be put in correspondence with the dyadic Green's functions given above. Thus, by putting it all together, the first term of Eq.~\eqref{Eq.VI.36} can be expressed as:
\begin{tcolorbox}[sharp corners,colback=red!5!white,colframe=red!50!white]
	\vspace{-0.35cm}
	\begin{equation}
		\bar{\mathcal{G}}_{\rm T-T\,(1,1)}^{(\omega^\pm)}=\frac{1}{32\pi k_0^5}\int\limits_{0}^{+\infty}{d\kappa_R\frac{\kappa_R}{|\tilde{k}_{z,2}|^2|\tilde{k}_{z,2}^\pm|^2} \sum_{i,j=\left\{\uparrow,\downarrow\right\}}{\bar{\mathcal{G}}_{\rm T-T\,(1,1)}^{(\omega^\pm)(i/j)}\bar{\Gamma}_{\rm T-T\,(1,1)}^{(\omega^\pm)(i/j)}}},
		\label{Eq.VI.40}
	\end{equation}
\end{tcolorbox}
\noindent where
\begin{align*}
	\bar{\mathcal{G}}_{\rm T-T\,(1,1)}^{(\omega^\pm)(\uparrow/\uparrow)}&=\left|t^{\rm (s)}\right|^2+\left|t^{\rm (p)}\right|^2\frac{{\rm T}^{(\omega^\pm)}}{(1\pm\tilde{\Omega})^2\left|\varepsilon(\omega)\right|\left|\varepsilon(\omega^\pm)\right|}\left\{\left[\tilde{k}_{z,2}^*\tilde{k}_{z,2}^{\pm*}+\kappa_R^2\right]\left[\tilde{k}_{z,2}\tilde{k}_{z,2}^{\pm}+\kappa_R^2\right]\right\},\\
	\bar{\Gamma}_{\rm T-T\,(1,1)}^{(\omega^\pm)(\uparrow/\uparrow)}&=\frac{2e^{-2k_0\text{Im}{[\tilde{k}_{z,1}]}z}}{|\tilde{k}_{z,2}-\tilde{k}_{z,2}^\pm|^2}\left[\frac{\text{Im}{[\tilde{k}_{z,2}]}+\text{Im}{[\tilde{k}_{z,2}^\pm]}}{4\text{Im}{[\tilde{k}_{z,2}]}\text{Im}{[\tilde{k}_{z,2}^\pm]}}-\frac{\text{Im}{[\tilde{k}_{z,2}]}+\text{Im}{[\tilde{k}_{z,2}^\pm]}}{|\tilde{k}_{z,2}-\tilde{k}_{z,2}^{\pm*}|^2}\right],\\\\
	\bar{\mathcal{G}}_{\rm T-T\,(1,1)}^{(\omega^\pm)(\uparrow/\downarrow)}&=\left|t^{\rm (s)}\right|^2+\left|t^{\rm (p)}\right|^2\frac{{\rm R}^{(\omega^\pm)}}{(1\pm\tilde{\Omega})^2\left|\varepsilon(\omega)\right|\left|\varepsilon(\omega^\pm)\right|}\left\{\left[\tilde{k}_{z,2}^*\tilde{k}_{z,2}^{\pm*}+\kappa_R^2\right]\left[\tilde{k}_{z,2}\tilde{k}_{z,2}^{\pm}-\kappa_R^2\right]\right\},\\
	\bar{\Gamma}_{\rm T-T\,(1,1)}^{(\omega^\pm)(\uparrow/\downarrow)}&=\frac{e^{-2k_0\text{Im}{[\tilde{k}_{z,1}]}z}}{2\text{Im}{[\tilde{k}_{z,2}^*]}[\tilde{k}_{z,2}-\tilde{k}_{z,2}^{\pm*}][\tilde{k}_{z,2}+\tilde{k}_{z,2}^\pm]},\\\\
	\bar{\mathcal{G}}_{\rm T-T\,(1,1)}^{(\omega^\pm)(\downarrow/\uparrow)}&=\left|t^{\rm (s)}\right|^2+\left|t^{\rm (p)}\right|^2\frac{{\rm R}^{(\omega^\pm)}}{(1\pm\tilde{\Omega})^2\left|\varepsilon(\omega)\right|\left|\varepsilon(\omega^\pm)\right|}\left\{\left[\tilde{k}_{z,2}^*\tilde{k}_{z,2}^{\pm *}-\kappa_R^2\right]\left[\tilde{k}_{z,2}\tilde{k}_{z,2}^{\pm}+\kappa_R^2\right]\right\},\\
	\bar{\Gamma}_{\rm T-T\,(1,1)}^{(\omega^\pm)(\downarrow/\uparrow)}&=\frac{e^{-2k_0\text{Im}{[\tilde{k}_{z,1}]}z}}{2\text{Im}{[\tilde{k}_{z,2}^*]}[\tilde{k}_{z,2}^*-\tilde{k}_{z,2}^{\pm}][\tilde{k}_{z,2}^*+\tilde{k}_{z,2}^{\pm *}]},\\\\
	\bar{\mathcal{G}}_{\rm T-T\,(1,1)}^{(\omega^\pm)(\downarrow/\downarrow)}&=\left|t^{\rm (s)}\right|^2+\left|t^{\rm (p)}\right|^2\frac{{\rm T}^{(\omega^\pm)}}{(1\pm\tilde{\Omega})^2\left|\varepsilon(\omega)\right|\left|\varepsilon(\omega^\pm)\right|}\left\{\left[\tilde{k}_{z,2}^*\tilde{k}_{z,2}^{\pm *}-\kappa_R^2\right]\left[\tilde{k}_{z,2}\tilde{k}_{z,2}^{\pm}-\kappa_R^2\right]\right\},\\
	\bar{\Gamma}_{\rm T-T\,(1,1)}^{(\omega^\pm)(\downarrow/\downarrow)}&=\frac{e^{-2\text{Im}{[\tilde{k}_{z,1}]}z}}{2\text{Im}{[\tilde{k}_{z,2}]}|\tilde{k}_{z,2}+\tilde{k}_{z,2}^\pm|^2},
\end{align*}
with
\begin{tcolorbox}[sharp corners,colback=blue!5!white,colframe=blue!50!white]
	\vspace{-0.35cm}
	\begin{subequations}
		\begin{align}
			{\rm T}^{(\omega^\pm)}&=\frac{1}{(1\pm\tilde{\Omega})^2\left|\varepsilon(\omega^\pm)\right|}\left\{|\tilde{k}_{z,1}|^2|\tilde{k}_{z,2}^\pm|^2+\kappa_R^2\left[|\tilde{k}_{z,2}^\pm|^2+|\tilde{k}_{z,1}|^2+\kappa_R^2\right]\right\},
			\label{Eq.VI.41a}\\
			{\rm R}^{(\omega^\pm)}&=\frac{1}{(1\pm\tilde{\Omega})^2\left|\varepsilon(\omega^\pm)\right|}\left\{|\tilde{k}_{z,1}|^2|\tilde{k}_{z,2}^\pm|^2+\kappa_R^2\left[|\tilde{k}_{z,2}^\pm|^2-|\tilde{k}_{z,1}|^2-\kappa_R^2\right]\right\}.
			\label{Eq.VI.41b}
		\end{align}
	\end{subequations}
\end{tcolorbox}
\noindent Proceeding similarly with the rest of contributions of Eq.~\eqref{Eq.VI.36}:
\begin{tcolorbox}[sharp corners,colback=red!5!white,colframe=red!50!white]
	\vspace{-0.35cm}
	\begin{equation}
		\bar{\mathcal{G}}_{\rm R-R\,(1,1)}^{(\omega^\pm)}=\frac{1}{32\pi k_0^5}\int\limits_{0}^{+\infty}{d\kappa_R\frac{\kappa_R}{|\tilde{k}_{z,2}|^2|\tilde{k}_{z,2}^\pm|^2} \bar{\mathcal{G}}_{\rm R-R\,(1,1)}^{(\omega^\pm)(\downarrow/\downarrow)}\bar{\Gamma}_{\rm R-R\,(1,1)}^{(\omega^\pm)} },
		\label{Eq.VI.42}
	\end{equation}
\end{tcolorbox}
\noindent where
\begin{align*}
	\bar{\mathcal{G}}_{\rm R-R\,(1,1)}^{(\omega^\pm)(\downarrow/\downarrow)}&=\left|t^{\rm (s)}\right|^2\left|r^{\rm (s)(\omega^\pm)}\right|^2+\left|t^{\rm (p)}\right|^2\left|r^{\rm (p)(\omega^\pm)}\right|^2\frac{{\rm T}^{(\omega^\pm)}}{(1\pm\tilde{\Omega})^2\left|\varepsilon(\omega)\right|\left|\varepsilon(\omega^\pm)\right|}\left\{\left[\tilde{k}_{z,2}^*\tilde{k}_{z,2}^{\pm *}-\kappa_R^2\right]\left[\tilde{k}_{z,2}\tilde{k}_{z,2}^{\pm}-\kappa_R^2\right]\right\},\\
	\bar{\Gamma}_{\rm R-R\,(1,1)}^{(\omega^\pm)}&=\frac{e^{-2k_0\text{Im}{[\tilde{k}_{z,1}]}z}}{2\text{Im}{[\tilde{k}_{z,2}^\pm]}|\tilde{k}_{z,2}+\tilde{k}_{z,2}^{\pm}|^2},
\end{align*}
\begin{tcolorbox}[sharp corners,colback=red!5!white,colframe=red!50!white]
	\vspace{-0.35cm}
	\begin{equation}
		\bar{\mathcal{G}}_{\rm T-R\,(1,1)}^{(\omega^\pm)}=\frac{1}{32\pi k_0^5} \int\limits_{0}^{+\infty}{d\kappa_R\frac{\kappa_R}{|\tilde{k}_{z,2}|^2|\tilde{k}_{z,2}^\pm|^2} \sum_{i=\left\{\uparrow,\downarrow\right\}}{\bar{\mathcal{G}}_{\rm T-R\,(1,1)}^{(\omega^\pm)(i/\downarrow)}\bar{\Gamma}_{\rm T-R\,(1,1)}^{(\omega^\pm)(i)}} },
		\label{Eq.VI.43}
	\end{equation}
\end{tcolorbox}
\noindent where
\begin{align*}
	\bar{\mathcal{G}}_{\rm T-R\,(1,1)}^{(\omega^\pm)(\uparrow/\downarrow)}&=\left|t^{\rm (s)}\right|^2r^{\rm (s)(\omega^\pm)}+\left|t^{\rm (p)}\right|^2r^{\rm (p)(\omega^\pm)}\frac{{\rm T}^{(\omega^\pm)}}{(1\pm\tilde{\Omega})^2\left|\varepsilon(\omega)\right|\left|\varepsilon(\omega^\pm)\right|}\left\{\left[\tilde{k}_{z,2}^*\tilde{k}_{z,2}^{\pm *}+\kappa_R^2\right]\left[\tilde{k}_{z,2}\tilde{k}_{z,2}^{\pm}-\kappa_R^2\right]\right\},\\
	\bar{\Gamma}_{\rm T-R\,(1,1)}^{(\omega^\pm)(\uparrow)}&=\frac{e^{-2k_0\text{Im}{[\tilde{k}_{z,1}]}z}}{2\text{Im}{[\tilde{k}_{z,2}^\pm]}[\tilde{k}_{z,2}+\tilde{k}_{z,2}^\pm][\tilde{k}_{z,2}^*-\tilde{k}_{z,2}^\pm]},\\\\
	\bar{\mathcal{G}}_{\rm T-R\,(1,1)}^{(\omega^\pm)(\downarrow/\downarrow)}&=\left|t^{\rm (s)}\right|^2r^{\rm (s)(\omega^\pm)}+\left|t^{\rm (p)}\right|^2r^{\rm (p)(\omega^\pm)}\frac{{\rm R}^{(\omega^\pm)}}{(1\pm\tilde{\Omega})^2\left|\varepsilon(\omega)\right|\left|\varepsilon(\omega^\pm)\right|}\left\{\left[\tilde{k}_{z,2}^*\tilde{k}_{z,2}^{\pm *}-\kappa_R^2\right]\left[\tilde{k}_{z,2}\tilde{k}_{z,2}^{\pm}-\kappa_R^2\right]\right\},\\
	\bar{\Gamma}_{\rm T-R\,(1,1)}^{(\omega^\pm)(\downarrow)}&=\frac{e^{-2k_0\text{Im}{[\tilde{k}_{z,1}]}z}}{i|\tilde{k}_{z,2}+\tilde{k}_{z,2}^\pm|^2[\tilde{k}_{z,2}^*-\tilde{k}_{z,2}^\pm]},
\end{align*}
\begin{tcolorbox}[sharp corners,colback=red!5!white,colframe=red!50!white]
	\vspace{-0.35cm}
	\begin{equation}
		\bar{\mathcal{G}}_{\rm Z-Z\,(1,1)}^{(\omega^\pm)}=\frac{1}{32\pi k_0^5} \int\limits_{0}^{+\infty}{d\kappa_R\frac{2\kappa_R^3(|\tilde{k}_{z,1}|^2+\kappa_R^2)}{|\tilde{k}_{z,2}|^2(1\pm\tilde{\Omega})^4\left|\varepsilon(\omega)\right|\left|\varepsilon(\omega^\pm)\right|^2}\left|t^{\rm (p)}\right|^2 \frac{e^{-2k_0\text{Im}{\left[\tilde{k}_{z,1}\right]z}}}{\text{Im}{[\tilde{k}_{z,2}]}} },
		\label{Eq.VI.44}
	\end{equation}
\end{tcolorbox}
\begin{tcolorbox}[sharp corners,colback=red!5!white,colframe=red!50!white]
	\vspace{-0.35cm}
	\begin{equation}
		\bar{\mathcal{G}}_{\rm T-Z\,(1,1)}^{(\omega^\pm)}=\frac{1}{32\pi k_0^5} \int\limits_{0}^{+\infty}{d\kappa_R\frac{2\kappa_R^3(|\tilde{k}_{z,1}|^2+\kappa_R^2)}{|\tilde{k}_{z,2}|^2(1\pm\tilde{\Omega})^4\left|\varepsilon(\omega)\right|\left|\varepsilon(\omega^\pm)\right|^2}\left|t^{\rm (p)}\right|^2 \sum_{i=\left\{\uparrow,\downarrow\right\}}{\bar{\mathcal{G}}_{\rm T-Z\,(1,1)}^{(\omega^\pm)(i)}\bar{\Gamma}_{\rm T-Z\,(1,1)}^{(\omega^\pm)(i)}} },
		\label{Eq.VI.45}
	\end{equation}
\end{tcolorbox}
\noindent where
\begin{align*}
	\bar{\mathcal{G}}_{\rm T-Z\,(1,1)}^{(\omega^\pm)(\uparrow)}&=\left[\tilde{k}_{z,2}^*\tilde{k}_{z,2}^{\pm *}+\kappa_R^2\right],\\
	\bar{\Gamma}_{\rm T-Z\,(1,1)}^{(\omega^\pm)(\uparrow)}&=\frac{e^{-2k_0\text{Im}{[\tilde{k}_{z,1}]}z}}{2\text{Im}{[\tilde{k}_{z,2}^{*}]}[\tilde{k}_{z,2}-\tilde{k}_{z,2}^{\pm*}]\tilde{k}_{z,2}^{\pm*}},\\\\
	\bar{\mathcal{G}}_{\rm T-Z\,(1,1)}^{(\omega^\pm)(\downarrow)}&=\left[\tilde{k}_{z,2}^*\tilde{k}_{z,2}^{\pm *}-\kappa_R^2\right],\\
	\bar{\Gamma}_{\rm T-Z\,(1,1)}^{(\omega^\pm)(\downarrow)}&=\frac{e^{-2k_0\text{Im}{[\tilde{k}_{z,1}]}z}}{2\text{Im}{[\tilde{k}_{z,2}^{*}]}[\tilde{k}_{z,2}^*+\tilde{k}_{z,2}^{\pm*}]\tilde{k}_{z,2}^{\pm*}};
\end{align*}
\begin{tcolorbox}[sharp corners,colback=red!5!white,colframe=red!50!white]
	\vspace{-0.35cm}
	\begin{equation}
		\bar{\mathcal{G}}_{\rm R-Z\,(1,1)}^{(\omega^\pm)}=\frac{1}{32\pi k_0^5} \int\limits_{0}^{+\infty}{d\kappa_R\frac{2\kappa_R^3(|\tilde{k}_{z,1}|^2+\kappa_R^2)}{|\tilde{k}_{z,2}|^2(1\pm\tilde{\Omega})^4\left|\varepsilon(\omega)\right|\left|\varepsilon(\omega^\pm)\right|^2}\left|t^{\rm (p)}\right|^2\bar{\mathcal{G}}_{\rm R-Z\,(1,1)}^{(\omega^\pm)(\downarrow)}\bar{\Gamma}_{\rm R-Z\,(1,1)}^{(\omega^\pm)}},
		\label{Eq.VI.46}
	\end{equation}
\end{tcolorbox}
\noindent where
\begin{align*}
	\bar{\mathcal{G}}_{\rm R-Z\,(1,1)}^{(\omega^\pm)(\downarrow)}&=r^{\rm (p)(\omega^\pm)*}\left[\tilde{k}_{z,2}^*\tilde{k}_{z,2}^{\pm *}-\kappa_R^2\right],\\
	\bar{\Gamma}_{\rm R-Z\,(1,1)}^{(\omega^\pm)}&=\frac{e^{-2k_0\text{Im}{[\tilde{k}_{z,1}]}z}}{-i[\tilde{k}_{z,2}-\tilde{k}_{z,2}^{\pm*}][\tilde{k}_{z,2}^*+\tilde{k}_{z,2}^{\pm*}]\tilde{k}_{z,2}^{\pm*}}.
\end{align*}

\subsubsection{Calculation of $\tilde{\mathcal{G}}_{1,1}^{(\omega^\pm)}{\rm :}$}
\label{SectSI.VI.C.II}

Now, following a similar procedure, we address the calculation of $\tilde{\mathcal{G}}_{1,1}^{(\omega^\pm)}$. In this case, the order of the matrices involved is to be:
\begin{equation*}
	[\hat{E}^{(+)}_{1,\alpha}]^\dagger\cdot\hat{E}^{(+)}_{1,\alpha}\propto G_{\alpha,\gamma}^*\hat{j}^\dagger_{1,\gamma} G_{\alpha,\delta}\hat{j}_{1,\delta}\propto G^*_{\alpha,\gamma}G_{\gamma,\beta}\hat{j}_{0,\beta} G_{\alpha,\delta}G^*_{\delta,\zeta}\hat{j}^\dagger_{0,\zeta}\propto G_{\alpha,\gamma}^*G_{\gamma,\beta} G_{\alpha,\delta}G^*_{\delta,\zeta}\delta_{\beta,\zeta}\propto G^*_{\alpha,\gamma}G_{\gamma,\beta} G_{\alpha,\delta}G^*_{\delta,\beta}.
\end{equation*}
In that way, it is easy to see that the four matrices can again be grouped in two pairs of customary products:
\begin{align*}
	{\bf G}^*({\bf r},\boldsymbol{\rho},\omega){\bf G}(\boldsymbol{\rho},\boldsymbol{\rho}'',\omega^\pm){\bf G}({\bf r},\boldsymbol{\rho}',\omega){\bf G}^*(\boldsymbol{\rho}',\boldsymbol{\rho}'',\omega^\pm)&=G^*_{\alpha,\gamma}({\bf r},\boldsymbol{\rho},\omega)G_{\gamma,\beta}(\boldsymbol{\rho},\boldsymbol{\rho}'',\omega^\pm)G_{\alpha,\delta}({\bf r},\boldsymbol{\rho}',\omega)G^*_{\delta,\beta}(\boldsymbol{\rho}',\boldsymbol{\rho}'',\omega^\pm)\\
	&=\left|\tilde{G}_{\alpha,\beta}({\bf r},\boldsymbol{\rho},\boldsymbol{\rho}'',\omega,\omega^\pm)\right|^2\left|\tilde{G}_{\alpha,\beta}({\bf r},\boldsymbol{\rho}',\boldsymbol{\rho}'',\omega,\omega^\pm)\right|^2,
\end{align*}
with $|\tilde{G}_{\alpha,\beta}({\bf r},\boldsymbol{\rho},\boldsymbol{\rho}'',\omega,\omega^\pm)|^2\equiv G^*_{\alpha,\gamma}({\bf r},\boldsymbol{\rho},\omega)G_{\gamma,\beta}(\boldsymbol{\rho},\boldsymbol{\rho}'',\omega^\pm)$, $|\tilde{G}_{\alpha,\beta}({\bf r},\boldsymbol{\rho}',\boldsymbol{\rho}'',\omega,\omega^\pm)|^2\equiv G_{\alpha,\delta}({\bf r},\boldsymbol{\rho}',\omega)G^*_{\delta,\beta}(\boldsymbol{\rho}',\boldsymbol{\rho}'',\omega^\pm)$, and noticing, once again, that the summation over $\gamma$ and $\delta$ is actually the customary matrix product. Inserting this latter expression into \eqref{Eq.VI.33b}, and applying the expansion given in Eq.~\eqref{Eq.VI.01}, it follows that:
\begin{align*}
	&\tilde{\mathcal{G}}_{1,1}^{(\omega^\pm)}=\iiint\limits_{\mathcal{V}}{d^3\boldsymbol{\rho}d^3\boldsymbol{\rho}'d^3\boldsymbol{\rho}''\text{Tr}{[{\bf G}^*({\bf r},\boldsymbol{\rho},\omega){\bf G}(\boldsymbol{\rho},\boldsymbol{\rho}'',\omega^\pm){\bf G}({\bf r},\boldsymbol{\rho}',\omega){\bf G}^*(\boldsymbol{\rho}',\boldsymbol{\rho}'',\omega^\pm)]}}\\
	&=\frac{k_0^2}{4\pi^2}\iiint\limits_{-\infty}^{\quad 0}{d\rho_zd\rho_z'd\rho_z'' \iint\limits_{-\infty}^{\quad +\infty}{d^2\boldsymbol{\kappa}_\parallel\text{Tr}{[\hat{\bf G}^*(\kappa_x,\kappa_y;\omega|z,\rho_z)\hat{\bf G}(\kappa_x^\pm,\kappa_y^\pm;\omega^\pm|\rho_z,\rho_z'')\hat{\bf G}(\kappa_x,\kappa_y;\omega|z,\rho_z')\hat{\bf G}^*(\kappa_x^\pm,\kappa_y^\pm;\omega^\pm|\rho_z',\rho_z'')]}}}.
\end{align*}
Hence:
\begin{tcolorbox}[sharp corners,colback=blue!5!white,colframe=blue!50!white]
	\vspace{-0.35cm}
	\begin{equation}
		\tilde{\mathcal{G}}_{1,1}^{(\omega^\pm)}=\frac{k_0^2}{4\pi^2}\iiint\limits_{-\infty}^{\quad 0}{d\rho_zd\rho_z'd\rho_z'' \int\limits_{0}^{+\infty}{d\kappa_R \kappa_R \int\limits_{0}^{2\pi}{d\kappa_\varphi \norm{\left|\hat{\tilde{{\bf G}}}_{z\leftarrow \rho_z\leftarrow \rho_z''}^{(\omega^\pm)[(\uparrow/\uparrow)(\updownarrow/\updownarrow)]}\right| \left|\hat{\tilde{{\bf G}}}_{z\leftarrow \rho_z'\leftarrow \rho_z''}^{(\omega^\pm)[(\uparrow/\uparrow)(\updownarrow/\updownarrow)]}\right|}_{\mathcal{F}}}}},
		\label{Eq.VI.47}
	\end{equation}
	where
	\begin{subequations}
		\begin{align}
			&\left|\hat{\tilde{{\bf G}}}_{z\leftarrow \rho_z\leftarrow \rho_z''}^{(\omega^\pm)[(\uparrow/\uparrow)(\updownarrow/\updownarrow)]}\right|\equiv\left|\hat{\tilde{{\bf G}}}_{z\leftarrow \rho_z\leftarrow \rho_z''}^{(\omega^\pm)[(\updownarrow/\updownarrow)]}\right|=\left[\hat{\bf G}^{(\uparrow/\uparrow)}(\kappa_x,\kappa_y;\omega|z,\rho_z)\right]^*\cdot\left[\hat{\bf G}^{(\updownarrow/\updownarrow)}(\kappa_x^\pm,\kappa_y^\pm;\omega^\pm|\rho_z,\rho_z'')\right],
			\label{Eq.VI.48a}\\
			&\left|\hat{\tilde{{\bf G}}}_{z\leftarrow \rho_z'\leftarrow \rho_z''}^{(\omega^\pm)[(\uparrow/\uparrow)(\updownarrow/\updownarrow)]}\right|\equiv \left|\hat{\tilde{{\bf G}}}_{z\leftarrow \rho_z'\leftarrow \rho_z''}^{(\omega^\pm)[(\updownarrow/\updownarrow)]}\right|=\hat{\bf G}^{(\uparrow/\uparrow)}(\kappa_x,\kappa_y;\omega|z,\rho_z')\cdot\left[\hat{\bf G}^{(\updownarrow/\updownarrow)}(\kappa_x^\pm,\kappa_y^\pm;\omega^\pm|\rho_z',\rho_z'')\right]^*.
			\label{Eq.VI.48b}
		\end{align}
	\end{subequations}
\end{tcolorbox}
\noindent Then, proceeding as previously,
\begin{align*}
	\left|\hat{\tilde{{\bf G}}}_{z\leftarrow \rho_z\leftarrow \rho_z''}^{(\omega^\pm)[\blue{(\updownarrow/\updownarrow)}]}\right|&=\frac{1}{4}\frac{1}{k_{z,2}^*k_{z,2}^{\pm}}\left[t^{\rm (s)*} \hat{\bf g}^{\rm (s)}+t^{\rm (p)*} \left(\left[\hat{\bf p}_{\rho_z}^{\uparrow}\right]^*\cdot\hat{\bf p}_{\rho_z}^{\rm (\omega^\pm)\blue{\updownarrow}}\right)\left(\left[\hat{\bf p}_{z}^{\uparrow}\right]^*\otimes\hat{\bf p}_{\rho_z''}^{\rm (\omega^\pm)\blue{\updownarrow}}\right)\right]e^{-i\left[k_{z,1}^*z-k_{z,2}^*\rho_z\right]}e^{ik_{z,2}^\pm\left|\rho_z-\rho_z''\right|}\\
	&+\frac{1}{4}\frac{1}{k_{z,2}^*k_{z,2}^\pm}\left[t^{\rm (s)*}r^{\rm (s)(\omega^\pm)}\hat{\bf g}^{\rm (s)}+t^{\rm (p)*}r^{\rm (p)(\omega^\pm)}\left(\left[\hat{\bf p}_{\rho_z}^{\uparrow}\right]^*\cdot\hat{\bf p}_{\rho_z}^{\rm (\omega^\pm)\blue{\downarrow}}\right)\left(\left[\hat{\bf p}_{z}^{\uparrow}\right]^*\otimes\hat{\bf p}_{\rho_z''}^{\rm (\omega^\pm)\blue{\uparrow}}\right)\right]e^{-i\left[k_{z,1}^*z-k_{z,2}^*\rho_z\right]}e^{-ik_{z,2}^\pm\left[\rho_z+\rho_z''\right]}\\
	&+\frac{i}{2}\frac{\delta{[\rho_z-\rho_z'']}}{k_{z,2}^*k_0^2(1\pm\tilde{\Omega})^2\varepsilon(\omega^\pm)}t^{\rm (p)*}\left(\left[\hat{\bf p}_{\rho_z}^{\uparrow}\right]^*\cdot \hat{\bf z}\right)\left(\left[{\bf p}_{z}^{\uparrow}\right]^*\otimes\hat{\bf z}\right)e^{-i\left[k_{z,1}^*z-k_{z,2}^*\rho_z\right]}\\
	&=\left|\hat{\tilde{{\bf T}}}_{z\leftarrow \rho_z\leftarrow \rho_z''}^{(\omega^\pm)[\blue{(\updownarrow/\updownarrow)}]}\right|+\left|\hat{\tilde{{\bf R}}}_{z\leftarrow \rho_z\leftarrow \rho_z''}^{(\omega^\pm)[\blue{(\downarrow/\uparrow)}]}\right|+\left|\hat{\tilde{{\bf Z}}}_{z\leftarrow \rho_z\leftarrow \rho_z''}^{(\omega^\pm)}\right|,
\end{align*}
\begin{align*}
	\left|\hat{\tilde{{\bf G}}}_{z\leftarrow \rho_z'\leftarrow \rho_z''}^{(\omega^\pm)[\red{(\updownarrow/\updownarrow)}]}\right|&=\frac{1}{4}\frac{1}{k_{z,2}k_{z,2}^{\pm*}}\left[t^{\rm (s)} \hat{\bf g}^{\rm (s)}+t^{\rm (p)} \left(\hat{\bf p}_{\rho_z'}^{\uparrow}\cdot\left[\hat{\bf p}_{\rho_z'}^{\rm (\omega^\pm)\red{\updownarrow}}\right]^*\right)\left(\hat{\bf p}_{z}^{\uparrow}\otimes\left[\hat{\bf p}_{\rho_z''}^{\rm (\omega^\pm)\red{\updownarrow}}\right]^*\right)\right]e^{i\left[k_{z,1}z-k_{z,2}\rho_z'\right]}e^{-ik_{z,2}^{\pm*}\left|\rho_z'-\rho_z''\right|}\\
	&+\frac{1}{4}\frac{1}{k_{z,2}k_{z,2}^{\pm*}}\left[t^{\rm (s)}r^{\rm (s)(\omega^\pm)*}\hat{\bf g}^{\rm (s)}+t^{\rm (p)}r^{\rm (p)(\omega^\pm)*}\left(\hat{\bf p}_{\rho_z'}^{\uparrow}\cdot\left[\hat{\bf p}_{\rho_z'}^{\rm (\omega^\pm)\red{\downarrow}}\right]^*\right)\left(\hat{\bf p}_{z}^{\uparrow}\otimes\left[\hat{\bf p}_{\rho_z''}^{\rm (\omega^\pm)\red{\uparrow}}\right]^*\right)\right]e^{i\left[k_{z,1}z-k_{z,2}\rho_z'\right]}e^{ik_{z,2}^{\pm*}\left[\rho_z'+\rho_z''\right]}\\
	&-\frac{i}{2}\frac{\delta{[\rho_z'-\rho_z'']}}{k_{z,2}k_0^2(1\pm\tilde{\Omega})^2\varepsilon^*(\omega^\pm)}t^{\rm (p)}\left(\hat{\bf p}_{\rho_z'}^{\uparrow}\cdot \hat{\bf z}\right)\left(\hat{\bf p}_{z}^{\uparrow}\otimes\hat{\bf z}\vphantom{\left(\hat{\bf p}_{\rho_z'}^{\uparrow}\cdot \hat{\bf z}\right)}\right)e^{i\left[k_{z,1}z-k_{z,2}\rho_z'\right]}\\
	&=\left|\hat{\tilde{{\bf T}}}_{z\leftarrow \rho_z'\leftarrow \rho_z''}^{(\omega^\pm)[\red{(\updownarrow/\updownarrow)}]}\right|+\left|\hat{\tilde{{\bf R}}}_{z\leftarrow \rho_z'\leftarrow \rho_z''}^{(\omega^\pm)[\red{(\downarrow/\uparrow)}]}\right|+\left|\hat{\tilde{{\bf Z}}}_{z\leftarrow \rho_z'\leftarrow \rho_z''}^{(\omega^\pm)}\right|.
\end{align*}
Therefore, as in the previous case, Eq.~\eqref{Eq.VI.47} can be recast as:
\begin{tcolorbox}[sharp corners,colback=red!5!white,colframe=red!50!white]
	\vspace{-0.35cm}
	\begin{equation}
		\tilde{\mathcal{G}}_{1,1}^{(\omega^\pm)}=\tilde{\mathcal{G}}_{\rm T-T\,(1,1)}^{(\omega^\pm)}+\tilde{\mathcal{G}}_{\rm R-R\,(1,1)}^{(\omega^\pm)}+2\text{Re}{\left[\tilde{\mathcal{G}}_{\rm T-R\,(1,1)}^{(\omega^\pm)}\right]}+\tilde{\mathcal{G}}_{\rm Z-Z\,(1,1)}^{(\omega^\pm)}+2\text{Re}{\left[\tilde{\mathcal{G}}_{\rm T-Z\,(1,1)}^{(\omega^\pm)}\right]}+2\text{Re}{\left[\tilde{\mathcal{G}}_{\rm R-Z\,(1,1)}^{(\omega^\pm)}\right]}.
		\label{Eq.VI.49}
	\end{equation}
\end{tcolorbox}
\noindent From this, we proceed with the calculation of Eq.~\eqref{Eq.VI.49} by analyzing separately each of the above contributions (cf. \hyperref[SectSI.Appendix.VI.B]{\!Appendix VI.B} for the demonstration of the $p$-like contribution to the (T-T)-term):
\begin{tcolorbox}[sharp corners,colback=red!5!white,colframe=red!50!white]
	\vspace{-0.35cm}
	\begin{equation}
		\tilde{\mathcal{G}}_{\rm T-T\,(1,1)}^{(\omega^\pm)}=\frac{1}{32\pi k_0^5}\int\limits_{0}^{+\infty}{d\kappa_R\frac{\kappa_R}{|\tilde{k}_{z,2}|^2|\tilde{k}_{z,2}^\pm|^2} \sum_{i,j=\left\{\uparrow,\downarrow\right\}}{\tilde{\mathcal{G}}_{\rm T-T\,(1,1)}^{(\omega^\pm)(i/j)}\tilde{\Gamma}_{\rm T-T\,(1,1)}^{(\omega^\pm)(i/j)}} },
		\label{Eq.VI.50}
	\end{equation}
\end{tcolorbox}
\noindent where
\begin{align*}
	\tilde{\mathcal{G}}_{\rm T-T\,(1,1)}^{(\omega^\pm)(\uparrow/\uparrow)}&=\left|t^{\rm (s)}\right|^2+\left|t^{\rm (p)}\right|^2\frac{{\rm T}^{(\omega^\pm)}}{(1\pm\tilde{\Omega})^2\left|\varepsilon(\omega)\right|\left|\varepsilon(\omega^\pm)\right|}\left\{\left[\tilde{k}_{z,2}^*\tilde{k}_{z,2}^{\pm}+\kappa_R^2\right]\left[\tilde{k}_{z,2}\tilde{k}_{z,2}^{\pm*}+\kappa_R^2\right]\right\},\\
	\tilde{\Gamma}_{\rm T-T\,(1,1)}^{(\omega^\pm)(\uparrow/\uparrow)}&=\frac{2e^{-2k_0\text{Im}{[\tilde{k}_{z,1}]}z}}{|\tilde{k}_{z,2}+\tilde{k}_{z,2}^{\pm*}|^2}\left[\frac{\text{Im}{[\tilde{k}_{z,2}]}+\text{Im}{[\tilde{k}_{z,2}^\pm]}}{4\text{Im}{[\tilde{k}_{z,2}]}\text{Im}{[\tilde{k}_{z,2}^\pm]}}-\frac{\text{Im}{[\tilde{k}_{z,2}]}+\text{Im}{[\tilde{k}_{z,2}^\pm]}}{|\tilde{k}_{z,2}+\tilde{k}_{z,2}^{\pm}|^2}\right],\\\\
	\tilde{\mathcal{G}}_{\rm T-T\,(1,1)}^{(\omega^\pm)(\uparrow/\downarrow)}&=\left|t^{\rm (s)}\right|^2+\left|t^{\rm (p)}\right|^2\frac{{\rm R}^{(\omega^\pm)}}{(1\pm\tilde{\Omega})^2\left|\varepsilon(\omega)\right|\left|\varepsilon(\omega^\pm)\right|}\left\{\left[\tilde{k}_{z,2}^*\tilde{k}_{z,2}^{\pm}+\kappa_R^2\right]\left[\tilde{k}_{z,2}\tilde{k}_{z,2}^{\pm*}-\kappa_R^2\right]\right\},\\
	\tilde{\Gamma}_{\rm T-T\,(1,1)}^{(\omega^\pm)(\uparrow/\downarrow)}&=\frac{e^{-2k_0\text{Im}{[\tilde{k}_{z,1}]}z}}{2\text{Im}{[\tilde{k}_{z,2}^*]}[\tilde{k}_{z,2}-\tilde{k}_{z,2}^{\pm*}][\tilde{k}_{z,2}+\tilde{k}_{z,2}^\pm]},\\\\
	\tilde{\mathcal{G}}_{\rm T-T\,(1,1)}^{(\omega^\pm)(\downarrow/\uparrow)}&=\left|t^{\rm (s)}\right|^2+\left|t^{\rm (p)}\right|^2\frac{{\rm R}^{(\omega^\pm)}}{(1\pm\tilde{\Omega})^2\left|\varepsilon(\omega)\right|\left|\varepsilon(\omega^\pm)\right|}\left\{\left[\tilde{k}_{z,2}^*\tilde{k}_{z,2}^{\pm}-\kappa_R^2\right]\left[\tilde{k}_{z,2}\tilde{k}_{z,2}^{\pm*}+\kappa_R^2\right]\right\},\\
	\tilde{\Gamma}_{\rm T-T\,(1,1)}^{(\omega^\pm)(\downarrow/\uparrow)}&=\frac{e^{-2k_0\text{Im}{[\tilde{k}_{z,1}]}z}}{2\text{Im}{[\tilde{k}_{z,2}^*]}[\tilde{k}_{z,2}^*-\tilde{k}_{z,2}^{\pm}][\tilde{k}_{z,2}^*+\tilde{k}_{z,2}^{\pm *}]},\\\\
	\tilde{\mathcal{G}}_{\rm T-T\,(1,1)}^{(\omega^\pm)(\downarrow/\downarrow)}&=\left|t^{\rm (s)}\right|^2+\left|t^{\rm (p)}\right|^2\frac{{\rm T}^{(\omega^\pm)}}{(1\pm\tilde{\Omega})^2\left|\varepsilon(\omega)\right|\left|\varepsilon(\omega^\pm)\right|}\left\{\left[\tilde{k}_{z,2}^*\tilde{k}_{z,2}^{\pm}-\kappa_R^2\right]\left[\tilde{k}_{z,2}\tilde{k}_{z,2}^{\pm*}-\kappa_R^2\right]\right\},\\
	\tilde{\Gamma}_{\rm T-T\,(1,1)}^{(\omega^\pm)(\downarrow/\downarrow)}&=\frac{e^{-2k_0\text{Im}{[\tilde{k}_{z,1}]}z}}{2\text{Im}{[\tilde{k}_{z,2}]}|\tilde{k}_{z,2}-\tilde{k}_{z,2}^{\pm*}|^2},
\end{align*}
\begin{tcolorbox}[sharp corners,colback=red!5!white,colframe=red!50!white]
	\vspace{-0.35cm}
	\begin{equation}
		\tilde{\mathcal{G}}_{\rm R-R\,(1,1)}^{(\omega^\pm)}=\frac{1}{32\pi k_0^5}\int\limits_{0}^{+\infty}{d\kappa_R\frac{\kappa_R}{|\tilde{k}_{z,2}|^2|\tilde{k}_{z,2}^\pm|^2} \tilde{\mathcal{G}}_{\rm R-R\,(1,1)}^{(\omega^\pm)(\downarrow/\downarrow)}\tilde{\Gamma}_{\rm R-R\,(1,1)}^{(\omega^\pm)} },
		\label{Eq.VI.51}
	\end{equation}
\end{tcolorbox}
\noindent where
\begin{align*}
	\tilde{\mathcal{G}}_{\rm R-R\,(1,1)}^{(\omega^\pm)(\downarrow/\downarrow)}&=\left|t^{\rm (s)}\right|^2\left|r^{\rm (s)(\omega^\pm)}\right|^2+\left|t^{\rm (p)}\right|^2\left|r^{\rm (p)(\omega^\pm)}\right|^2\frac{{\rm T}^{(\omega^\pm)}}{(1\pm\tilde{\Omega})\left|\varepsilon(\omega)\right|\left|\varepsilon(\omega^\pm)\right|}\left\{\left[\tilde{k}_{z,2}^*\tilde{k}_{z,2}^{\pm}-\kappa_R^2\right]\left[\tilde{k}_{z,2}\tilde{k}_{z,2}^{\pm*}-\kappa_R^2\right]\right\},\\
	\tilde{\Gamma}_{\rm R-R\,(1,1)}^{(\omega^\pm)}&=\frac{e^{-2k_0\text{Im}{[\tilde{k}_{z,1}]}z}}{2\text{Im}{[\tilde{k}_{z,2}^\pm]}|\tilde{k}_{z,2}-\tilde{k}_{z,2}^{\pm *}|^2},
\end{align*}
\begin{tcolorbox}[sharp corners,colback=red!5!white,colframe=red!50!white]
	\vspace{-0.35cm}
	\begin{equation}
		\tilde{\mathcal{G}}_{\rm T-R\,(1,1)}^{(\omega^\pm)}=\frac{1}{32\pi k_0^5} \int\limits_{0}^{+\infty}{d\kappa_R\frac{\kappa_R}{|\tilde{k}_{z,2}|^2|\tilde{k}_{z,2}^\pm|^2} \sum_{i=\left\{\uparrow,\downarrow\right\}}{\tilde{\mathcal{G}}_{\rm T-R\,(1,1)}^{(\omega^\pm)(i/\downarrow)}\tilde{\Gamma}_{\rm T-R\,(1,1)}^{(\omega^\pm)(i)}} },
		\label{Eq.VI.52}
	\end{equation}
\end{tcolorbox}
\noindent where
\begin{align*}
	\tilde{\mathcal{G}}_{\rm T-R\,(1,1)}^{(\omega^\pm)(\uparrow/\downarrow)}&=\left|t^{\rm (s)}\right|^2r^{\rm (s)(\omega^\pm)*}+\left|t^{\rm (p)}\right|^2r^{\rm (p)(\omega^\pm)*}\frac{{\rm T}^{(\omega^\pm)}}{(1\pm\tilde{\Omega})^2\left|\varepsilon(\omega)\right|\left|\varepsilon(\omega^\pm)\right|}\left\{\left[\tilde{k}_{z,2}^*\tilde{k}_{z,2}^{\pm}+\kappa_R^2\right]\left[\tilde{k}_{z,2}\tilde{k}_{z,2}^{\pm*}-\kappa_R^2\right]\right\},\\
	\tilde{\Gamma}_{\rm T-R\,(1,1)}^{(\omega^\pm)(\uparrow)}&=\frac{e^{-2k_0\text{Im}{[\tilde{k}_{z,1}]}z}}{2\text{Im}{[\tilde{k}_{z,2}^\pm]}[\tilde{k}_{z,2}-\tilde{k}_{z,2}^{\pm*}][\tilde{k}_{z,2}^*+\tilde{k}_{z,2}^{\pm*}]},\\\\
	\tilde{\mathcal{G}}_{\rm T-R\,(1,1)}^{(\omega^\pm)(\downarrow/\downarrow)}&=\left|t^{\rm (s)}\right|^2r^{\rm (s)(\omega^\pm)*}+\left|t^{\rm (p)}\right|^2r^{\rm (p)(\omega^\pm)*}\frac{{\rm R}^{(\omega^\pm)}}{(1\pm\tilde{\Omega})^2\left|\varepsilon(\omega)\right|\left|\varepsilon(\omega^\pm)\right|}\left\{\left[\tilde{k}_{z,2}^*\tilde{k}_{z,2}^{\pm}-\kappa_R^2\right]\left[\tilde{k}_{z,2}\tilde{k}_{z,2}^{\pm*}-\kappa_R^2\right]\right\},\\
	\tilde{\Gamma}_{\rm T-R\,(1,1)}^{(\omega^\pm)(\downarrow)}&=\frac{e^{-2k_0\text{Im}{[\tilde{k}_{z,1}]}z}}{i|\tilde{k}_{z,2}-\tilde{k}_{z,2}^{\pm*}|^2[\tilde{k}_{z,2}^*+\tilde{k}_{z,2}^{\pm*}]}.
\end{align*}
with ${\rm T}^{(\omega^\pm)}$ and ${\rm R}^{(\omega^\pm)}$ given, respectively, in Eq.~\eqref{Eq.VI.41a} and \eqref{Eq.VI.41b}.
\begin{tcolorbox}[sharp corners,colback=red!5!white,colframe=red!50!white]
	\vspace{-0.35cm}
	\begin{equation}
		\tilde{\mathcal{G}}_{\rm Z-Z\,(1,1)}^{(\omega^\pm)}=\frac{1}{32\pi k_0^5}\int\limits_{0}^{+\infty}{d\kappa_R\frac{2\kappa_R^3(|\tilde{k}_{z,1}|^2+\kappa_R^2)}{|\tilde{k}_{z,2}|^2(1\pm\tilde{\Omega})^4\left|\varepsilon(\omega)\right|\left|\varepsilon(\omega^\pm)\right|^2}\left|t^{\rm (p)}\right|^2 \frac{e^{-2k_0\text{Im}{\left[\tilde{k}_{z,1}\right]z}}}{\text{Im}{[\tilde{k}_{z,2}]}} },
		\label{Eq.VI.53}
	\end{equation}
\end{tcolorbox}
\begin{tcolorbox}[sharp corners,colback=red!5!white,colframe=red!50!white]
	\vspace{-0.35cm}
	\begin{equation}
		\tilde{\mathcal{G}}_{\rm T-Z\,(1,1)}^{(\omega^\pm)}=\frac{1}{32\pi k_0^5} \int\limits_{0}^{+\infty}{d\kappa_R\frac{2\kappa_R^3(|\tilde{k}_{z,1}|^2+\kappa_R^2)}{|\tilde{k}_{z,2}|^2(1\pm\tilde{\Omega})^4\left|\varepsilon(\omega)\right|\left|\varepsilon(\omega^\pm)\right|^2}\left|t^{\rm (p)}\right|^2\sum_{i=\left\{\uparrow,\downarrow\right\}}{\tilde{\mathcal{G}}_{\rm T-Z\,(1,1)}^{(\omega^\pm)(i)}\tilde{\Gamma}_{\rm T-Z\,(1,1)}^{(\omega^\pm)(i)}} },
		\label{Eq.VI.54}
	\end{equation}
\end{tcolorbox}
\noindent where
\begin{align*}
	\tilde{\mathcal{G}}_{\rm T-Z\,(1,1)}^{(\omega^\pm)(\uparrow)}&=\left[\tilde{k}_{z,2}^*\tilde{k}_{z,2}^{\pm}+\kappa_R^2\right],\\
	\tilde{\Gamma}_{\rm T-Z\,(1,1)}^{(\omega^\pm)(\uparrow)}&=\frac{e^{-2k_0\text{Im}{[\tilde{k}_{z,1}]}z}}{2\text{Im}{[\tilde{k}_{z,2}]}[\tilde{k}_{z,2}+\tilde{k}_{z,2}^{\pm}]\tilde{k}_{z,2}^{\pm}},\\\\
	\tilde{\mathcal{G}}_{\rm T-Z\,(1,1)}^{(\omega^\pm)(\downarrow)}&=\left[\tilde{k}_{z,2}^*\tilde{k}_{z,2}^{\pm}-\kappa_R^2\right],\\
	\tilde{\Gamma}_{\rm T-Z\,(1,1)}^{(\omega^\pm)(\downarrow)}&=\frac{e^{-2k_0\text{Im}{[\tilde{k}_{z,1}]}z}}{2\text{Im}{[\tilde{k}_{z,2}]}[\tilde{k}_{z,2}^*-\tilde{k}_{z,2}^{\pm}]\tilde{k}_{z,2}^{\pm}},
\end{align*}
\begin{tcolorbox}[sharp corners,colback=red!5!white,colframe=red!50!white]
	\vspace{-0.35cm}
	\begin{equation}
		\tilde{\mathcal{G}}_{\rm R-Z\,(1,1)}^{(\omega^\pm)}=\frac{1}{32\pi k_0^5}\int\limits_{0}^{+\infty}{d\kappa_R\frac{2\kappa_R^3(|\tilde{k}_{z,1}|^2+\kappa_R^2)}{|\tilde{k}_{z,2}|^2(1\pm\tilde{\Omega})^4\left|\varepsilon(\omega)\right|\left|\varepsilon(\omega^\pm)\right|^2}\left|t^{\rm (p)}\right|^2\tilde{\mathcal{G}}_{\rm R-Z\,(1,1)}^{(\omega^\pm)(\downarrow)}\tilde{\Gamma}_{\rm R-Z\,(1,1)}^{(\omega^\pm)}},
		\label{Eq.VI.55}
	\end{equation}
\end{tcolorbox}
\noindent where
\begin{align*}
	\tilde{\mathcal{G}}_{\rm R-Z\,(1,1)}^{(\omega^\pm)(\downarrow)}&=r^{\rm (p)(\omega^\pm)}\left[\tilde{k}_{z,2}^*\tilde{k}_{z,2}^{\pm}-\kappa_R^2\right],\\
	\tilde{\Gamma}_{\rm R-Z\,(1,1)}^{(\omega^\pm)}&=\frac{e^{-2k_0\text{Im}{[\tilde{k}_{z,1}]}z}}{i[\tilde{k}_{z,2}^*-\tilde{k}_{z,2}^{\pm}][\tilde{k}_{z,2}+\tilde{k}_{z,2}^{\pm}]\tilde{k}_{z,2}^{\pm}}.
\end{align*}

\subsubsection{Calculation of $\bar{\bar{\mathcal{G}}}_{0,2}^{(\omega^\pm)}{\rm :}$}
\label{SectSI.VI.C.III}

Finally, we proceed with the calculation of $\bar{\bar{\mathcal{G}}}_{0,2}^{(\omega^\pm)}$. Again, we start by checking the order of the matrices involved:
\begin{equation*}
	[\hat{E}^{(+)}_{0,\alpha}]^\dagger\cdot\hat{E}^{(+)}_{2,\alpha}\propto G_{\alpha,\beta}^*\hat{j}^\dagger_{0,\beta} G_{\alpha,\gamma}\hat{j}_{2,\gamma}\propto G^*_{\alpha,\beta}\hat{j}^\dagger_{0,\beta} G_{\alpha,\gamma}G_{\gamma,\delta}\hat{j}_{1,\delta}\propto G^*_{\alpha,\beta}G_{\alpha,\gamma} G_{\gamma,\delta}G_{\delta,\zeta}\delta_{\beta,\zeta}\propto G^*_{\alpha,\beta}G_{\alpha,\gamma} G_{\gamma,\delta}G_{\delta,\beta}.
\end{equation*}
In this way we can see that the four matrices can again be grouped in two sets, now consisting in a single matrix and an ordinary product of three matrices:
\begin{align*}
	{\bf G}^*({\bf r},\boldsymbol{\rho},\omega){\bf G}({\bf r},\boldsymbol{\rho}',\omega){\bf G}(\boldsymbol{\rho}',\boldsymbol{\rho}'',\omega^\pm){\bf G}^*(\boldsymbol{\rho}'',\boldsymbol{\rho},\omega)&=G^*_{\alpha,\beta}({\bf r},\boldsymbol{\rho},\omega)G_{\alpha,\gamma}({\bf r},\boldsymbol{\rho}',\omega)G_{\gamma,\delta}(\boldsymbol{\rho}',\boldsymbol{\rho}'',\omega^\pm)G_{\delta,\beta}(\boldsymbol{\rho}'',\boldsymbol{\rho},\omega)\\
	&=\bar{\bar{G}}^*_{\alpha,\beta}({\bf r},\boldsymbol{\rho},\omega)\bar{\bar{G}}_{\alpha,\beta}({\bf r},\boldsymbol{\rho},\boldsymbol{\rho}',\boldsymbol{\rho}'',\omega,\omega^\pm),
\end{align*}
with $\bar{\bar{G}}^*_{\alpha,\beta}({\bf r},\boldsymbol{\rho},\omega)\equiv G^*_{\alpha,\beta}({\bf r},\boldsymbol{\rho},\omega)$, $\bar{\bar{G}}_{\alpha,\beta}({\bf r},\boldsymbol{\rho},\boldsymbol{\rho}',\boldsymbol{\rho}'',\omega,\omega^\pm)\equiv G_{\alpha,\gamma}({\bf r},\boldsymbol{\rho}',\omega)G_{\gamma,\delta}(\boldsymbol{\rho}',\boldsymbol{\rho}'',\omega^\pm)G_{\delta,\beta}(\boldsymbol{\rho}'',\boldsymbol{\rho},\omega)$, and noticing, once again, that the summations over $\gamma$ and $\delta$ stand for the ordinary matrix product. Inserting this latter expression into \eqref{Eq.VI.30}, and applying the expansion given in Eq.~\eqref{Eq.VI.01}, it follows that:
\begin{align*}
	&\bar{\bar{\mathcal{G}}}_{0,2}^{(\omega^\pm)}=\iiint\limits_{\mathcal{V}}{d^3\boldsymbol{\rho}d^3\boldsymbol{\rho}'d^3\boldsymbol{\rho}''\text{Tr}{[{\bf G}^*({\bf r},\boldsymbol{\rho},\omega){\bf G}({\bf r},\boldsymbol{\rho}',\omega){\bf G}(\boldsymbol{\rho}',\boldsymbol{\rho}'',\omega^\pm){\bf G}^*(\boldsymbol{\rho}'',\boldsymbol{\rho},\omega)]}}\\
	&=\frac{k_0^2}{4\pi^2}\iiint\limits_{-\infty}^{\quad 0}{d\rho_zd\rho_z'd\rho_z'' \iint\limits_{-\infty}^{\quad +\infty}{d^2\boldsymbol{\kappa}_\parallel \text{Tr}{[\hat{\bf G}^*(\kappa_x,\kappa_y;\omega|z,\rho_z)\hat{\bf G}(\kappa_x,\kappa_y;\omega|z,\rho_z')\hat{\bf G}(\kappa_x^\pm,\kappa_y^\pm;\omega^\pm|\rho_z',\rho_z'')\hat{\bf G}(\kappa_x,\kappa_y;\omega|\rho_z'',\rho_z)}]}}.
\end{align*}
Hence:
\begin{tcolorbox}[sharp corners,colback=blue!5!white,colframe=blue!50!white]
	\vspace{-0.35cm}
	\begin{equation}
		\bar{\bar{\mathcal{G}}}_{0,2}^{(\omega^\pm)}=\frac{k_0^2}{4\pi^2}\iiint\limits_{-\infty}^{\quad 0}{d\rho_zd\rho_z'd\rho_z'' \int\limits_{0}^{+\infty}{d\kappa_R \kappa_R \int\limits_{0}^{2\pi}{d\kappa_\varphi \norm{\left[\hat{\bar{\bar{{\bf G}}}}_{z\leftarrow \rho_z}^{[(\uparrow/\uparrow)]}\right]^*\hat{\bar{\bar{{\bf G}}}}_{z\leftarrow \rho_z'\leftarrow \rho_z'' \leftarrow \rho_z}^{(\omega^\pm)[(\uparrow/\uparrow)(\updownarrow/\updownarrow)(\updownarrow/\updownarrow)]}}_{\mathcal{F}}}}},
		\label{Eq.VI.56}
	\end{equation}
	where
	\begin{subequations}
		\begin{align}
			&\!\!\!\!\!\!\left[\hat{\bar{\bar{{\bf G}}}}_{z\leftarrow \rho_z}^{[(\uparrow/\uparrow)]}\right]^*\equiv\left[\hat{\bar{\bar{{\bf G}}}}_{z\leftarrow \rho_z}\right]^*=\left[\hat{{\bf G}}^{(\uparrow/\uparrow)}(\kappa_x,\kappa_y;\omega|z,\rho_z)\right]^*,
			\label{Eq.VI.57a}\\
			&\!\!\!\!\hat{\bar{\bar{{\bf G}}}}_{z\leftarrow \rho_z'\leftarrow \rho_z'' \leftarrow \rho_z}^{(\omega^\pm)[(\uparrow/\uparrow)(\updownarrow/\updownarrow)(\updownarrow/\updownarrow)]}\equiv \hat{\bar{\bar{{\bf G}}}}_{z\leftarrow \rho_z'\leftarrow \rho_z'' \leftarrow \rho_z}^{(\omega^\pm)[(\updownarrow/\updownarrow)(\updownarrow/\updownarrow)]}=\hat{\bf G}^{(\uparrow/\uparrow)}(\kappa_x,\kappa_y;\omega|z,\rho_z')\!\cdot\!\hat{\bf G}^{(\updownarrow/\updownarrow)}(\kappa_x^\pm,\kappa_y^\pm;\omega^\pm|\rho_z',\rho_z'')\!\cdot\!\hat{\bf G}^{(\updownarrow/\updownarrow)}(\kappa_x,\kappa_y;\omega|\rho_z'',\rho_z).
			\label{Eq.VI.57b}
		\end{align}
	\end{subequations}
\end{tcolorbox}
\noindent Then, from the notation given in Eqs.~\eqref{Eq.VI.05}--\eqref{Eq.VI.12d}, in this case it follows that:
\begin{align*}
	\left[\hat{\bar{\bar{{\bf G}}}}_{z\leftarrow \rho_z}\right]^*&=\frac{-i}{2}\frac{1}{k_{z,2}^*}\left[t^{\rm (s)} \hat{\bf g}^{\rm (s)}+t^{\rm (p)}\left(\hat{\bf p}_z^\uparrow\otimes\hat{\bf p}_{\rho_z}^\uparrow\right)\right]^*e^{-i\left[k_{z,1}^*z-k_{z,2}^*\rho_z\right]},\\\\
	\hat{\bar{\bar{{\bf G}}}}_{z\leftarrow \rho_z'\leftarrow \rho_z'' \leftarrow \rho_z}^{(\omega^\pm)[\red{(\updownarrow/\updownarrow)}\blue{(\updownarrow/\updownarrow)}]}&=\hat{\bar{\bar{{\bf TT}}}}_{z\leftarrow \rho_z'\leftarrow \rho_z'' \leftarrow \rho_z}^{(\omega^\pm)[\red{(\updownarrow/\updownarrow)}\blue{(\updownarrow/\updownarrow)}]}+\hat{\bar{\bar{{\bf TR}}}}_{z\leftarrow \rho_z'\leftarrow \rho_z'' \leftarrow \rho_z}^{(\omega^\pm)[\red{(\updownarrow/\updownarrow)}\blue{(\downarrow/\uparrow)}]}+\hat{\bar{\bar{{\bf TZ}}}}_{z\leftarrow \rho_z'\leftarrow \rho_z'' \leftarrow \rho_z}^{(\omega^\pm)[\red{(\updownarrow/\updownarrow)}]}\\
	&+\hat{\bar{\bar{{\bf RT}}}}_{z\leftarrow \rho_z'\leftarrow \rho_z'' \leftarrow \rho_z}^{(\omega^\pm)[\red{(\downarrow/\uparrow)}\blue{(\updownarrow/\updownarrow)}]}+\hat{\bar{\bar{{\bf RR}}}}_{z\leftarrow \rho_z'\leftarrow \rho_z'' \leftarrow \rho_z}^{(\omega^\pm)[\red{(\downarrow/\uparrow)}\blue{(\downarrow/\uparrow)}]}+\hat{\bar{\bar{{\bf RZ}}}}_{z\leftarrow \rho_z'\leftarrow \rho_z'' \leftarrow \rho_z}^{(\omega^\pm)[\red{(\downarrow/\uparrow)}]}\\
	&+\hat{\bar{\bar{{\bf ZT}}}}_{z\leftarrow \rho_z'\leftarrow \rho_z'' \leftarrow \rho_z}^{(\omega^\pm)[\blue{(\updownarrow/\updownarrow)}]}+\hat{\bar{\bar{{\bf ZR}}}}_{z\leftarrow \rho_z'\leftarrow \rho_z'' \leftarrow \rho_z}^{(\omega^\pm)[\blue{(\downarrow/\uparrow)}]}+\hat{\bar{\bar{{\bf ZZ}}}}_{z\leftarrow \rho_z'\leftarrow \rho_z'' \leftarrow \rho_z}^{(\omega^\pm)},
\end{align*}
with
\begin{align*}
	\hat{\bar{\bar{{\bf TT}}}}_{z\leftarrow \rho_z'\leftarrow \rho_z'' \leftarrow \rho_z}^{(\omega^\pm)[\red{(\updownarrow/\updownarrow)}\blue{(\updownarrow/\updownarrow)}]}&=\frac{-i}{8}\frac{1}{k_{z,2}^2k_{z,2}^\pm}\left[t^{\rm (s)}\hat{\bf g}^{\rm (s)}+t^{\rm (p)}\left(\hat{\bf p}_{\rho_z'}^\uparrow\cdot\hat{\bf p}_{\rho_z'}^{(\omega^\pm)\red{\updownarrow}}\right)\left(\hat{\bf p}_{\rho_z''}^{(\omega^\pm)\red{\updownarrow}}\cdot\hat{\bf p}_{\rho_z''}^{\blue{\updownarrow}}\right)\left(\hat{\bf p}_z^\uparrow\otimes\hat{\bf p}_{\rho_z}^{\blue{\updownarrow}}\right) \right]\bar{\bar{\Gamma}}_{\rm T-T\,(0,2)}^{(\omega^\pm)(\red{\updownarrow}/\blue{\updownarrow})},\\
	\hat{\bar{\bar{{\bf TR}}}}_{z\leftarrow \rho_z'\leftarrow \rho_z'' \leftarrow \rho_z}^{(\omega^\pm)[\red{(\updownarrow/\updownarrow)}\blue{(\downarrow/\uparrow)}]}&=\frac{-i}{8}\frac{1}{k_{z,2}^2k_{z,2}^\pm}\left[t^{\rm (s)}r^{\rm (s)}\hat{\bf g}^{\rm (s)}+t^{\rm (p)}r^{\rm (p)}\left(\hat{\bf p}_{\rho_z'}^\uparrow\cdot\hat{\bf p}_{\rho_z'}^{(\omega^\pm)\red{\updownarrow}}\right)\left(\hat{\bf p}_{\rho_z''}^{(\omega^\pm)\red{\updownarrow}}\cdot\hat{\bf p}_{\rho_z''}^{\blue{\downarrow}}\right)\left(\hat{\bf p}_z^\uparrow\otimes\hat{\bf p}_{\rho_z}^{\blue{\uparrow}}\vphantom{\left(\hat{\bf p}_{\rho_z''}^{(\omega^\pm)\red{\updownarrow}}\cdot\hat{\bf p}_{\rho_z''}^{\blue{\downarrow}}\right)}\right) \right]\bar{\bar{\Gamma}}_{\rm T-R\,(0,2)}^{(\omega^\pm)(\red{\updownarrow})},\\
	\hat{\bar{\bar{{\bf TZ}}}}_{z\leftarrow \rho_z'\leftarrow \rho_z'' \leftarrow \rho_z}^{(\omega^\pm)[\red{(\updownarrow/\updownarrow)}]}&=\frac{1}{4}\frac{1}{k_{z,2}k_{z,2}^\pm k_0^2\varepsilon(\omega)}\left[t^{\rm (p)}\left(\hat{\bf p}_{\rho_z'}^\uparrow\cdot\hat{\bf p}_{\rho_z'}^{(\omega^\pm)\red{\updownarrow}}\right)\left(\hat{\bf p}_{\rho_z''}^{(\omega^\pm)\red{\updownarrow}}\cdot\hat{\bf z}\right)\left(\hat{\bf p}_z^\uparrow\otimes\hat{\bf z}\vphantom{\left(\hat{\bf p}_{\rho_z''}^{(\omega^\pm)\red{\updownarrow}}\cdot\hat{\bf z}\right)}\right) \right]\bar{\bar{\Gamma}}_{\rm T-Z\,(0,2)}^{(\omega^\pm)(\red{\updownarrow})},
\end{align*}
\begin{align*}
	\bar{\bar{\Gamma}}_{\rm T-T\,(0,2)}^{(\omega^\pm)(\red{\updownarrow}/\blue{\updownarrow})}&=e^{i\left[k_{z,1}z-k_{z,2}\rho_z'\right]}\red{e^{ik_{z,2}^\pm\left|\rho_z'-\rho_z''\right|}}\blue{e^{ik_{z,2}\left|\rho_z''-\rho_z\right|}},\\
	\bar{\bar{\Gamma}}_{\rm T-R\,(0,2)}^{(\omega^\pm)(\red{\updownarrow})}&=e^{i\left[k_{z,1}z-k_{z,2}\rho_z'\right]}\red{e^{ik_{z,2}^\pm\left|\rho_z'-\rho_z''\right|}}\blue{e^{-ik_{z,2}\left[\rho_z''+\rho_z\right]}},\\
	\bar{\bar{\Gamma}}_{\rm T-Z\,(0,2)}^{(\omega^\pm)(\red{\updownarrow})}&=e^{i\left[k_{z,1}z-k_{z,2}\rho_z'\right]}\red{e^{ik_{z,2}^\pm\left|\rho_z'-\rho_z''\right|}}\blue{\delta{[\rho_z''-\rho_z]}},
\end{align*}

\begin{align*}
	\hat{\bar{\bar{{\bf RT}}}}_{z\leftarrow \rho_z'\leftarrow \rho_z'' \leftarrow \rho_z}^{(\omega^\pm)[\red{(\downarrow/\uparrow)}\blue{(\updownarrow/\updownarrow)}]}&=\frac{-i}{8}\frac{1}{k_{z,2}^2k_{z,2}^\pm}\left[t^{\rm (s)}r^{\rm (s)(\omega^\pm)}\hat{\bf g}^{\rm (s)}+t^{\rm (p)}r^{\rm (p)(\omega^\pm)}\left(\hat{\bf p}_{\rho_z'}^\uparrow\cdot\hat{\bf p}_{\rho_z'}^{(\omega^\pm)\red{\downarrow}}\right)\left(\hat{\bf p}_{\rho_z''}^{(\omega^\pm)\red{\uparrow}}\cdot\hat{\bf p}_{\rho_z''}^{\blue{\updownarrow}}\right)\left(\hat{\bf p}_z^\uparrow\otimes\hat{\bf p}_{\rho_z}^{\blue{\updownarrow}}\right) \right]\bar{\bar{\Gamma}}_{\rm R-T\,(0,2)}^{(\omega^\pm)(\blue{\updownarrow})},\\
	\hat{\bar{\bar{{\bf RR}}}}_{z\leftarrow \rho_z'\leftarrow \rho_z'' \leftarrow \rho_z}^{(\omega^\pm)[\red{(\downarrow/\uparrow)}\blue{(\downarrow/\uparrow)}]}&=\frac{-i}{8}\frac{1}{k_{z,2}^2k_{z,2}^\pm}\left[t^{\rm (s)}r^{\rm (s)(\omega^\pm)}r^{\rm (s)}\hat{\bf g}^{\rm (s)}+t^{\rm (p)}r^{\rm (p)(\omega^\pm)}r^{\rm (p)}\left(\hat{\bf p}_{\rho_z'}^\uparrow\cdot\hat{\bf p}_{\rho_z'}^{(\omega^\pm)\red{\downarrow}}\right)\left(\hat{\bf p}_{\rho_z''}^{(\omega^\pm)\red{\uparrow}}\cdot\hat{\bf p}_{\rho_z''}^{\blue{\downarrow}}\right)\left(\hat{\bf p}_z^\uparrow\otimes\hat{\bf p}_{\rho_z}^{\blue{\uparrow}}\vphantom{\left(\hat{\bf p}_{\rho_z''}^{(\omega^\pm)\red{\uparrow}}\cdot\hat{\bf p}_{\rho_z''}^{\blue{\downarrow}}\right)}\right)\right]\bar{\bar{\Gamma}}_{\rm R-R\,(0,2)}^{(\omega^\pm)},\\
	\hat{\bar{\bar{{\bf RZ}}}}_{z\leftarrow \rho_z'\leftarrow \rho_z'' \leftarrow \rho_z}^{(\omega^\pm)[\red{(\downarrow/\uparrow)}]}&=\frac{1}{4}\frac{1}{k_{z,2}k_{z,2}^\pm k_0^2\varepsilon(\omega)}\left[t^{\rm (p)}r^{\rm (p)(\omega^\pm)}\left(\hat{\bf p}_{\rho_z'}^\uparrow\cdot\hat{\bf p}_{\rho_z'}^{(\omega^\pm)\red{\downarrow}}\right)\left(\hat{\bf p}_{\rho_z''}^{(\omega^\pm)\red{\uparrow}}\cdot\hat{\bf z}\right)\left(\hat{\bf p}_z^\uparrow\otimes\hat{\bf z}\vphantom{\left(\hat{\bf p}_{\rho_z''}^{(\omega^\pm)\red{\uparrow}}\cdot\hat{\bf z}\right)}\right)\right]\bar{\bar{\Gamma}}_{\rm R-Z\,(0,2)}^{(\omega^\pm)},
\end{align*}

\begin{align*}
	\bar{\bar{\Gamma}}_{\rm R-T\,(0,2)}^{(\omega^\pm)(\blue{\updownarrow})}&=e^{i\left[k_{z,1}z-k_{z,2}\rho_z'\right]}\red{e^{-ik_{z,2}^\pm\left[\rho_z'+\rho_z''\right]}}\blue{e^{ik_{z,2}\left|\rho_z''-\rho_z\right|}},\\
	\bar{\bar{\Gamma}}_{\rm R-R\,(0,2)}^{(\omega^\pm)}&=e^{i\left[k_{z,1}z-k_{z,2}\rho_z'\right]}\red{e^{-ik_{z,2}^\pm\left[\rho_z'+\rho_z''\right]}}\blue{e^{-ik_{z,2}\left[\rho_z''+\rho_z\right]}},\\
	\bar{\bar{\Gamma}}_{\rm R-Z\,(0,2)}^{(\omega^\pm)}&=e^{i\left[k_{z,1}z-k_{z,2}\rho_z'\right]}\red{e^{-ik_{z,2}^\pm\left[\rho_z'+\rho_z''\right]}}\blue{\delta{[\rho_z''-\rho_z]}},
\end{align*}

\begin{align*}
	\hat{\bar{\bar{{\bf ZT}}}}_{z\leftarrow \rho_z'\leftarrow \rho_z'' \leftarrow \rho_z}^{(\omega^\pm)[\blue{(\updownarrow/\updownarrow)}]}&=\frac{1}{4}\frac{1}{k_{z,2}^2k_0^2(1\pm\tilde{\Omega})^2\varepsilon(\omega^\pm)}\left[t^{\rm (p)}\left(\hat{\bf p}_{\rho_z'}^\uparrow\cdot\hat{\bf z}\right)\left(\hat{\bf z}\cdot\hat{\bf p}_{\rho_z''}^{\blue{\updownarrow}}\right)\left(\hat{\bf p}_z^\uparrow\otimes\hat{\bf p}_{\rho_z}^{\blue{\updownarrow}}\right) \right]\bar{\bar{\Gamma}}_{\rm Z-T\,(0,2)}^{(\omega^\pm)(\blue{\updownarrow})},\\
	\hat{\bar{\bar{{\bf ZR}}}}_{z\leftarrow \rho_z'\leftarrow \rho_z'' \leftarrow \rho_z}^{(\omega^\pm)[\blue{(\downarrow/\uparrow)}]}&=\frac{1}{4}\frac{1}{k_{z,2}^2k_0^2(1\pm\tilde{\Omega})^2\varepsilon(\omega^\pm)}\left[t^{\rm (p)}r^{\rm (p)}\left(\hat{\bf p}_{\rho_z'}^\uparrow\cdot\hat{\bf z}\right)\left(\hat{\bf z}\cdot\hat{\bf p}_{\rho_z''}^{\blue{\downarrow}}\right)\left(\hat{\bf p}_z^\uparrow\otimes\hat{\bf p}_{\rho_z}^{\blue{\uparrow}}\vphantom{\left(\hat{\bf z}\cdot\hat{\bf p}_{\rho_z''}^{\blue{\downarrow}}\right)}\right)\right]\bar{\bar{\Gamma}}_{\rm Z-R\,(0,2)}^{(\omega^\pm)},\\
	\hat{\bar{\bar{{\bf ZZ}}}}_{z\leftarrow \rho_z'\leftarrow \rho_z'' \leftarrow \rho_z}^{(\omega^\pm)}&=\frac{i}{2}\frac{1}{k_{z,2}k_0^4(1\pm\tilde{\Omega})^2\varepsilon(\omega)\varepsilon(\omega^\pm)}\left[t^{\rm (p)}\left(\hat{\bf p}_{\rho_z'}^\uparrow\cdot\hat{\bf z}\right)\left(\hat{\bf z}\cdot\hat{\bf z}\vphantom{\left(\hat{\bf p}_{\rho_z'}^\uparrow\cdot\hat{\bf z}\right)}\right)\left(\hat{\bf p}_z^\uparrow\otimes\hat{\bf z}\vphantom{\left(\hat{\bf p}_{\rho_z'}^\uparrow\cdot\hat{\bf z}\right)}\right)\right]\bar{\bar{\Gamma}}_{\rm Z-Z\,(0,2)}^{(\omega^\pm)},
\end{align*}

\begin{align*}
	\bar{\bar{\Gamma}}_{\rm Z-T\,(0,2)}^{(\omega^\pm)(\blue{\updownarrow})}&=e^{i\left[k_{z,1}z-k_{z,2}\rho_z'\right]}\red{\delta{[\rho_z'-\rho_z'']}}\blue{e^{ik_{z,2}\left|\rho_z''-\rho_z\right|}},\\
	\bar{\bar{\Gamma}}_{\rm Z-R\,(0,2)}^{(\omega^\pm)}&=e^{i\left[k_{z,1}z-k_{z,2}\rho_z'\right]}\red{\delta{[\rho_z'-\rho_z'']}}\blue{e^{-ik_{z,2}\left[\rho_z''+\rho_z\right]}},\\
	\bar{\bar{\Gamma}}_{\rm Z-Z\,(0,2)}^{(\omega^\pm)}&=e^{i\left[k_{z,1}z-k_{z,2}\rho_z'\right]}\red{\delta{[\rho_z'-\rho_z'']}}\blue{\delta{[\rho_z''-\rho_z]}}.
\end{align*}
According to the above expressions, we rewrite Eq.~\eqref{Eq.VI.56} in the following way:
\begin{tcolorbox}[sharp corners,colback=red!5!white,colframe=red!50!white]
	\vspace{-0.25cm}
	\begin{equation}
		\bar{\bar{\mathcal{G}}}_{0,2}^{(\omega^\pm)}=\bar{\bar{\mathcal{G}}}_{\rm T-T\,(0,2)}^{(\omega^\pm)}+\bar{\bar{\mathcal{G}}}_{\rm R-R\,(0,2)}^{(\omega^\pm)}+\bar{\bar{\mathcal{G}}}_{\rm T-R\,(0,2)}^{(\omega^\pm)}+\bar{\bar{\mathcal{G}}}_{\rm R-T\,(0,2)}^{(\omega^\pm)}+\bar{\bar{\mathcal{G}}}_{\rm Z-Z\,(0,2)}^{(\omega^\pm)}+\bar{\bar{\mathcal{G}}}_{\rm T-Z\,(0,2)}^{(\omega^\pm)}+\bar{\bar{\mathcal{G}}}_{\rm Z-T\,(0,2)}^{(\omega^\pm)}+\bar{\bar{\mathcal{G}}}_{\rm R-Z\,(0,2)}^{(\omega^\pm)}+\bar{\bar{\mathcal{G}}}_{\rm Z-R\,(0,2)}^{(\omega^\pm)}.
		\label{Eq.VI.58}
	\end{equation}
\end{tcolorbox}
\noindent Now, we will proceed with the calculation of Eq.~\eqref{Eq.VI.58} by analyzing separately each of the contributions (cf.~\hyperref[SectSI.Appendix.VI.C]{\!Appendix VI.C} for the demonstration of the $p$-like contribution to the (T-T)-term):
\begin{tcolorbox}[sharp corners,colback=red!5!white,colframe=red!50!white]
	\vspace{-0.35cm}
	\begin{equation}
		\bar{\bar{\mathcal{G}}}_{\rm T-T\,(0,2)}^{(\omega^\pm)}=\frac{1}{32\pi k_0^5}\int\limits_{0}^{+\infty}{d\kappa_R\frac{-\kappa_R}{|\tilde{k}_{z,2}|^2\tilde{k}_{z,2}\tilde{k}_{z,2}^\pm} \sum_{i,j=\left\{\uparrow,\downarrow\right\}}{\bar{\bar{\mathcal{G}}}_{\rm T-T\,(0,2)}^{(\omega^\pm)(i/j)}\bar{\bar{\Gamma}}_{\rm T-T\,(0,2)}^{(\omega^\pm)(i/j)}} },
		\label{Eq.VI.59}
	\end{equation}
\end{tcolorbox}
\noindent where
\begin{align*}
	\bar{\bar{\mathcal{G}}}_{\rm T-T\,(0,2)}^{(\omega^\pm)(\uparrow/\uparrow)}&=\left|t^{\rm (s)}\right|^2+\left|t^{\rm (p)}\right|^2\frac{{\rm T}}{(1\pm\tilde{\Omega})^2\varepsilon(\omega)\varepsilon(\omega^\pm)}\left[\tilde{k}_{z,2}\tilde{k}_{z,2}^{\pm}+\kappa_R^2\right]\left[\tilde{k}_{z,2}\tilde{k}_{z,2}^{\pm}+\kappa_R^2\right],\\
	\bar{\bar{\Gamma}}_{\rm T-T\,(0,2)}^{(\omega^\pm)(\uparrow/\uparrow)}&=\frac{e^{-2k_0\text{Im}{[\tilde{k}_{z,1}]}z}}{4i[\text{Im}{[\tilde{k}_{z,2}]}]^2[\tilde{k}_{z,2}^*-\tilde{k}_{z,2}^{\pm}]},
\end{align*}
\begin{align*}
	\bar{\bar{\mathcal{G}}}_{\rm T-T\,(0,2)}^{(\omega^\pm)(\uparrow/\downarrow)}&=\left|t^{\rm (s)}\right|^2+\left|t^{\rm (p)}\right|^2\frac{{\rm R}}{(1\pm\tilde{\Omega})^2\varepsilon(\omega)\varepsilon(\omega^\pm)}\left[\tilde{k}_{z,2}\tilde{k}_{z,2}^{\pm}+\kappa_R^2\right]\left[\tilde{k}_{z,2}\tilde{k}_{z,2}^{\pm}-\kappa_R^2\right],\\
	\bar{\bar{\Gamma}}_{\rm T-T\,(0,2)}^{(\omega^\pm)(\uparrow/\downarrow)}&=\frac{e^{-2k_0\text{Im}{[\tilde{k}_{z,1}]}z}}{4[\tilde{k}_{z,2}-\tilde{k}_{z,2}^\pm]}\left[\frac{4i}{[\tilde{k}_{z,2}+\tilde{k}_{z,2}^\pm][\tilde{k}_{z,2}^*-\tilde{k}_{z,2}^\pm]}+\frac{1}{\tilde{k}_{z,2}\text{Im}{[\tilde{k}_{z,2}]}}\right],\\\\
	\bar{\bar{\mathcal{G}}}_{\rm T-T\,(0,2)}^{(\omega^\pm)(\downarrow/\uparrow)}&=\left|t^{\rm (s)}\right|^2+\left|t^{\rm (p)}\right|^2\frac{{\rm T}}{(1\pm\tilde{\Omega})^2\varepsilon(\omega)\varepsilon(\omega^\pm)}\left[\tilde{k}_{z,2}\tilde{k}_{z,2}^{\pm}-\kappa_R^2\right]\left[\tilde{k}_{z,2}\tilde{k}_{z,2}^{\pm}-\kappa_R^2\right],\\
	\bar{\bar{\Gamma}}_{\rm T-T\,(0,2)}^{(\omega^\pm)(\downarrow/\uparrow)}&=\frac{e^{-2k_0\text{Im}{[\tilde{k}_{z,1}]}z}}{4i\text{Im}{[\tilde{k}_{z,2}]}\text{Im}{[\tilde{k}_{z,2}^*]}[\tilde{k}_{z,2}+\tilde{k}_{z,2}^{\pm}]},\\\\
	\bar{\bar{\mathcal{G}}}_{\rm T-T\,(0,2)}^{(\omega^\pm)(\downarrow/\downarrow)}&=\left|t^{\rm (s)}\right|^2+\left|t^{\rm (p)}\right|^2\frac{{\rm R}}{(1\pm\tilde{\Omega})^2\varepsilon(\omega)\varepsilon(\omega^\pm)}\left[\tilde{k}_{z,2}\tilde{k}_{z,2}^{\pm}-\kappa_R^2\right]\left[\tilde{k}_{z,2}\tilde{k}_{z,2}^{\pm}+\kappa_R^2\right],\\
	\bar{\bar{\Gamma}}_{\rm T-T\,(0,2)}^{(\omega^\pm)(\downarrow/\downarrow)}&=\frac{e^{-2k_0\text{Im}{[\tilde{k}_{z,1}]}z}}{4\text{Im}{[\tilde{k}_{z,2}^*]}[\tilde{k}_{z,2}+\tilde{k}_{z,2}^{\pm}]k_{z,2}},
\end{align*}
with
\begin{tcolorbox}[sharp corners,colback=blue!5!white,colframe=blue!50!white]
	\vspace{-0.35cm}
	\begin{subequations}
		\begin{align}
			{\rm T}&=\frac{1}{\left|\varepsilon(\omega)\right|}\left\{|\tilde{k}_{z,1}|^2|\tilde{k}_{z,2}|^2+\kappa_R^2\left[|\tilde{k}_{z,2}|^2+|\tilde{k}_{z,1}|^2+\kappa_R^2\right]\right\},
			\label{Eq.VI.60a}\\
			{\rm R}&=\frac{1}{\left|\varepsilon(\omega)\right|}\left\{|\tilde{k}_{z,1}|^2|\tilde{k}_{z,2}|^2+\kappa_R^2\left[|\tilde{k}_{z,2}|^2-|\tilde{k}_{z,1}|^2-\kappa_R^2\right]\right\},
			\label{Eq.VI.60b}
		\end{align}
	\end{subequations}
\end{tcolorbox}
\begin{tcolorbox}[sharp corners,colback=red!5!white,colframe=red!50!white]
	\vspace{-0.35cm}
	\begin{equation}
		\bar{\bar{\mathcal{G}}}_{\rm R-R\,(0,2)}^{(\omega^\pm)}=\frac{1}{32\pi k_0^5}\int\limits_{0}^{+\infty}{d\kappa_R\frac{-\kappa_R}{|\tilde{k}_{z,2}|^2\tilde{k}_{z,2}\tilde{k}_{z,2}^\pm} \bar{\bar{\mathcal{G}}}_{\rm R-R\,(0,2)}^{(\omega^\pm)(\downarrow/\uparrow)}\bar{\bar{\Gamma}}_{\rm R-R\,(0,2)}^{(\omega^\pm)} },
		\label{Eq.VI.61}
	\end{equation}
\end{tcolorbox}
\noindent where
\begin{align*}
	\bar{\bar{\mathcal{G}}}_{\rm R-R\,(0,2)}^{(\omega^\pm)(\downarrow/\uparrow)}&=\left|t^{\rm (s)}\right|^2r^{\rm (s)(\omega^\pm)}r^{\rm (s)}+\left|t^{\rm (p)}\right|^2r^{\rm (p)(\omega^\pm)}r^{\rm (p)}\frac{{\rm T}}{(1\pm\tilde{\Omega})^2\varepsilon(\omega)\varepsilon(\omega^\pm)}\left[\tilde{k}_{z,2}\tilde{k}_{z,2}^{\pm}-\kappa_R^2\right]\left[\tilde{k}_{z,2}\tilde{k}_{z,2}^{\pm}-\kappa_R^2\right],\\
	\bar{\bar{\Gamma}}_{\rm R-R\,(0,2)}^{(\omega^\pm)}&=\frac{e^{-2k_0\text{Im}{[\tilde{k}_{z,1}]}z}}{2\text{Im}{[\tilde{k}_{z,2}^*]}[\tilde{k}_{z,2}+\tilde{k}_{z,2}^{\pm}]^2},
\end{align*}
\begin{tcolorbox}[sharp corners,colback=red!5!white,colframe=red!50!white]
	\vspace{-0.35cm}
	\begin{equation}
		\bar{\bar{\mathcal{G}}}_{\rm T-R\,(0,2)}^{(\omega^\pm)}=\frac{1}{32\pi k_0^5} \int\limits_{0}^{+\infty}{d\kappa_R\frac{-\kappa_R}{|\tilde{k}_{z,2}|^2\tilde{k}_{z,2}\tilde{k}_{z,2}^\pm} \sum_{i=\left\{\uparrow,\downarrow\right\}}{\bar{\bar{\mathcal{G}}}_{\rm T-R\,(0,2)}^{(\omega^\pm)(i/\uparrow)}\bar{\bar{\Gamma}}_{\rm T-R\,(0,2)}^{(\omega^\pm)(i)}} },
		\label{Eq.VI.62}
	\end{equation}
\end{tcolorbox}
\noindent where
\begin{align*}
	\bar{\bar{\mathcal{G}}}_{\rm T-R\,(0,2)}^{(\omega^\pm)(\uparrow/\uparrow)}&=\left|t^{\rm (s)}\right|^2r^{\rm (s)}+\left|t^{\rm (p)}\right|^2r^{\rm (p)}\frac{{\rm T}}{(1\pm\tilde{\Omega})^2\varepsilon(\omega)\varepsilon(\omega^\pm)}\left[\tilde{k}_{z,2}\tilde{k}_{z,2}^{\pm}+\kappa_R^2\right]\left[\tilde{k}_{z,2}\tilde{k}_{z,2}^{\pm}-\kappa_R^2\right],\\
	\bar{\bar{\Gamma}}_{\rm T-R\,(0,2)}^{(\omega^\pm)(\uparrow)}&=\frac{e^{-2k_0\text{Im}{[\tilde{k}_{z,1}]}z}}{4\text{Im}{[\tilde{k}_{z,2}^*]}[\tilde{k}_{z,2}+\tilde{k}_{z,2}^{\pm}]\tilde{k}_{z,2}},
\end{align*}
\begin{align*}
	\bar{\bar{\mathcal{G}}}_{\rm T-R\,(0,2)}^{(\omega^\pm)(\downarrow/\uparrow)}&=\left|t^{\rm (s)}\right|^2r^{\rm (s)}+\left|t^{\rm (p)}\right|^2r^{\rm (p)}\frac{{\rm T}}{(1\pm\tilde{\Omega})^2\varepsilon(\omega)\varepsilon(\omega^\pm)}\left[\tilde{k}_{z,2}\tilde{k}_{z,2}^{\pm}-\kappa_R^2\right]\left[\tilde{k}_{z,2}\tilde{k}_{z,2}^{\pm}+\kappa_R^2\right],\\
	\bar{\bar{\Gamma}}_{\rm T-R\,(0,2)}^{(\omega^\pm)(\downarrow)}&=\frac{e^{-2k_0\text{Im}{[\tilde{k}_{z,1}]}z}}{4\text{Im}{[\tilde{k}_{z,2}^*]}[\tilde{k}_{z,2}+\tilde{k}_{z,2}^{\pm}]\tilde{k}_{z,2}},
\end{align*}
\begin{tcolorbox}[sharp corners,colback=red!5!white,colframe=red!50!white]
	\vspace{-0.35cm}
	\begin{equation}
		\bar{\bar{\mathcal{G}}}_{\rm R-T\,(0,2)}^{(\omega^\pm)}=\frac{1}{32\pi k_0^5} \int\limits_{0}^{+\infty}{d\kappa_R\frac{-\kappa_R}{|\tilde{k}_{z,2}|^2\tilde{k}_{z,2}\tilde{k}_{z,2}^\pm} \sum_{i=\left\{\uparrow,\downarrow\right\}}{\bar{\bar{\mathcal{G}}}_{\rm R-T\,(0,2)}^{(\omega^\pm)(\downarrow/i)}\bar{\bar{\Gamma}}_{\rm R-T\,(0,2)}^{(\omega^\pm)(i)}} },
		\label{Eq.VI.63}
	\end{equation}
\end{tcolorbox}
\noindent where
\begin{align*}
	\bar{\bar{\mathcal{G}}}_{\rm R-T\,(0,2)}^{(\omega^\pm)(\downarrow/\uparrow)}&=\left|t^{\rm (s)}\right|^2r^{\rm (s)(\omega^\pm)}+\left|t^{\rm (p)}\right|^2r^{\rm (p)(\omega^\pm)}\frac{{\rm T}}{(1\pm\tilde{\Omega})^2\varepsilon(\omega)\varepsilon(\omega^\pm)}\left[\tilde{k}_{z,2}\tilde{k}_{z,2}^{\pm}-\kappa_R^2\right]\left[\tilde{k}_{z,2}\tilde{k}_{z,2}^{\pm}+\kappa_R^2\right],\\
	\bar{\bar{\Gamma}}_{\rm R-T\,(0,2)}^{(\omega^\pm)(\uparrow)}&=\frac{e^{-2k_0\text{Im}{[\tilde{k}_{z,1}]}z}}{2\text{Im}{[\tilde{k}_{z,2}]}[\tilde{k}_{z,2}+\tilde{k}_{z,2}^{\pm}][\tilde{k}_{z,2}^*-\tilde{k}_{z,2}^{\pm}]},\\\\
	\bar{\bar{\mathcal{G}}}_{\rm R-T\,(0,2)}^{(\omega^\pm)(\downarrow/\downarrow)}&=\left|t^{\rm (s)}\right|^2r^{\rm (s)(\omega^\pm)}+\left|t^{\rm (p)}\right|^2r^{\rm (p)(\omega^\pm)}\frac{{\rm R}}{(1\pm\tilde{\Omega})^2\varepsilon(\omega)\varepsilon(\omega^\pm)}\left[\tilde{k}_{z,2}\tilde{k}_{z,2}^{\pm}-\kappa_R^2\right]\left[\tilde{k}_{z,2}\tilde{k}_{z,2}^{\pm}-\kappa_R^2\right],\\
	\bar{\bar{\Gamma}}_{\rm R-T\,(0,2)}^{(\omega^\pm)(\downarrow)}&=\frac{e^{-2k_0\text{Im}{[\tilde{k}_{z,1}]}z}}{i[\tilde{k}_{z,2}+\tilde{k}_{z,2}^{\pm}]^2[\tilde{k}_{z,2}^{\pm}-\tilde{k}_{z,2}^*]},
\end{align*}
\begin{tcolorbox}[sharp corners,colback=red!5!white,colframe=red!50!white]
	\vspace{-0.35cm}
	\begin{equation}
		\bar{\bar{\mathcal{G}}}_{\rm Z-Z\,(0,2)}^{(\omega^\pm)}=\frac{1}{32\pi k_0^5} \int\limits_{0}^{+\infty}{d\kappa_R \frac{2\kappa_R^3(|\tilde{k}_{z,1}|^2+\kappa_R^2)}{|\tilde{k}_{z,2}|^2(1\pm\tilde{\Omega})^2\left|\varepsilon(\omega)\right|\varepsilon(\omega)\varepsilon(\omega^\pm)}\left|t^{\rm (p)}\right|^2 \frac{e^{-2k_0\text{Im}{\left[\tilde{k}_{z,1}\right]z}}}{\text{Im}{[\tilde{k}_{z,2}]}} },
		\label{Eq.VI.64}
	\end{equation}
\end{tcolorbox}
\begin{tcolorbox}[sharp corners,colback=red!5!white,colframe=red!50!white]
	\vspace{-0.35cm}
	\begin{equation}
		\bar{\bar{\mathcal{G}}}_{\rm T-Z\,(0,2)}^{(\omega^\pm)}=\frac{1}{32\pi k_0^5} \int\limits_{0}^{+\infty}{d\kappa_R \frac{2\kappa_R^3(|\tilde{k}_{z,1}|^2+\kappa_R^2)}{|\tilde{k}_{z,2}|^2(1\pm\tilde{\Omega})^2 \left|\varepsilon(\omega)\right|\varepsilon(\omega)\varepsilon(\omega^\pm)}\left|t^{\rm (p)}\right|^2 \sum_{i=\left\{\uparrow,\downarrow\right\}}{\bar{\bar{\mathcal{G}}}_{\rm T-Z\,(0,2)}^{(\omega^\pm)(i)}\bar{\bar{\Gamma}}_{\rm T-Z\,(0,2)}^{(\omega^\pm)(i)}} },
		\label{Eq.VI.65}
	\end{equation}
\end{tcolorbox}
\noindent where
\begin{align*}
	\bar{\bar{\mathcal{G}}}_{\rm T-Z\,(0,2)}^{(\omega^\pm)(\uparrow)}&=\left[\tilde{k}_{z,2}\tilde{k}_{z,2}^{\pm}+\kappa_R^2\right],\\
	\bar{\bar{\Gamma}}_{\rm T-Z\,(0,2)}^{(\omega^\pm)(\uparrow)}&=\frac{e^{-2k_0\text{Im}{[\tilde{k}_{z,1}]}z}}{2\text{Im}{[\tilde{k}_{z,2}^*]}[\tilde{k}_{z,2}^*-\tilde{k}_{z,2}^{\pm}]\tilde{k}_{z,2}^{\pm}},\\\\
	\bar{\bar{\mathcal{G}}}_{\rm T-Z\,(0,2)}^{(\omega^\pm)(\downarrow)}&=\left[\tilde{k}_{z,2}\tilde{k}_{z,2}^{\pm}-\kappa_R^2\right],\\
	\bar{\bar{\Gamma}}_{\rm T-Z\,(0,2)}^{(\omega^\pm)(\downarrow)}&=\frac{e^{-2k_0\text{Im}{[\tilde{k}_{z,1}]}z}}{2\text{Im}{[\tilde{k}_{z,2}^*]}[\tilde{k}_{z,2}+\tilde{k}_{z,2}^{\pm}]\tilde{k}_{z,2}^{\pm}},
\end{align*}
\begin{tcolorbox}[sharp corners,colback=red!5!white,colframe=red!50!white]
	\vspace{-0.35cm}
	\begin{equation}
		\bar{\bar{\mathcal{G}}}_{\rm Z-T\,(0,2)}^{(\omega^\pm)}=\frac{1}{32\pi k_0^5} \int\limits_{0}^{+\infty}{d\kappa_R \frac{2\kappa_R^3}{|\tilde{k}_{z,2}|^2(1\pm\tilde{\Omega})^2\sqrt{\varepsilon(\omega)}\sqrt{\varepsilon(\omega^\pm)}\varepsilon(\omega^\pm)}\left|t^{\rm (p)}\right|^2\sum_{i=\left\{\uparrow,\downarrow\right\}}{\bar{\bar{\mathcal{G}}}_{\rm Z-T\,(0,2)}^{(\omega^\pm)(i)}\bar{\bar{\Gamma}}_{\rm Z-T\,(0,2)}^{(\omega^\pm)(i)}} },
		\label{Eq.VI.66}
	\end{equation}
\end{tcolorbox}
\noindent where
\begin{align*}
	\bar{\bar{\mathcal{G}}}_{\rm Z-T\,(0,2)}^{(\omega^\pm)(\uparrow)}&=\frac{{\rm T}}{(1\pm\tilde{\Omega})},\\
	\bar{\bar{\Gamma}}_{\rm Z-T\,(0,2)}^{(\omega^\pm)(\uparrow)}&=\frac{e^{-2k_0\text{Im}{[\tilde{k}_{z,1}]}z}}{4i[\text{Im}{[\tilde{k}_{z,2}]}]^2\tilde{k}_{z,2}},
\end{align*}
\begin{align*}
	\bar{\bar{\mathcal{G}}}_{\rm Z-T\,(0,2)}^{(\omega^\pm)(\downarrow)}&=\frac{{\rm R}}{(1\pm\tilde{\Omega})},\\
	\bar{\bar{\Gamma}}_{\rm Z-T\,(0,2)}^{(\omega^\pm)(\downarrow)}&=\frac{e^{-2k_0\text{Im}{[\tilde{k}_{z,1}]}z}}{4\text{Im}{[\tilde{k}_{z,2}^*]}\tilde{k}_{z,2}^2},
\end{align*}
\begin{tcolorbox}[sharp corners,colback=red!5!white,colframe=red!50!white]
	\vspace{-0.35cm}
	\begin{equation}
		\!\!\!\!\!\!\!\!\!\!\bar{\bar{\mathcal{G}}}_{\rm R-Z\,(0,2)}^{(\omega^\pm)}=\frac{1}{32\pi k_0^5} \int\limits_{0}^{+\infty}{d\kappa_R\frac{2\kappa_R^3(|\tilde{k}_{z,1}|^2+\kappa_R^2)}{|\tilde{k}_{z,2}|^2(1\pm\tilde{\Omega})^2\left|\varepsilon(\omega)\right|\varepsilon(\omega)\varepsilon(\omega^\pm)}\left|t^{\rm (p)}\right|^2\frac{r^{\rm (p)(\omega^\pm)}(\tilde{k}_{z,2}\tilde{k}_{z,2}^{\pm}-\kappa_R^2)e^{-2k_0\text{Im}{[\tilde{k}_{z,1}]}z}}{i[\tilde{k}_{z,2}+\tilde{k}_{z,2}^{\pm}][\tilde{k}_{z,2}^*-\tilde{k}_{z,2}^{\pm}]\tilde{k}_{z,2}^{\pm}} },
		\label{Eq.VI.67}
	\end{equation}
\end{tcolorbox}
\begin{tcolorbox}[sharp corners,colback=red!5!white,colframe=red!50!white]
	\vspace{-0.35cm}
	\begin{equation}
		\bar{\bar{\mathcal{G}}}_{\rm Z-R\,(0,2)}^{(\omega^\pm)}=\frac{1}{32\pi k_0^5} \int\limits_{0}^{+\infty}{d\kappa_R\frac{2\kappa_R^3}{|\tilde{k}_{z,2}|^2(1\pm\tilde{\Omega})^2\varepsilon(\omega)\varepsilon(\omega^\pm)}\left|t^{\rm (p)}\right|^2\frac{r^{\rm (p)}{\rm T}e^{-2k_0\text{Im}{[\tilde{k}_{z,1}]}z}}{4\text{Im}{[\tilde{k}_{z,2}^*]}\tilde{k}_{z,2}^2} }.
		\label{Eq.VI.68}
	\end{equation}
\end{tcolorbox}

\section{APPENDIX}
\label{SectSI.Appendix}
\renewcommand\theequation{S.App.\arabic{equation}}
\setcounter{equation}{0}

\subsection{Appendix VI.A: Calculation of Frobenius inner product associated to $\bar{\mathcal{G}}_{1,1}^{(\omega^\pm)}$}
\label{SectSI.Appendix.VI.A}
\begin{align*}
	\norm{\left[\hat{\bar{{\bf G}}}_{z\leftarrow \rho_z\leftarrow \rho_z''}^{(\omega^\pm)[(\uparrow/\uparrow)\blue{(\updownarrow/\updownarrow)}]}\right]^*\hat{\bar{{\bf G}}}_{z\leftarrow \rho_z'\leftarrow \rho_z''}^{(\omega^\pm)[(\uparrow/\uparrow)\red{(\updownarrow/\updownarrow)}]}}^{\rm (p)}_{\mathcal{F}}
	&\propto \norm{\hat{\bf g}^{\rm(p)(\uparrow/\uparrow)*}_{z\leftarrow \rho_z}\blue{\hat{\bf g}^{\rm(p)(\omega^\pm)(\updownarrow/\updownarrow)*}_{\rho_z\leftarrow \rho_z''}}\hat{\bf g}^{\rm(p)(\uparrow/\uparrow)}_{z\leftarrow \rho_z'}\red{\hat{\bf g}^{\rm(p)(\omega^\pm)(\updownarrow/\updownarrow)}_{\rho_z'\leftarrow \rho_z''}}}_{\mathcal{F}}\\
	&=\norm{\left(\hat{\bf p}_z^{\uparrow}\otimes\hat{\bf p}_{\rho_z}^\uparrow\right)^*\blue{\left(\hat{\bf p}_{\rho_z}^{(\omega^\pm)\updownarrow}\otimes\hat{\bf p}_{\rho_z''}^{(\omega^\pm)\updownarrow}\right)^*}\left(\hat{\bf p}_z^{\uparrow}\otimes\hat{\bf p}_{\rho_z'}^\uparrow\right)\red{\left(\hat{\bf p}_{\rho_z'}^{(\omega^\pm)\updownarrow}\otimes\hat{\bf p}_{\rho_z''}^{(\omega^\pm)\updownarrow}\right)}}_{\mathcal{F}}\\
	&=\left(\hat{\bf p}_{\rho_z}^\uparrow\cdot \blue{\hat{\bf p}_{\rho_z}^{(\omega^\pm)\updownarrow}}\right)^*\left(\hat{\bf p}_{\rho_z'}^\uparrow\cdot \red{\hat{\bf p}_{\rho_z'}^{(\omega^\pm)\updownarrow}}\right)\left(\left[\hat{\bf p}_z^{\uparrow}\right]^*\cdot\hat{\bf p}_{z}^{\uparrow}\right)\left(\blue{\left[\hat{\bf p}_{\rho_z''}^{(\omega^\pm)\updownarrow}\right]^*}\cdot\red{\hat{\bf p}_{\rho_z''}^{(\omega^\pm)\updownarrow}}\right).
\end{align*}

\subsection{Appendix VI.B: Calculation of Frobenius inner product associated to $\tilde{\mathcal{G}}_{1,1}^{(\omega^\pm)}$}
\label{SectSI.Appendix.VI.B}
\begin{align*}
	\norm{\left|\hat{\tilde{{\bf G}}}_{z\leftarrow \rho_z\leftarrow \rho_z''}^{(\omega^\pm)[(\uparrow/\uparrow)\blue{(\updownarrow/\updownarrow)}]}\right|\left|\hat{\tilde{{\bf G}}}_{z\leftarrow \rho_z'\leftarrow \rho_z''}^{(\omega^\pm)[(\uparrow/\uparrow)\red{(\updownarrow/\updownarrow)}]}\right|}^{\rm (p)}_{\mathcal{F}}
	&\propto \norm{\hat{\bf g}^{\rm(p)(\uparrow/\uparrow)*}_{z\leftarrow \rho_z}\blue{\hat{\bf g}^{\rm(p)(\omega^\pm)(\updownarrow/\updownarrow)}_{\rho_z\leftarrow \rho_z''}}\hat{\bf g}^{\rm(p)(\uparrow/\uparrow)}_{z\leftarrow \rho_z'}\red{\hat{\bf g}^{\rm(p)(\omega^\pm)(\updownarrow/\updownarrow)*}_{\rho_z'\leftarrow \rho_z''}}}_{\mathcal{F}}\\
	&=\norm{\left(\hat{\bf p}_z^{\uparrow}\otimes\hat{\bf p}_{\rho_z}^\uparrow\right)^*\blue{\left(\hat{\bf p}_{\rho_z}^{(\omega^\pm)\updownarrow}\otimes\hat{\bf p}_{\rho_z''}^{(\omega^\pm)\updownarrow}\right)}\left(\hat{\bf p}_z^{\uparrow}\otimes\hat{\bf p}_{\rho_z'}^\uparrow\right)\red{\left(\hat{\bf p}_{\rho_z'}^{(\omega^\pm)\updownarrow}\otimes\hat{\bf p}_{\rho_z''}^{(\omega^\pm)\updownarrow}\right)^*}}_{\mathcal{F}}\\
	&=\left(\left[\hat{\bf p}_{\rho_z}^\uparrow\right]^*\cdot \blue{\hat{\bf p}_{\rho_z}^{(\omega^\pm)\updownarrow}}\right)\left(\hat{\bf p}_{\rho_z'}^\uparrow\cdot \red{\left[\hat{\bf p}_{\rho_z'}^{(\omega^\pm)\updownarrow}\right]^*}\right)\left(\left[\hat{\bf p}_{z}^{\uparrow}\right]^*\cdot\hat{\bf p}_z^{\uparrow}\right)\left(\blue{\hat{\bf p}_{\rho_z''}^{(\omega^\pm)\updownarrow}}\cdot\red{\left[\hat{\bf p}_{\rho_z''}^{(\omega^\pm)\updownarrow}\right]^*}\right).
\end{align*}

\subsection{Appendix VI.C: Calculation of Frobenius inner product associated to $\bar{\bar{\mathcal{G}}}_{0,2}^{(\omega^\pm)}$}
\label{SectSI.Appendix.VI.C}
\begin{align*}
	\norm{\left[\hat{\bar{\bar{{\bf G}}}}^{[(\uparrow/\uparrow)]}_{z\leftarrow \rho_z}\right]^*\hat{\bar{\bar{{\bf G}}}}_{z\leftarrow \rho_z'\leftarrow \rho_z'' \leftarrow \rho_z}^{(\omega^\pm)[(\uparrow/\uparrow)\blue{(\updownarrow/\updownarrow)}\red{(\updownarrow/\updownarrow)}]}}^{\rm (p)}_{\mathcal{F}}
	&\propto \norm{\hat{\bf g}^{\rm(p)(\uparrow/\uparrow)*}_{z\leftarrow \rho_z}\hat{\bf g}^{\rm(p)(\uparrow/\uparrow)}_{z\leftarrow \rho_z'}\blue{\hat{\bf g}^{\rm(p)(\omega^\pm)(\updownarrow/\updownarrow)}_{\rho_z'\leftarrow \rho_z''}}\red{\hat{\bf g}^{\rm(p)(\updownarrow/\updownarrow)}_{\rho_z''\leftarrow \rho_z}}}_{\mathcal{F}}\\
	&=\norm{\left(\hat{\bf p}_z^{\uparrow}\otimes\hat{\bf p}_{\rho_z}^\uparrow\right)^*\left(\hat{\bf p}_{z}^{\uparrow}\otimes\hat{\bf p}_{\rho_z'}^{\uparrow}\right)\blue{\left(\hat{\bf p}_{\rho_z'}^{(\omega^\pm)\updownarrow}\otimes\hat{\bf p}_{\rho_z''}^{(\omega^\pm)\updownarrow}\right)}\red{\left(\hat{\bf p}_{\rho_z''}^{\updownarrow}\otimes\hat{\bf p}_{\rho_z}^{\updownarrow}\right)}}_{\mathcal{F}}\\
	&=\left(\hat{\bf p}_{\rho_z'}^{\uparrow}\cdot\blue{\hat{\bf p}_{\rho_z'}^{(\omega^\pm)\updownarrow}}\right)\left(\blue{\hat{\bf p}_{\rho_z''}^{(\omega^\pm)\updownarrow}}\cdot\red{\hat{\bf p}_{\rho_z''}^{\updownarrow}}\right)\left(\left[\hat{\bf p}_z^{\uparrow}\right]^*\cdot\hat{\bf p}_{z}^{\uparrow}\right)\left(\left[\hat{\bf p}_{\rho_z}^\uparrow\right]^*\cdot\red{\hat{\bf p}_{\rho_z}^{\updownarrow}}\right).
\end{align*}



{\small
\begin{thebibliography}{90}
\setlength{\itemsep}{1ex}
\label{Sect.Ref}

\bibitem{Zheludev2015} Zheludev, N.~I. Obtaining optical properties on demand. \href{https://doi.org/10.1126/science.aac4360}{{\em Science} \textbf{348,} 973 (2015)}.

\bibitem{Engheta2021} Engheta, N. Metamaterials with high degrees of freedom: space, time, and more. \href{https://doi.org/10.1515/nanoph-2020-0414}{{\em Nanophotonics} \textbf{10,} 639 (2021)}.

\bibitem{Galiffi2022} Galiffi, E. {\em et al.} Photonics of time-varying media. \href{https://doi.org/10.1117/1.AP.4.1.014002}{{\em Adv. Photonics} \textbf{4,} 014002 (2022)}.

\bibitem{Yin2022} Yin, S., Galiffi, E. \& Al\`u, A. Floquet metamaterials. \href{https://doi.org/10.1186/s43593-022-00015-1}{{\em eLight} \textbf{2,} 8 (2022)}.

\bibitem{Yuan2022} Yuan, L. \& Fan, S. Temporal modulation brings metamaterials into new era. \href{https://doi.org/10.1038/s41377-022-00870-0}{{\em Light Sci. Appl.} \textbf{11,} 173 (2022)}.

\bibitem{Li2018} Li, W. \& Fan, S. Nanophotonic control of thermal radiation for energy applications [Invited]. \href{https://doi.org/10.1364/OE.26.015995}{{\em Opt. Express} \textbf{26,} 15995 (2018)}.

\bibitem{Baranov2019} Baranov, D.~G. {\em et al.} Nanophotonic engineering of far-field thermal emitters. \href{https://doi.org/10.1038/s41563-019-0363-y}{{\em Nat. Mater.} \textbf{18,} 920 (2019)}.

\bibitem{Li2021B} Li, Y. {\em et al.} Transforming heat transfer with thermal metamaterials and devices. \href{https://doi.org/10.1038/s41578-021-00283-2}{{\em Nat. Rev. Mater.} \textbf{6,} 488 (2021)}.

\bibitem{Modest} Modest, M.~F. {\em Radiative Heat Transfer} (Elsevier, 2013).

\bibitem{Boriskina2017} Boriskina, S.~V., Zandavi, H., Song, B., Huang, Y. \& Chen, G. Heat is the new light. \href{https://opg.optica.org/opn/abstract.cfm?uri=opn-28-11-26}{{\em Opt. Photonics News} \textbf{28,} 26 (2017)}.

\bibitem{Raman2014} Raman, A.~P., Anoma, M.~A., Zhu, L., Rephaeli, E. \& Fan, S. Passive radiative cooling below ambient air temperature under direct sunlight. \href{https://doi.org/10.1038/nature13883}{{\em Nature} \textbf{515,} 540 (2014)}.

\bibitem{Bierman2016} Bierman, D.~M. {\em et al.} Enhanced photovoltaic energy conversion using thermally based spectral shaping. \href{https://doi.org/10.1038/nenergy.2016.68}{{\em Nat. Energy} \textbf{1,} 16068 (2016)}.

\bibitem{Ilic2016} Ilic, O. {\em et al.} Tailoring high-temperature radiation and the resurrection of the incandescent source. \href{https://doi.org/10.1038/nnano.2015.309}{{\em Nat. Nanotechnol.} \textbf{11,} 320 (2016)}.

\bibitem{Lochbaum2017} Lochbaum, A. {\em et al.} On-chip narrowband thermal emitter for mid-IR optical gas sensing. \href{https://doi.org/10.1021/acsphotonics.6b01025}{{\em ACS Photonics} \textbf{4,} 1371 (2017)}.

\bibitem{Li2012} Li, N. {\em et al.} Colloquium: Phononics: Manipulating heat flow with electronic analogs and beyond. \href{https://doi.org/10.1103/RevModPhys.84.1045}{{\em Rev. Mod. Phys.} \textbf{84,} 1045 (2012)}.

\bibitem{Biehs2016} Biehs, S.-A. \&  Ben-Abdallah, P. Revisiting super-Planckian thermal emission in the far-field regime. \href{https://doi.org/10.1103/PhysRevB.93.165405}{{\em Phys. Rev. B} \textbf{93,} 165405 (2016)}.

\bibitem{Hurtado2018} Fern\'andez-Hurtado, V. {\em et al.} Super-Planckian far-field radiative heat transfer. \href{https://doi.org/10.1103/PhysRevB.97.045408}{{\em Phys. Rev. B} \textbf{97,} 045408 (2018)}.

\bibitem{Hadad2016} Hadad, Y., Soric, J.~C. \& Al\`u, A. Breaking temporal symmetries for emission and absorption. \href{https://doi.org/10.1073/pnas.1517363113}{{\em Proc. Natl Acad. Sci. USA} \textbf{113,} 3471 (2016)}.

\bibitem{Greffet2018} Greffet, J.-J., Bouchon, P., Brucoli, G., \& Marquier, F. Light emission by nonequilibrium bodies: Local Kirchhoff law. \href{https://doi.org/10.1103/PhysRevX.8.021008}{{\em Phys. Rev. X} \textbf{8,} 021008 (2018)}.

\bibitem{Boriskina2016} Boriskina, S.~V. {\em et al.} Heat meets light on the nanoscale. \href{https://doi.org/10.1515/nanoph-2016-0010}{{\em Nanophotonics} \textbf{5,} 134 (2016)}.

\bibitem{Inoue2015} Inoue, T., De Zoysa, M., Asano, T. \& Noda, S. Realization of narrowband thermal emission with optical nanostructures. \href{https://doi.org/10.1364/OPTICA.2.000027}{{\em Optica} \textbf{2,} 27 (2015)}.

\bibitem{Shen2014} Shen, Y. {\em et al.} Optical broadband angular selectivity. \href{https://doi.org/10.1126/science.1249799}{{\em Science} \textbf{343,} 1499 (2014)}.

\bibitem{Ikeda2008} Ikeda, K. {\em et al.} Controlled thermal emission of polarized infrared waves from arrayed plasmon nanocavities. \href{https://doi.org/10.1063/1.2834903}{{\em Appl. Phys. Lett.} \textbf{92,} 021117 (2008)}.

\bibitem{Liu2011} Liu, X. {\em et al.} Taming the blackbody with infrared metamaterials as selective thermal emitters. \href{https://doi.org/10.1103/PhysRevLett.107.045901}{{\em Phys. Rev. Lett.} \textbf{107,} 045901 (2011)}.

\bibitem{Liu2015} Liu, X. \& Zhang, Z. Near-field thermal radiation between metasurfaces. \href{https://doi.org/10.1021/acsphotonics.5b00298}{{\em ACS Photonics} \textbf{2,} 1320 (2015)}.

\bibitem{Luo2004} Luo, C., Narayanaswamy, A., Chen, G. \& Joannopoulos, J.~D. Thermal radiation from photonic crystals: A direct calculation. \href{https://doi.org/10.1103/PhysRevLett.93.213905}{{\em Phys. Rev. Lett.} \textbf{93,} 213905 (2004)}.

\bibitem{Greffet2002} Greffet, J.-J. {\em et al.} Coherent emission of light by thermal sources. \href{https://doi.org/10.1038/416061a}{{\em Nature} \textbf{416,} 61 (2002)}.

\bibitem{Yu2013} Yu, Z. {\em et al.} Enhancing far-field thermal emission with thermal extraction. \href{https://doi.org/10.1038/ncomms2765}{{\em Nat. Commun.} \textbf{4,} 1730 (2013)}.

\bibitem{Cuevas2018} Cuevas, J.~C. \& Garc\'ia-Vidal, F.~J. Radiative heat transfer. \href{https://doi.org/10.1021/acsphotonics.8b01031}{{\em ACS Photonics} \textbf{5,} 3896 (2018)}.

\bibitem{Carminati1999} Carminati, R. \& Greffet, J.-J. Near-field effects in spatial coherence of thermal sources. \href{https://doi.org/10.1103/PhysRevLett.82.1660}{{\em Phys. Rev. Lett.} \textbf{82,} 1660 (1999)}.

\bibitem{Shchegrov2000} Shchegrov, A.~V., Joulain, K., Carminati, R. \& Greffet, J.-J. Near-field spectral effects due to electromagnetic surface excitations. \href{https://doi.org/10.1103/PhysRevLett.85.1548}{{\em Phys. Rev. Lett.} \textbf{85,} 1548 (2000)}.

\bibitem{Joulain2005} Joulain, K., Mulet, J.-P., Marquier, F., Carminati, R. \& Greffet, J.-J. Surface electromagnetic waves thermally excited: Radiative heat transfer, coherence properties and Casimir forces revisited in the near field. \href{https://doi.org/10.1016/j.surfrep.2004.12.002}{{\em Surf. Sci. Rep.} \textbf{57,} 59 (2005)}.

\bibitem{Novotny} Novotny, L. \& Hecht, B. {\em Principles of Nano-Optics} (Cambridge University Press, 2012).

\bibitem{Galiffi2020} Galiffi, E. {\em et al.} Wood anomalies and surface-wave excitation with a time grating. \href{https://doi.org/10.1103/PhysRevLett.125.127403}{{\em Phys. Rev. Lett.} \textbf{125,} 127403 (2020)}.

\bibitem{Solis2021} Sol\'is, D.~M. \& Engheta, N. Functional analysis of the polarization response in linear time-varying media: A generalization of the Kramers-Kronig relations. \href{https://doi.org/10.1103/PhysRevB.103.144303}{{\em Phys. Rev. B} \textbf{103,} 144303 (2021)}.

\bibitem{Buddhiraju2020} Buddhiraju, S., Li, W., \& Fan, S. Photonic refrigeration from time-modulated thermal emission. \href{https://doi.org/10.1103/PhysRevLett.124.077402}{{\em Phys. Rev. Lett.} \textbf{124,} 077402 (2020)}.

\bibitem{Alcazar2021} Fern\'andez-Alc\'azar, L.~J., Kononchuk, R., Li, H., \& Kottos, T. Extreme nonreciprocal near-field thermal radiation via Floquet photonics. \href{https://doi.org/10.1103/PhysRevLett.126.204101}{{\em Phys. Rev. Lett.} \textbf{126,} 204101 (2021)}.

\bibitem{Coppens2017} Coppens, Z.~J. \& Valentine, J.~G. Spatial and temporal modulation of thermal emission. \href{https://doi.org/10.1002/adma.201701275}{{\em Adv. Mater.} \textbf{29,} 1701275 (2017)}.

\bibitem{Gong2021} Gong, T. Corrado, M.~R., Mahbub, A.~R., Shelden, C. \& Munday, J. Recent progress in engineering the Casimir effect -- applications to nanophotonics, nanomechanics, and chemistry. \href{https://doi.org/10.1515/nanoph-2020-0425}{{\em Nanophotonics} \textbf{10,} 523 (2021)}.

\bibitem{Dodonov2020} Dodonov, V. Fifty years of the dynamical Casimir effect. \href{https://doi.org/10.3390/physics2010007}{{\em Physics} \textbf{2,} 67 (2020)}.

\bibitem{Sloan2021} Sloan, J., Rivera, N., Joannopoulos, J.~D. \& Solja\v{c}i\'c, M. Casimir light in dispersive nanophotonics. \href{https://doi.org/10.1103/PhysRevLett.127.053603}{{\em Phys. Rev. Lett.} \textbf{127,} 053603 (2021)}.

\bibitem{Nation2012} Nation, P.~D., Johansson, J.~R., Blencow, M.~P. \& Nori, F. Colloquium: Stimulating uncertainty: Amplifying the quantum vacuum with superconducting circuits. \href{https://doi.org/10.1103/RevModPhys.84.1}{{\em Rev. Mod. Phys.} \textbf{84,}, 1 (2012)}.

\bibitem{Rytov} Rytov, S.~M., Kravtsov, Y.~A. \& Tatarskii, V.~I. {\em Principles of Statistical Radiophysics} (Springer, 1989).

\bibitem{Nyquist1928} Nyquist, H. Thermal agitation of electric charge in conductors. \href{https://doi.org/10.1103/PhysRev.32.110}{{\em Phys. Rev.} \textbf{32,} 110 (1928)}.

\bibitem{Callen1951} Callen, H.~B. \& Welton, T.~A. Irreversibility and generalized noise. \href{https://doi.org/10.1103/PhysRev.83.34}{{\em Phys. Rev.} \textbf{83,} 34 (1951)}.

\bibitem{Kubo1966} Kubo, R. The fluctuation-dissipation theorem. \href{https://doi.org/10.1088/0034-4885/29/1/306}{{\em Rep. Prog. Phys.} \textbf{29,} 255 (1966)}.

\bibitem{Vogel} Vogel, W. \& Welsch, D.-G. {\em Quantum Optics} (Wiley-VCH, 2006).

\bibitem{Scheel2008} Scheel, S. \& Buhmann, S.~Y. Macroscopic quantum electrodynamics -- Concepts and application. \href{http://www.physics.sk/aps/pubs/2008/aps-08-05/aps-08-05.pdf}{{\em Acta Phys. Slovaca} \textbf{58,} 675 (2008)}.

\bibitem{Rivera2020} Rivera, N. \& Kaminer, I. Light-matter interactions with photonic quasiparticles. \href{https://doi.org/10.1038/s42254-020-0224-2}{{\em Nat. Rev. Phys.} \textbf{2,} 538 (2020)}.

\bibitem{SupplementaryInformation} See {\em Supplementary Information} for further details on the time-dependent Hamiltonian~(Sec.~I), the Heisenberg equations of motion of the polaritonic operators~(Sec.~II), the electric field and current density operators~(Sec.~III~and~IV), the correlations~(Sec.~V), and the thermal emission spectra of a planar slabs~(Sec.~VI).

\bibitem{Boyd} Boyd, R.~W. {\em Nonlinear Optics} (Academic Press, 2019).

\bibitem{Loudon} Loudon, R. {\em The Quantum Theory of Light} (Oxford Univ. Press, 2000).

\bibitem{Delga2014} Delga, A., Feist, J., Bravo-Abad, J. \& Garc\'ia-Vidal, F.~J. Quantum emitters near a metal nanoparticle: Strong coupling and quenching. \href{https://doi.org/10.1103/PhysRevLett.112.253601}{{\em Phys. Rev. Lett.} \textbf{112,} 253601 (2014)}.

\bibitem{Liberal2019} Liberal, I., Ederra, I. \& Ziolkowski, R.~W. Control of a quantum emitter's bandwidth by managing its reactive power. \href{https://doi.org/10.1103/PhysRevA.100.023830}{{\em Phys. Rev. A} \textbf{100,} 023830 (2019)}.

\bibitem{Mandel} Mandel, L. \& Wolf, E. {\em Optical Coherence and Quantum Optics} (Cambridge Univ. Press, 1995).

\bibitem{Glauber1963} Glauber, R.~J. The quantum theory of optical coherence. \href{https://doi.org/10.1103/PhysRev.130.2529}{{\em Phys. Rev.} \textbf{130,} 2529 (1963)}.

\bibitem{Mandel1965} Mandel, L. \& Wolf, E. Coherence properties of optical fields. \href{https://doi.org/10.1103/RevModPhys.37.231}{{\em Rev. Mod. Phys.} \textbf{37,} 231 (1965)}.

\bibitem{Pendry2022} Pendry, J.~B., Galiffi, E. \& Huidobro, P.~A. Photon conservation in trans-luminal metamaterials. \href{https://doi.org/10.1364/OPTICA.462488}{{\em Optica} \textbf{9,} 724 (2022)}.

\bibitem{Caloz2022} Caloz, C., Deck-L\'eger, Z.-L., Bahrami, A., C\'espedes, O. \& Li, Z. Generalized space-time engineered modulation (GSTEM) metamaterials. Preprint at \href{https://doi.org/10.48550/arXiv.2207.06539}{https://arxiv.org/abs/2207.06539}.

\bibitem{Hopfield1958} Hopfield, J.~J. Theory of the contribution of excitons to the complex dielectric constant of crystals. \href{https://doi.org/10.1103/PhysRev.112.1555}{{\em Phys. Rev.} \textbf{112,} 1555 (1958)}.

\bibitem{Jacob2010} Jacob, Z. {\em et al.} Engineering photonic density of states using metamaterials. \href{https://doi.org/10.1007/s00340-010-4096-5}{{\em Appl. Phys. B} \textbf{100,} 215 (2010)}.

\bibitem{Mason2011} Mason, J.~A., Smith, S. \&  Wasserman, D. Strong absorption and selective thermal emission from a midinfrared metamaterial, \href{https://doi.org/10.1063/1.3600779}{{\em Appl. Phys. Lett.} \textbf{98,} 241105 (2011)}.

\bibitem{Agarwal1975} Agarwal, G.~S. Quantum electrodynamics in the presence of dielectrics and conductors. I. Electromagnetic-field response functions and black-body fluctuations in finite geometries. \href{https://doi.org/10.1103/PhysRevA.11.230}{{\em Phys. Rev. A} \textbf{11,} 230 (1975)}.

\bibitem{Mandel1966} Mandel, L. Antinormally ordered correlations and quantum counters. \href{https://doi.org/10.1103/PhysRev.152.438}{{\em Phys. Rev.} \textbf{152,} 438 (1966)}.

\bibitem{Pendry1997} Pendry, J.~B. Shearing the vacuum - quantum friction. \href{https://doi.org/10.1088/0953-8984/9/47/001}{{\em J. Phys. Condens. Matter} \textbf{9,} 10301 (1997)}.

\bibitem{Volokitin2011} Volokitin, A.~I. \& Persson, B.~N.~J. Quantum friction. \href{https://doi.org/10.1103/PhysRevLett.106.094502}{{\em Phys. Rev. Lett.} \textbf{106,} 094502 (2011)}.

\bibitem{Intravaia2014} Intravaia, F., Behunin, R.~O. \& Dalvit, D.~A.~R. Quantum friction and fluctuation theorems. \href{https://doi.org/10.1103/PhysRevA.89.050101}{{\em Phys. Rev. A} \textbf{89,} 050101 (2014)}.

\bibitem{Manjavacas2010A} Manjavacas A. \& Garc\'ia de Abajo, F.~J. Thermal and vacuum friction acting on rotating particles. \href{https://doi.org/10.1103/PhysRevA.82.063827}{{\em Phys. Rev. A} \textbf{82,} 063827 (2010)}.

\bibitem{Manjavacas2010B} Manjavacas A. \& Garc\'ia de Abajo, F.~J. Vacuum friction in rotating particles. \href{https://doi.org/10.1103/PhysRevLett.105.113601}{{\em Phys. Rev. Lett.} \textbf{105,} 113601 (2010)}.

\bibitem{Zhao2012} Zhao, R., Manjavacas, A., Garc\'ia de Abajo, F.~J. \& Pendry, J.~B. Rotational quantum friction. \href{https://doi.org/10.1103/PhysRevLett.109.123604}{{\em Phys. Rev. Lett.} \textbf{109,} 123604 (2012)}.

\bibitem{Sipe1987} Sipe, J.~E. New Green-function formalism for surface optics. \href{https://doi.org/10.1364/JOSAB.4.000481}{{\em J. Opt. Soc. Am. B} \textbf{4,} 481 (1987)}.

\bibitem{Caldwell2015} Caldwell, J.~D. {\em et al.} Low-loss, infrared and terahertz nanophotonics using surface phonon polaritons. \href{https://doi.org/10.1515/nanoph-2014-0003}{{\em Nanophotonics} \textbf{4,} 44 (2015)}.

\bibitem{Liberal2018} Liberal, I. \& Engheta, N. Manipulating thermal emission with spatially static fluctuating fields in arbitrarily shaped epsilon-near-zero bodies. \href{https://doi.org/10.1073/pnas.1718264115}{{\em Proc. Natl Acad. Sci. USA} \textbf{115,} 2878 (2018)}.

\bibitem{Liberal2017A} Liberal, I. \& Engheta, N. Near-zero refractive index photonics. \href{https://doi.org/10.1038/nphoton.2017.13}{{\em Nat. Photon.} \textbf{11,} 149 (2017)}.

\bibitem{Liberal2017B} Liberal, I. \& Engheta, N. Zero-index structures as an alternative platform for quantum optics. \href{https://doi.org/10.1073/pnas.1611924114}{{\em Proc. Natl Acad. Sci. USA} \textbf{114,} 822 (2017)}.

\end{thebibliography}}


\clearpage
{\small
	\begin{thebibliography}{90}
		\setlength{\itemsep}{1ex}
		\label{SectSI.Refs}
		
		\bibitem{LoudonSI} Loudon, R. {\em The Quantum Theory of Light} (Oxford Univ. Press, 2000).
		
		\bibitem{BoydSI} Boyd, R.~W. {\em Nonlinear Optics} (Academic Press, 2019).
		
		\bibitem{VogelSI} Vogel, W. \& Welsch, D.-G. {\em Quantum Optics} (Wiley-VCH, 2006).
		
		\bibitem{Delga2014SI} Delga, A., Feist, J., Bravo-Abad, J. \& Garc\'ia-Vidal, F.~J. Quantum emitters near a metal nanoparticle: Strong coupling and quenching. \href{https://doi.org/10.1103/PhysRevLett.112.253601}{{\em Phys. Rev. Lett.} \textbf{112,} 253601 (2014)}.
		
		\bibitem{Liberal2019SI} Liberal, I., Ederra, I. \& Ziolkowski, R.~W. Control of a quantum emitter's bandwidth by managing its reactive power. \href{https://doi.org/10.1103/PhysRevA.100.023830}{{\em Phys. Rev. A} \textbf{100,} 023830 (2019)}.
		
		\bibitem{Scheel2008SI} Scheel, S. \& Buhmann, S.~Y. Macroscopic quantum electrodynamics -- Concepts and application. \href{http://www.physics.sk/aps/pubs/2008/aps-08-05/aps-08-05.pdf}{{\em Acta Phys. Slovaca} \textbf{58,} 675 (2008)}.
		
		\bibitem{Sloan2021SI} Sloan, J., Rivera, N., Joannopoulos, J.~D. \& Solja\v{c}i\'c, M. Casimir light in dispersive nanophotonics. \href{https://doi.org/10.1103/PhysRevLett.127.053603}{{\em Phys. Rev. Lett.} \textbf{127,} 053603 (2021)}.
		
		\bibitem{NovotnySI} Novotny, L. \& Hecht, B. {\em Principles of Nano-Optics} (Cambridge University Press, 2012).
		
		\bibitem{MandelSI} Mandel, L. \& Wolf, E. {\em Optical Coherence and Quantum Optics} (Cambridge Univ. Press, 1995).
		
		\bibitem{footnote01SI} Notice that, even though strictly speaking the time-modulation would break down the linear behavior of the medium, according to the Hamiltonian given in Eq.~\eqref{Eq.I.01}, we are actually assuming a linear response to relate the polarization to the external electric field.
		
		\bibitem{Joulain2005SI} Joulain, K., Mulet, J.-P., Marquier, F., Carminati, R. \& Greffet, J.-J. Surface electromagnetic waves thermally excited: Radiative heat transfer, coherence properties and Casimir forces revisited in the near field. \href{https://doi.org/10.1016/j.surfrep.2004.12.002}{{\em Surf. Sci. Rep.} \textbf{57,} 59 (2005)}.
		
		\bibitem{footnote02SI} It should be noted that, when analyzing the contributions resulting from the time-modulation we will need to deal with some auxiliary coordinates of the same medium associated to different currents; they will be denoted with the same coordinates by adding some tildes.
		
		\bibitem{Sipe1987SI} Sipe, J.~E. New Green-function formalism for surface optics. \href{https://doi.org/10.1364/JOSAB.4.000481}{{\em J. Opt. Soc. Am. B} \textbf{4,} 481 (1987)}.
		
		\bibitem{footnote03SI} The change to cylindrical coordinates in the wave vector is so that $\kappa_x=\kappa_R\cos{\kappa_\varphi}$, $\kappa_y=\kappa_R\sin{\kappa_\varphi}$, and $\kappa_z=\kappa_z$, with $\kappa_x^2+\kappa_y^2=\kappa_R^2$.
		
		\bibitem{Carminati1999SI} Carminati, R. \& Greffet, J.-J. Near-field effects in spatial coherence of thermal sources. \href{https://doi.org/10.1103/PhysRevLett.82.1660}{{\em Phys. Rev. Lett.} \textbf{82,} 1660 (1999)}.
		
		\bibitem{Shchegrov2000SI} Shchegrov, A.~V., Joulain, K., Carminati, R. \& Greffet, J.-J. Near-field spectral effects due to electromagnetic surface excitations. \href{https://doi.org/10.1103/PhysRevLett.85.1548}{{\em Phys. Rev. Lett.} \textbf{85,} 1548 (2000)}.
		
		\bibitem{footnote04SI} Notice that the frequency-shifted longitudinal wave vector can be expressed as $k_{z,i}^\pm=k_0(1\pm\tilde{\Omega})\sqrt{\varepsilon_i(\omega^\pm)}\kappa_{z,i}^\pm$, where, according to Eq.~\eqref{Eq.VI.02}, it can be recast as $k_{z,i}^\pm=k_0\tilde{k}_{z,i}^\pm$ with $\tilde{k}_{z,i}^\pm=(1\pm\tilde{\Omega})\sqrt{\varepsilon_i(\omega^\pm)}\kappa_{z,i}^\pm$, and $\kappa_{z,i}^\pm=\sqrt{1-\kappa_R^2/[(1\pm\tilde{\Omega})^2\varepsilon_i(\omega^\pm)]}$.
\end{thebibliography}}
\end{document}